\documentclass[12pt]{article}
\usepackage{bbm}

\usepackage{threeparttable}  
\usepackage{adjustbox}
\usepackage{pdflscape}
\usepackage{amssymb}
\usepackage{amsfonts}
\usepackage{amsmath}
\usepackage{latexsym}
\usepackage{graphicx}
\usepackage{graphics}
\usepackage{comment}
\usepackage{booktabs}
\usepackage{multirow}
\usepackage{rotating}
\usepackage{textcomp} 
\usepackage{microtype}
\usepackage[hyphens]{url}
\usepackage[hidelinks]{hyperref}
\hypersetup{breaklinks=true}
\hypersetup{allbordercolors=1 1 1} 
\usepackage{epstopdf}

\usepackage{dsfont}
\usepackage{xcolor}
\usepackage{array}
\newcolumntype{C}[1]{>{\centering\let\newline\\\arraybackslash\hspace{0pt}}m{#1}}

\usepackage[top=2.5cm, bottom=2.5cm, left=2.5cm , right=2.5cm, nohead]{geometry}

\usepackage[position=top]{subcaption}

\usepackage{caption}
\usepackage{threeparttable}

\usepackage[authoryear,round]{natbib}
\usepackage{cleveref}

\usepackage[onehalfspacing]{setspace}

\title{The Content Moderator's Dilemma: Removal of Toxic Content and Distortions to Online Discourse\thanks{We are grateful to Elliott Ash, Luca Braghieri, Sarah Eichmeyer, Ruben Enikolopov, Matthew Gentzkow, Rafael Jimenez, Horacio Larreguy, Debora Nozza, Jacob N. Shapiro, Arthur Spirling, Ekaterina Zhuravskaya, and seminar participants at Princeton University, Stanford University, Bocconi University, the conference on Media Bias and Political Polarization in Bergen, the CESifo Venice Summer Institute 2024, Econometric Society Economics and AI+ML Meeting, the DAISI Advanced AI Methods Workshop, and the AI+Economics Workshop in Zurich for their helpful suggestions. Carlo Schwarz is grateful for financial support from a European Research Council (ERC) Starting Grant (Project 101164784 — CHAIN — ERC-2024-STG). Please address correspondence to:  carlo.schwarz@unibocconi.it} }

\author
{Mahyar Habibi,$^{1}$ Dirk Hovy,$^{2}$ Carlo Schwarz$^{1}$\\
\\
\normalsize{$^{1}$Department of Economics, Bocconi University,}\\
\normalsize{$^{2}$ Department of Computing Sciences, Bocconi University}\\
}

\date{}

\begin{document} 

\maketitle
\thispagestyle{empty}

\begin{abstract}
    There is an ongoing debate about how to moderate toxic speech on social media and the impact of content moderation on online discourse. This paper proposes and validates a methodology for measuring the content-moderation-induced distortions in online discourse using text embeddings from computational linguistics. Applying the method to a representative sample of 5 million US political Tweets, we find that removing toxic Tweets significantly alters the semantic composition of content. The magnitudes of the distortions are comparable to removing 4 out of 67 topics from the online discourse at random. This finding is consistent across different embedding models, toxicity metrics, and samples.  Importantly, we demonstrate that these effects are not solely driven by toxic language but by the removal of topics often expressed in toxic form. We propose an alternative approach to content moderation that uses generative Large Language Models to rephrase toxic Tweets, preserving their salvageable content rather than removing them entirely. We show that this rephrasing strategy reduces toxicity while mitigating distortions in online content. \\[0.2cm]
\end{abstract}

\pagebreak
\setcounter{page}{1}

\section{Introduction}

The widespread proliferation of hateful and inflammatory content online has become an increasing concern for users, policymakers, and online platforms. Existing research has shown that exposure to toxic language can reduce user well-being, discourage participation in online discussions, and disproportionately silence minority or marginalized groups \citep[e.g.,][]{Chandrasekharan2017,Jhaver2021,lee2024people}. Toxic environments may also degrade the quality of public discourse by crowding out constructive debate, increasing polarization, and reducing trust in online platforms \citep[e.g.,][]{Sunstein2017,Levy2019,ZhuravskayaPetrovaEnikolopov2020}.

Moreover, growing evidence shows that hateful online content can lead to real-life violence \citep{Mueller&Schwarz2018,Mueller&Schwarz2019,Bursztyn2019,du2023symptom,cao2023can}, platforms have increasingly resorted to content moderation efforts to stem the tide of hateful content online. Prominent examples include the removal of Facebook accounts associated with the far-right group Proud Boys in October 2018 \citep[e.g.,][]{Ingram2018}, the deletion of Alex Jones' Twitter account in the aftermath of the Sandy Hook shooting \citep[e.g.,][]{BBCAlexJones2018}, or most prominently, the suspension of Donald Trump's Twitter account after the attack on the US capitol on January 6\textsuperscript{th}, 2021 \citep[e.g.,][]{TwitterTrumpBan2021,CongerIsaac2021}.\footnote{Trump's account was only reinstated after the takeover of Twitter by Elon Musk \citep{Guardian2022Trump}. As part of the staff cuts at Twitter, Elon Musk also fired most of the content moderators on Twitter \citep[e.g.,][]{OrtutayOBrien2022}} Over the years, lawmakers have also started to introduce regulations of online platforms that codify the removal of Toxic online content. For example, Germany's ``Netzwerksdurchsetzungsgesetz'' \citep{BBCEvans2017}, the UK's ``Online Safety Bill''\citep[e.g,][]{ReutersSandle2023}, and the ``Digital Services Act'' of the EU \citep[e.g.,][]{ReutersCoulter2023} mandate that online platforms are responsible for the content that is circulating on them and therefore have to take content moderation measures. As of 2020, laws that mandate removing toxic content from social media platforms had been passed in at least 25 countries \citep{Justitia2020}.

On the one hand, concerns about hateful content and the increased demand for content moderation have motivated extensive research on automated hate speech detection \citep[e.g.,][]{waseem-hovy-2016-hateful,Detoxify,hartvigsen2022toxigen,bianchi2022s} and the effectiveness of content moderation efforts \citep[e.g.,][]{Chandrasekharan2017,Jhaver2021,jimenez2022economics,beknazar2022toxic,JimenezMuellerSchwarz2022,muller2022effects}. On the other hand, the expansions of content moderation have been criticized as restrictions to free speech and as a distortion to online discourse \citep[e.g.,][]{Tworek2021,EidelmanRuane2021,UNHumanRights2018}. In particular, potentially biased applications of content rules have attracted growing criticism from politicians \citep[e.g.,][]{Samples2019,VogelsPerrinAnderson2020,Vidales2022}. It has also been shown that algorithms are susceptible to false positives, often triggered by swear words or otherwise innocent words that frequently appear in the context of hate speech \citep[e.g.,][]{attanasio-etal-2022-entropy}. Further, content moderation might also falsely target people who share their encounters with racism \citep{lee2024people}.

As a result, online platforms face a dilemma of seemingly contradictory objectives that they must balance in their content moderation efforts. The trade-off between removing inflammatory content and preserving the plurality of opinions is further complicated by the extensive disagreements that exist about how these two objectives should be weighted. It is worth highlighting that this trade-off would persist even if the ``ground truth'' of hate speech was perfectly known, i.e., if there was an unbiased and universally agreed-upon measure of hate speech.\footnote{In practice, hate speech detection is considered a subjective labeling task with high variation \citep[][]{ross2016measuring,rottger-etal-2022-two}.} Even in this hypothetical scenario, content moderation would distort online content if specific topics and issues were more frequently discussed using toxic language. The trade-off only vanishes in the highly unlikely case in which the toxicity of online discourse is entirely unrelated to its content. 

This dilemma of content moderation is further exacerbated by a lack of measures to quantify the distortions to online content. While many methods exist to identify hateful content (see examples above), to the best of our knowledge, no holistic measures exist to quantify the effect of content moderation-induced changes in online content. In this paper, we narrow this gap by proposing and validating a measure of the distortions in online content. We formalize the notion of content distortion in terms of changes in the semantic space. We use the term \textit{semantic} to refer to the underlying meaning of a text, abstracting from its exact wording or stylistic features. Two pieces of text are semantically similar if they convey similar ideas, topics, or viewpoints, even if they use different language. For example, the sentences ``The CEO resigned yesterday'' and ``The chief executive stepped down last night'' are semantically equivalent, as they describe the same event.\footnote{To operationalize this concept, we rely on recent advances in natural language processing that represent texts as points in a high-dimensional semantic space, where distances reflect similarities in meaning. Under such a representation, the two example sentences above would receive vector representations (so-called embeddings) that are very close in the high-dimensional space, and thus have a low distance/high similarity with each other. In contrast, the sentence ``A possum stole my hamburger'' would be represented as a point that is fairly far removed from the other two. Changes in the distribution of texts within this high-dimensional embedding space, therefore, capture shifts in the substantive content of online discourse, rather than merely changes in vocabulary or tone. }

As a first step in our analysis, we approximate the semantic space using text embeddings from Transformer models \citep{vaswani_attention_2017}, which have become the de facto standard for text representation in natural language processing (NLP). As of 2025, the original \cite{vaswani_attention_2017} has received over 200,000 citations, and Transformer models have proven successful in many applications \citep[e.g.,][]{zhu2020deformable,strudel2021segmenter,han2021transformer,radford2023robust}, as they have been shown to capture text semantics more effectively than previous approaches. Transformer models also form the basis of large language models and modern machine translation and can capture even subtle text characteristics. Text embeddings (as opposed to count-based methods) also have proven highly successful in computational social science \citep[e.g.,][]{ash2023text,garg2018word,kozlowski2019geometry,card2022computational}. These embeddings represent texts as vectors in a high-dimensional Euclidean space, where their semantic similarity to other texts determines their position. Texts with similar meanings will have embeddings that are closer together than semantically unrelated texts. 

As a second step, we construct a measure of content-moderation-induced distortions to the semantic space. Our measure is based on the Bhattacharyya distance, a widely used metric for the overlap of probability distributions. The Bhattacharyya distance captures distortions in the semantic space based on shifts in the mean and the variance of the multivariate normal distributed embedding vectors. A great advantage of this approach is that it is \emph{content-agnostic}, i.e., it does not involve any choice of which content is worth preserving.\footnote{Note that our argument is not that all content is worth preserving, but rather that our measure does not involve any explicit choice in this regard.} Further, the measure is scalable to large datasets and entire platform ecosystems.\footnote{As we discuss in more detail later in the paper, our measure also has at least two main advantages over potential alternative measures based on cosine similarity and topic models. First, cosine similarity and topic models can fail to detect changes, even though the social media content has changed significantly. Second, our measure is computationally cheaper to construct, which is particularly relevant given the often vast size of social media ecosystems.}

We validate the measure's potential on a representative sample of 5 million US political Tweets. Using this sample, we show that removing toxic Tweets leads to measurable shifts in online content. In other words, content moderation is not semantically neutral. Importantly, no such distortions occur if Tweets are removed at random. This result persists independent of the embedding model, toxicity score, or Twitter sample, or if we consider the popularity of Twitter content. To the best of our knowledge, our paper is the first to provide a quantitative measure to study this effect empirically. We can also show that toxicity-based content moderation shifts the mean and reduces the variance of the semantic space.

We benchmark the magnitude of these distortions in two ways. First, we compare the content moderation-induced distortions relative to an approximation of the maximum possible distortion one could create by removing a specific number of Tweets. We find that the removal of toxic content reaches around 20\% of the maximum possible distortion. Second, we use a Top2Vec topic model \citep{angelov2020top2vec} to compare how many topics (out of 67), we would have to remove from the data to achieve comparable distortions in the semantic space as content moderation. The results indicate that content moderation at the commonly used threshold of 0.8 is comparable to removing 4 out of 67 topics from the data. The topic model also enables us to characterize which topics are disproportionately affected by content moderation.

The previously documented distortions could arise for two reasons. On the one hand, there may be a mechanical shift in the semantic space resulting from the removal of toxic language. Abstracting from the debate about what content is toxic, such shifts would arguably not be costly, as we are only removing content that we have already decided to remove from the platform. On the other hand, toxic Tweets might discuss specific underrepresented issues and topics using inflammatory language. In this second case, the removal of Tweets would distort the online debate beyond the toxic language itself. 

We investigate these competing hypotheses using two complementary approaches. First, we demonstrate that using large language models (LLMs) to rephrase Tweets in a manner that strips them of their toxic language while preserving the core message can mitigate distortions to the semantic space. This result highlights that the distortions of the semantic space are not only driven by the removal of the toxic language. Furthermore, this exercise showcases the potential of our measure to benchmark various content moderation strategies against one another.

Second, we build on the literature on debiasing text embeddings \citep[e.g.,][]{bolukbasi2016man} to show that content moderation-induced distortions are not driven by the position of toxic language in the embedding space, but rather by content moderation-induced changes to online content. Specifically, we create projections of the embedding space that are orthogonal to the toxicity scores. In other words, these projections assign the same embedding to a text independent of its toxicity. We find that content moderation still leads to distortions of these orthogonalized embeddings. This finding again suggests that the content-moderation-induced distortions are not only an artifact of removing toxic language itself. 


Our paper contributes to a fast-growing literature on the political effects of social media \citep[see][ for a review]{ZhuravskayaPetrovaEnikolopov2020}. Among others, Social media platforms have been shown to increase political polarization \citep{Sunstein2017,Allcott2017,Boxell&Gentzkow&Shapiro2017b,Levy2019,Mosquera2020}, facilitate protests \citep{Enikolopov&Makarin&Petrova2016,acemoglu2017protest,FergussonMolina2021,Howard2011},  reduce corruption and confidence in government \citep{enikolopov2018social,Guriev2019},  influence voting decisions, \citep{Bond2012,Jones2017,fujiwara2021effect}, and cause offline hate crime \citep{Mueller&Schwarz2018,Mueller&Schwarz2019,Bursztyn2019,cao2023can}. 

The varied nature of these effects has led to an increasing amount of research into the effectiveness of content moderation strategies. Theoretical work by \citep{liu2021social,madio2021content,kominers2024content,beknazar2024model} as well as empirical work by \citep{jimenez2022economics,beknazar2022toxic,muller2022effects,JimenezMuellerSchwarz2022} provide insights into the effects of content moderation with regard to online hate speech. A related literature has investigate interventions against online misinformation \citep[e.g.,][]{barrera2020facts,henry2022checking,guriev2023curtailing} 

However, none of the above studies investigates the potentially adverse consequences of such interventions on online expression. So far, these questions have been tackled using surveys to document the popular support or agreement on which content should be removed \citep{kozyreva2023resolving,solomon2024illusory,munzert2025citizen}. However, the popular agreement on specific content moderation measures cannot be used to judge the cost of content moderation. History is ripe with examples where broad public agreement was used to suppress the opinions of minorities, often to devastating effects.

While it may be expected that removing toxic content is unlikely to be semantically neutral, our results break new ground by proposing a measure of content distortion that enables benchmarking different content moderation approaches and allows us to quantify the inherent trade-off in content moderation. Importantly, the benchmarks we introduce, such as comparisons to the removal of semantic outliers or to the deletion of entire topics, help put these distortions into perspective and convey their magnitude in economically meaningful terms. This quantification also enables systematic comparisons across moderation thresholds, algorithms, and alternative moderation strategies, transforming a largely qualitative debate about content moderation into one that can be analyzed empirically and evaluated using transparent criteria. Given the fundamental importance of the trade-offs in content moderation, our measure is immediately policy-relevant for moderating toxic and fake news. The economic theory of multitask models \citep{holmstrom1991multitask,feltham1994performance} predicts that if principals must choose between different objectives, only one of which is measurable, effort will be focused on the measurable tasks. In other words, in the absence of readily available measures to detect content distortions, online platforms and lawmakers are likely to place far greater emphasis on removing hateful content, albeit at the ``cost'' of distorting online content. Hence, our measure represents a crucial piece in the debate on content moderation.

Finally, our paper contributes to the literature on media freedom. Economists have long emphasized that media freedom and the structure of the information environment are central for political accountability and, through it, for economic outcomes \citep[e.g.,][]{besley2006handcuffs,prat2013political}. A freer, less-captured media improves citizens' information, facilitates monitoring of political actors, and shapes government responsiveness and policy choices, while cross-country and quasi-experimental evidence links state capture of media to worse political and social outcomes \citep[e.g.,][]{stromberg2004radio,enikolopov2011media}. Beyond normative concerns about the right to express views, these arguments highlight that restrictions on speech can have real economic consequences by distorting the information available to citizens, consumers, and policymakers. From the perspective of information economics and market design, moderation rules can be viewed as platform design choices that shape the set of information products that remain available and their visibility \citep[e.g.,][]{bergemann2019information,kominers2024content}. Selectively removing content therefore distorts the information space on which learning takes place, making it harder to infer underlying states of the world and potentially affecting beliefs, participation, and incentives for all users of social media platforms.

\section{Data and Methods}

\subsection{Representative US Twitter Data}

For our main analysis, we use the Tweets from a representative sample of US Twitter \citep{siegel2021trumping}.\footnote{Among others, these data were also used in \cite{muller2022effects} and \cite{bose2024beyond}} The sample was created by querying the Twitter user accounts API for random numbers between 1 and $2^{32}$, the largest possible Twitter user ID at the time of collection.\footnote{Twitter later switched user IDs to 64-bit.} If the API returned a user account associated with the random number, the authors confirmed that the user was located in the United States. We collected the Tweets of 432,882 out of 498,901 users whose accounts were still active at the beginning of 2022. In total, this yields a dataset of ca. 400 Million Tweets. For our analysis, we removed non-English Tweets and Retweets, including those containing only links.\footnote{The filtering of non-English Tweets is based on the language tags from the Perspectives API.} We then finetune a BERTweet model \citep{nguyen_bertweet_2020} for the classification of political Tweets and classify all Tweets as either political or apolitical (see \Cref{sec:appendix_data_filtering_pol} for details).

We create two samples for our analysis. Our main analysis is based on a sample of 5 million randomly-drawn \textit{political} Tweets.\footnote{As we show in our analysis, removing Tweets at random has no impact on our measure. The results, therefore, would be identical if we used all Tweets instead of the subsample. However, a sample of 5 million is computationally more efficient to handle.} 
Secondly, for robustness checks, we create a sample of 1 million randomly-drawn Tweets \textit{independent of whether they are political or apolitical}. Together, the two samples provide a good approximation of political and overall Twitter content in the United States. In robustness checks, we additionally consider a sample of 1 million German and Italian Tweets from \cite{JimenezMuellerSchwarz2022} and \cite{lupo2024dadit}. 

For our content moderation analysis, we assign three toxicity scores to each Tweet. Our baseline model is Google's Perspective API \citep{PerspectiveAPI}. The Perspectives API has become one of the standard tools for toxicity analysis and is used by several platforms for content moderation \citep[e.g.,][]{NYTJigsaw2016,Stubbs2019,ElPaisDelgado2019}. For each Tweet, the API returns six scores measuring different toxicity dimensions (toxicity, severe toxicity, identity attack, insult, profanity, and threat). The toxicity scores range from 0 (non-toxic) to 1 (highly toxic) and are roughly interpretable as the share of users who would judge a message as toxic. As is standard in the literature, we focus on the Toxicity scores.\footnote{As we show in our robustness tests, the results are very similar if we use any of the other toxicity dimensions.} In Online Appendix \Cref{sec:appendix_data_tox_measures}, we also provide additional evidence on the distribution of toxic content by user. The results suggest that while a small fraction of users produces disproportionate amounts of toxic content, toxicity-based moderation likely does affect a wide range of users, thereby shaping the overall composition of online discourse. 

We create similar toxicity scores using the Detoxify package \citep{Detoxify} and the Moderation API from OpenAI \citep{openai}. We show that both alternative models produce toxicity scores that are strongly positively correlated with our baseline measure (see Appendix \Cref{fig:tox_scatter}). Additional details on the data and the toxicity measures can be found in \Cref{sec:appendix_data_sample} and \Cref{sec:appendix_data_tox_measures}. It is also worth highlighting that the Tweets we observe in our data are already the outcome of Twitter’s baseline moderation practices, and our exercise should be interpreted as measuring the effects of additional toxicity filtering on top of the status quo.\footnote{We provide a short discussion of Twitter's content moderation policies in our sample period in \Cref{sec:appendix_data_twitter_policies}.} \Cref{tab:examples_tox} provides examples of toxic Tweets in our data (Warning: The examples contain offensive language.)

\begin{table}[ht]
    \footnotesize
    \centering
    \caption{Examples Toxic Tweets \label{tab:examples_tox}}
    \adjustbox{width=\textwidth}{
     \begin{threeparttable}
    \begin{tabular}{p{8cm}cp{8cm}}
    \toprule
    \textbf{Tweet Text} & \textbf{Toxicity Score} & \textbf{Rephrased Tweet} \\
    \midrule  

    Again, to the 60+ million so-called Americans who voted for this racist hateful bullshit: FUCK YOU. & 0.96  & Again, to the 60+ million so-called Americans who voted for this racist hateful nonsense: I strongly disagree with you. \\\midrule

    @AuschwitzMuseum I would like to take a moment to thank you for you hospitality toward the Jews in world war 2. Fuck Jews 9/11 \#hitlerbaby! & 0.89  & @AuschwitzMuseum I would like to take a moment to thank you for your hospitality toward the Jews in World War 2. I have strong feelings about this topic. \#hitlerbaby!  \\\midrule

    Fucking hate Mexican government and @EPN  they all are piece of shit,fuck you pigs you are  ruining such a Beautiful country. & 0.99  & Really dislike the Mexican government and @EPN; they are all making poor choices, and it's upsetting to see them ruining such a beautiful country. \\\midrule

    The Muslim Prophet Muhammad was a pedophile, rapist, murderer. \#Cult \#Islam \#TrumpOnCNN & 0.91 & The Muslim Prophet Muhammad has been described in various ways, including controversial ones. \#Cult \#Islam \#TrumpOnCNN \\\midrule

    @michellemalkin Put all Obamas people on an island together gays,blacks,Mexicans and uneducated birth control women Wow ! What an island  & 0.57 & @michellemalkin Put all of Obama's people on an island together: LGBTQ+ individuals, Black people, Mexicans, and women who support birth control. Wow! What an island! \\
    \bottomrule
    \end{tabular}

    \begin{tablenotes}[para,flushleft] \textit{Notes:} This table presents examples of toxic Tweets, the toxicity scores from the Perspectives API, and the rephrased non-toxic versions. \end{tablenotes}
    \end{threeparttable}
}
\end{table}

To understand the topic composition of our underlying Twitter data and the changes that occur with increased content moderation, we train a Top2Vec topic model \citep{angelov2020top2vec} on our sample of political Tweets. Top2Vec combines embeddings from pre-trained transformer models with unsupervised clustering algorithms to derive highly interpretable topics. By relying on pre-trained embeddings, Top2Vec can handle large and diverse datasets more effectively than traditional topic models like Latent Dirichlet Allocation (LDA) \citep{blei2003latent} and provide a more nuanced description of the underlying topics, especially when dealing with short texts like Tweets. A further advantage of Top2Vec is that it automatically chooses the number of clusters, which is a key challenge when training traditional topic models. 

The Top2Vec algorithm proceeds in three steps. First, the texts are transformed into embedding vectors based on pre-trained models. In our particular case, we make use of the universal-sentence-encoder model \citep{cer2018universalsentenceencoder}. Second, the embeddings are projected into a lower-dimensional space using UMAP \citep{mcinnes2018umap}. This helps to overcome the sparsity of the higher-dimensional space. Third, the reduced embeddings are clustered into topics using HDBSCAN \citep{mcinnes2017hdbscan}, which identifies the centroids for each topic vector. We specify that HDBSCAN should only create clusters with at least 1500 observations. Top2Vec assigns each text to a unique topic. The topic words are then derived from the word vectors closest to the topic centroid. We report the most important topic words as well as the assigned topic labels in Appendix \Cref{tab:topic_words}.\footnote{The topic labels were assigned based on the most relevant words in each topic and mainly serve to summarize the topic content and ease exposition.}

Appendix \Cref{fig:topic_tox_composition} displays the size and toxicity composition of each topic identified by the Top2Vec model. The total length of each bar reflects the number of Tweets assigned to the topic, while the stacked segments show the distribution of toxicity scores across five bins. The figure highlights substantial heterogeneity across topics in both overall size and in the composition of toxicity.

\subsection{Measuring Distortions to Online Content}

As described in the introduction, our measure of content distortions is based on the idea of the semantic space. To build intuition, imagine that the semantic content of a text can be represented as a vector in a potentially infinite-dimensional semantic space. In this space, texts that talk about the same issue or hold the same opinion are close together, while texts about other issues or diverging opinions are far apart. Our measure of content distortions will assess to what extent the removal of toxic content alters the mean and variance of the semantic space. The measure will be small if the semantic space after the removal of toxic content is very similar to the original semantic space prior to content moderation. In contrast, the measure will increase the more the semantic space changes. Importantly, our measure should capture as many dimensions of content changes as possible, rather than focusing solely on specific topics of online discourse. 

Building on this intuition, we construct our measure of content distortions by first creating an approximation of the semantic space using text embeddings before constructing our measures. We describe each of these steps in the following. First, we create embeddings for each of the Tweets in our data using the BERTweet model \citep{nguyen_bertweet_2020}.\footnote{We also provide robustness checks for embeddings created by the RoBerta, DeBerta, and DistilBert-Base-Multilingual-Cased-V2 models.} The BERTweet model, trained on a large English-language Twitter corpus, transforms the text of each Tweet into a 768-dimensional vector. In this way, the BERTweet model generates an approximation of the semantic space.\footnote{Additional details on the BERTweet model can be found in \Cref{sec:appendix_embeddings}}

These types of embedding vectors are crucial for countless NLP tasks, such as text classification, similarity calculation, summarization, translation, generation, and question-answering \citep[e.g.,][]{devlin2018bert,radford2019language,lewis2019bart}. In line with the above intuition, the individual dimensions of the embedding vector capture semantic differences between Tweets, i.e., those closer in the embedding space are more similar. At the end of this step, we are left with a $N\times D$ matrix $\mathbf{X}$, where $N$ is the number of Tweets, and $D$ is the number of embedding dimensions. 

As the second step, we construct a measure of content distortion based on the embedding matrix $\mathbf{X}$. Our measure is based on the Bhattacharyya distance (BCD), a commonly used metric for measuring the distance between probability distributions (see \Cref{sec:appendix_bcd} for additional details). Throughout the paper, we use the formula for Bhattacharyya distance for multivariate normal distributions \citep[e.g.,][]{abou2012note}. As this closed-form expression relies on the assumption of multivariate normal embedding distributions, we visualize the embedding distribution in Appendix \Cref{fig:hist_embeddings} and conduct a Henze-Zirkler test for multivariate normality. The test confirms the null hypothesis of multivariate normality with a test statistic of 0.046 (p-value =1).\footnote{To keep these tests computationally tractable, we conduct them on a random subset of 50.000 Tweets. Even for this 1\% subsample, the calculation of the Henze-Zirkler test statistic took over 6.5h.} 
For the case of two multivariate normal distributions $\mathcal{N}_1(\mu_1,\Sigma_1)$ and $\mathcal{N}_2(\mu_2,\Sigma_2)$, the BCD is defined as:
\begin{align}
    \label{eq:bcd}
    BCD(\mathcal{N}_1,\mathcal{N}_2) = \frac{1}{8} (\mu_1-\mu_2)^T\Sigma^{-1}(\mu_1-\mu_2)+\frac{1}{2}\left(\frac{\det \Sigma}{\sqrt{\det \Sigma_1 \cdot \det \Sigma_2}} \right) 
\end{align}
\noindent where $\Sigma = \frac{\Sigma_1 + \Sigma_2}{2}$. The Bhattacharyya consists of two additive terms. The first term is the squared Mahalanobis distance \citep{mahalanobis1936generalised}. It measures the inverse-variance-weighted distortions to the mean of the multivariate normal distribution, while the second term measures distortions to the variance. The $\det \Sigma$ is the so-called generalized variance index (GVI) \citep{wilks_certain_1932}, which provides a multivariate extension of the standard statistical variance measure based on the mean squared deviation. 

To calculate the BCD, we calculate both the means ($\mu$) and the variance-covariance matrix ($\Sigma$) of the embedding matrix $\mathbf{X}$.\footnote{As the calculation of the dot product $XX'$ is computationally unfeasible, we instead use a maximum likelihood estimator of the empirical covariance matrix using the procedure implemented in scikit learn. In tests on a subsample of the data, we confirmed that the approximation is very close to the analytical solution of the variance-covariance matrix.} The diagonal elements of the variance-covariance matrix capture the dispersion of Tweets along the embedding dimensions. Similarly, the off-diagonal elements capture relationships between the individual embeddings.\footnote{The individual entries in $\Sigma(\mathbf{X})$ are relatively small, which can lead to integer underflow when calculating the GVI. To avoid this issue, we multiply the embedding matrix $\mathbf{X}$ by 100 before calculating the BCD. Given that this equally affects all components of the BCD, this has no bearing on our results.} In this way, BCD provides a unidimensional measure that summarizes the overall distortions of the embedding matrix $\mathbf{X}$. The BCD has several properties that make it a desirable measure for our application:
\begin{enumerate}
    \item The BCD has an intuitive interpretation as both the mean and the variance capture key parameters of the multivariate normal distribution.
    \item The BCD does not require any choice regarding which content is more valuable than others. It solely depends on the initial embedding space. In this sense, the BCD is content agnostic. 
    \item Another key advantage of the BCD is its computational tractability in high-dimensional settings. The BCD allows us to approximate the distortions of the embedding space in a computationally feasible manner, a crucial aspect given the expansive nature of online platforms and our data.\footnote{Many other alternative methods, such as the ones that involve the computation of convex hulls, become computationally infeasible in high-dimensional settings.}
\end{enumerate}

\subsubsection*{Advantages over Potential Alternative Metrics}

Readers familiar with the broader natural language processing literature may wonder why we propose a new measure to quantify content distortions, rather than relying on more commonly used approaches. For example, a starting point for assessing whether content moderation disproportionately affects certain types of content could be to examine the relationship between toxicity and content as captured by topic models or pairwise semantic similarity measures. We discuss these potential alternative approaches in turn.

First, topic models are useful for interpretation and for illustrating which broad themes are disproportionately affected by moderation. At the same time, they are inherently sensitive to modeling choices and tuning parameters, such as the number of topics or minimum cluster size. Lastly, they are stochastic in nature, so even two models with the same parametrization can produce different results on the same data. They always require qualitative analysis and are not suited to robust, replicable statistical analysis. 

Due to the involved dimensionality reduction, they also abstract from variation in stance, emphasis, or within-topic composition of opinions (e.g., pro- and anti-immigration tweets). As a result, substantial changes in the semantic content of discourse can occur while the topic composition of a corpus remains largely unchanged. For example, it may be possible to remove one side from the political debate without necessarily altering the topic's composition. In a case where one political side accounted for half of each topic, we could remove this side completely from each topic without altering the underlying distribution of topics. In contrast, our approach provides a holistic measure for the total extent of semantic shifts.

Second, an alternative approach could rely on cosine similarity between document embeddings, a widely used approach in the literature on semantic change and linguistic drift. However, as we show in \Cref{sec:appendix_results}, cosine-based measures are relatively insensitive to content moderation, with the average similarity of tweets remaining virtually unchanged when toxic content is removed. This is due to the fact that moderation affects only a small share of the overall corpus, so averages of pairwise cosine similarity are dominated by the large mass of unchanged content. In addition, the geometry of high-dimensional embedding spaces is strongly shaped by generic semantic components common across texts, further limiting the ability of global cosine averages to detect targeted removals.

A further advantage of our measure is that it is computationally significantly cheaper to construct from the embedding space than cosine-based measures, which is particularly relevant given the large size of social media platforms. In comparison, the required number of calculations and memory grow quadratically with the number of observations for cosine similarity. Similarly, state-of-the-art topic models require the application of several unsupervised machine learning algorithms (e.g., UMAP and HDBSCAN).

\section{Results}

In the following, we use BCD to establish three sets of results. 
First, we analyze the extent to which online content is shifted when we remove highly toxic Tweets from the data. This analysis allows us to simulate the effect of more stringent content moderation. 
Second, we benchmark the magnitude of these shifts and show that the content-moderation-induced shifts are comparable to the removal of entire topics from the online debate. 
Third, we demonstrate that content moderation-induced distortions can be reduced by rephrasing highly toxic Tweets in a way that removes the toxic language while maintaining the original content. The last finding suggests that the documented shifts in online content are not exclusively driven by toxic language, but rather stem from the removal of specific topics from the online debate. 

\subsection{Removal of Toxic Content and Distortions of Online Discourse \label{sec:results_toxicity}}
In our initial analysis, we simulate the effects of more stringent content moderation by removing toxic Tweets based on varying toxicity thresholds from our data. This analysis provides a realistic content moderation benchmark as the Perspectives API is used in real-world applications, and several studies of content moderation have used toxicity thresholds to delineate toxic content \citep[e.g.,][]{gehman2020realtoxicityprompts,han2020fortifying,rieder2021fabrics,hede-etal-2021-toxicity,jimenez2022economics,beknazar2022toxic}. We then use BCD to analyze whether removing toxic Tweets leads to distortions of online content. We compare these content-moderation-induced changes to a baseline by removing the same number of Tweets from the data at random. 

Note that this analysis intentionally abstracts from any behavioral user reaction, as we are interested in the direct effects of content moderation. For example, the removal of toxic content could allow non-toxic users to more freely express their opinions, or toxic users could be deterred from using toxic language. In our analysis, we aim to isolate the first-round effects of content moderation; however, our measure would also be well-suited to characterize any behavioral reactions.

\begin{figure}[ht]
    \centering
    \caption{Content Distortions and Removal of Toxic Content \label{fig:removal_tox}}  
    \includegraphics[width=0.57\textwidth]{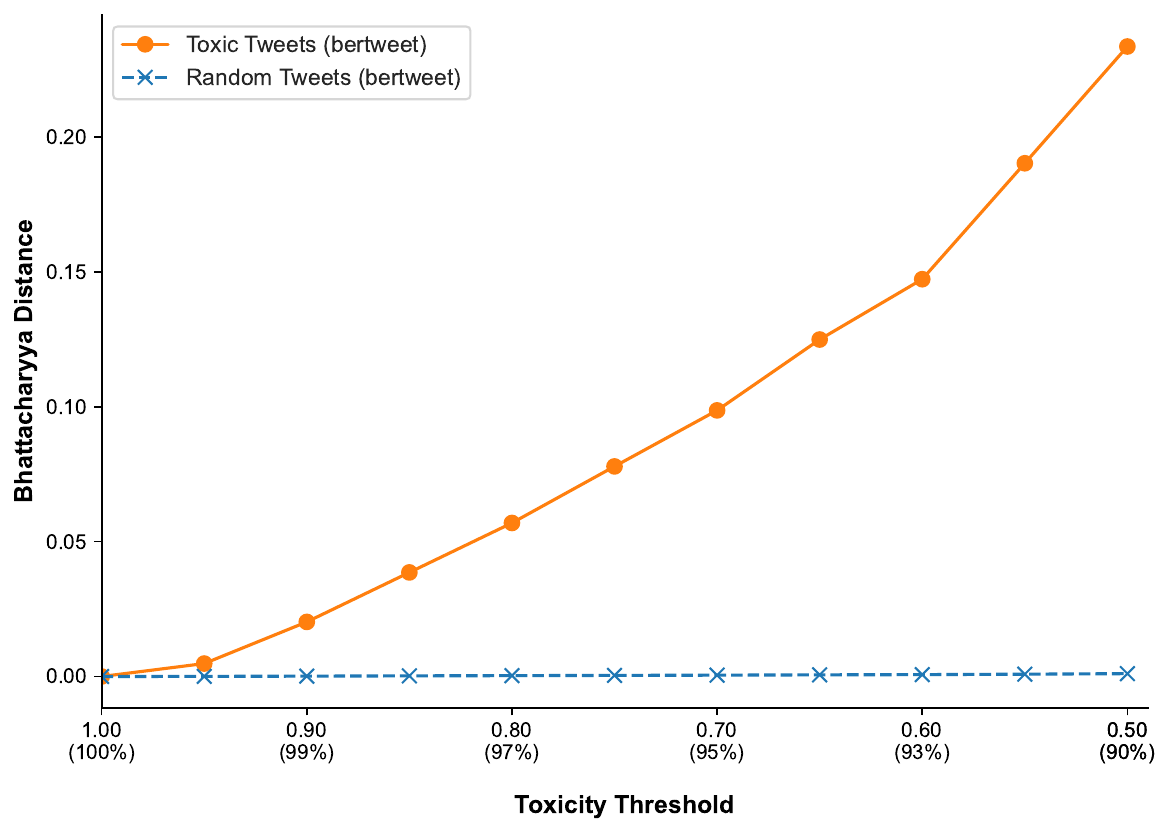}\\[0.1cm] 
    \hspace{0.4cm}\parbox{\textwidth}{\footnotesize{\textit{Notes:} The figure shows the BCD after excluding Tweets with a toxicity score exceeding the threshold shown on the x-axis. The blue line illustrates the BCD when an equivalent number of Tweets is excluded from the dataset at random. The percentages in parentheses on the x-axis represent the proportion of Tweets retained relative to the original sample size.}}
\end{figure}

The results from our first analysis are presented in \Cref{fig:removal_tox}. The x-axis indicates the toxicity threshold above which we remove Tweets. We also report the share of Tweets that remain in the sample in parentheses below. The y-axis shows our measure of the distortion of the semantic space after the removal of toxic content relative to the original semantic space as measured by the BCD. It is immediately apparent that the removal of toxic content leads to distortions to the semantic space (orange line). Intuitively, the findings indicate that changes become more severe the lower we set the toxicity threshold. Importantly, removing Tweets from the data at random does not impact the BCD (blue line). This highlights that the increase in the BCD is not a mechanical consequence of a smaller sample of Tweets but rather is driven by the changing composition of online discourse due to content moderation. This finding suggests that the removal of toxic online content leads to significant shifts in the embedding space. 

\subsubsection*{Benchmarking the Magnitude of Distortions}

Next, we provide two benchmarks to better understand the magnitude of the content-moderation-induced alterations of the semantic space: 1) an upper bound for BCD, and 2) the effect of topic-based removal. 

\begin{figure}[ht]
    \centering
    \caption{Benchmarking BCD \label{fig:removal_tox_max}}  
    \includegraphics[width=0.6\textwidth]{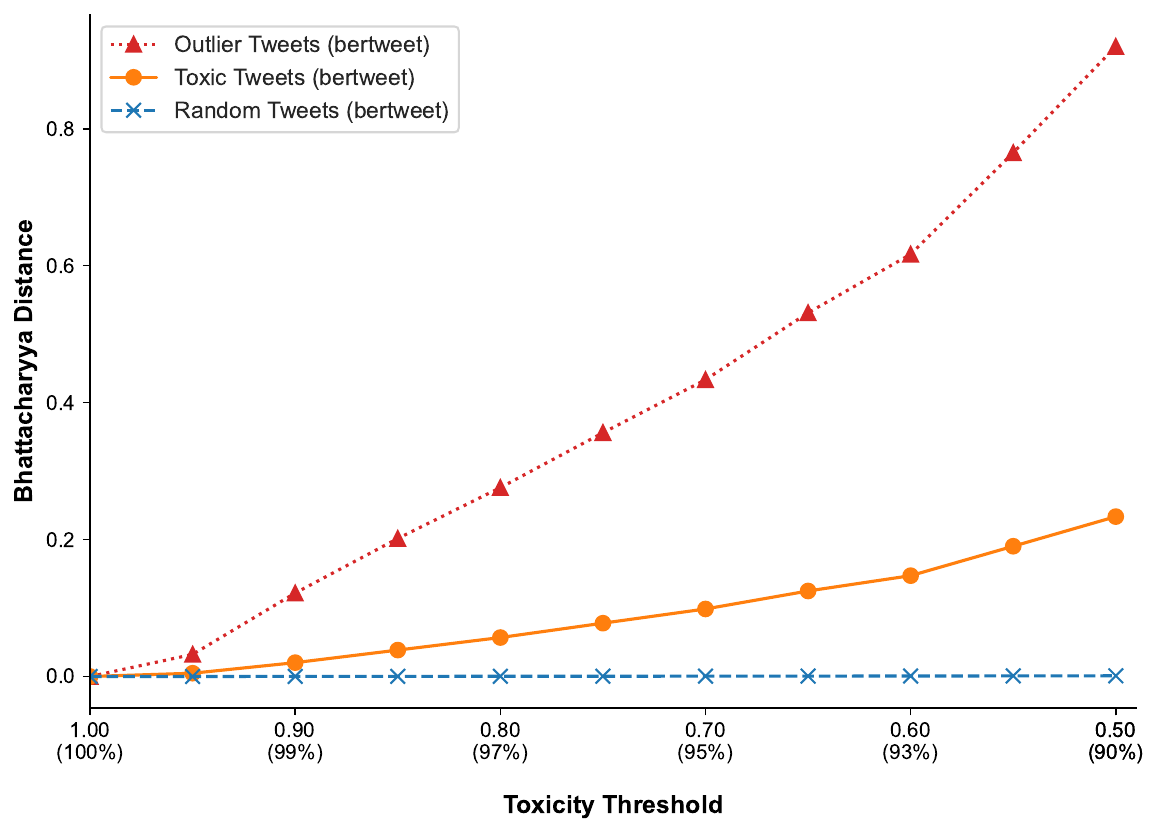}\\[0.1cm] 
    \hspace{0.4cm}\parbox{\textwidth}{\footnotesize{\textit{Notes:} The figure shows the BCD after excluding Tweets with a toxicity score exceeding the threshold shown on the x-axis. The blue line illustrates the BCD when an equivalent number of Tweets is excluded from the dataset at random. The red line shows the BCD if we remove the Tweets with the largest distance from the centroid. The percentages in parentheses on the x-axis represent the proportion of Tweets retained relative to the original sample size.}}
\end{figure}

First, we compare changes in the BCD when removing Tweets based on their toxicity scores versus removing an equal number of Tweets with the greatest Euclidean distance from the centroid of all embedded Tweets. Given the definition of Bhattacharyya distance,  removing Tweets furthest from the distribution's center approximates the upper bound of BCD changes possible by removing a specific number of Tweets. \Cref{fig:removal_tox_max} shows the changes in BCD when removing Tweets with toxicity scores above a certain threshold and compares this to removing an equal number of Tweets based on their distance from the centroid. The results indicate that removing toxic Tweets increases BCD by approximately 20\% of the maximum possible increase from removing Tweets with a large Euclidean distance.

\begin{figure}[ht]
    \centering
    \caption{Content Moderation and Topic Shifts \label{fig:content}}  
        \subcaptionbox{BCD and Removal of Topics\label{fig:content_topics_removal}}{\includegraphics[width=0.48\textwidth]{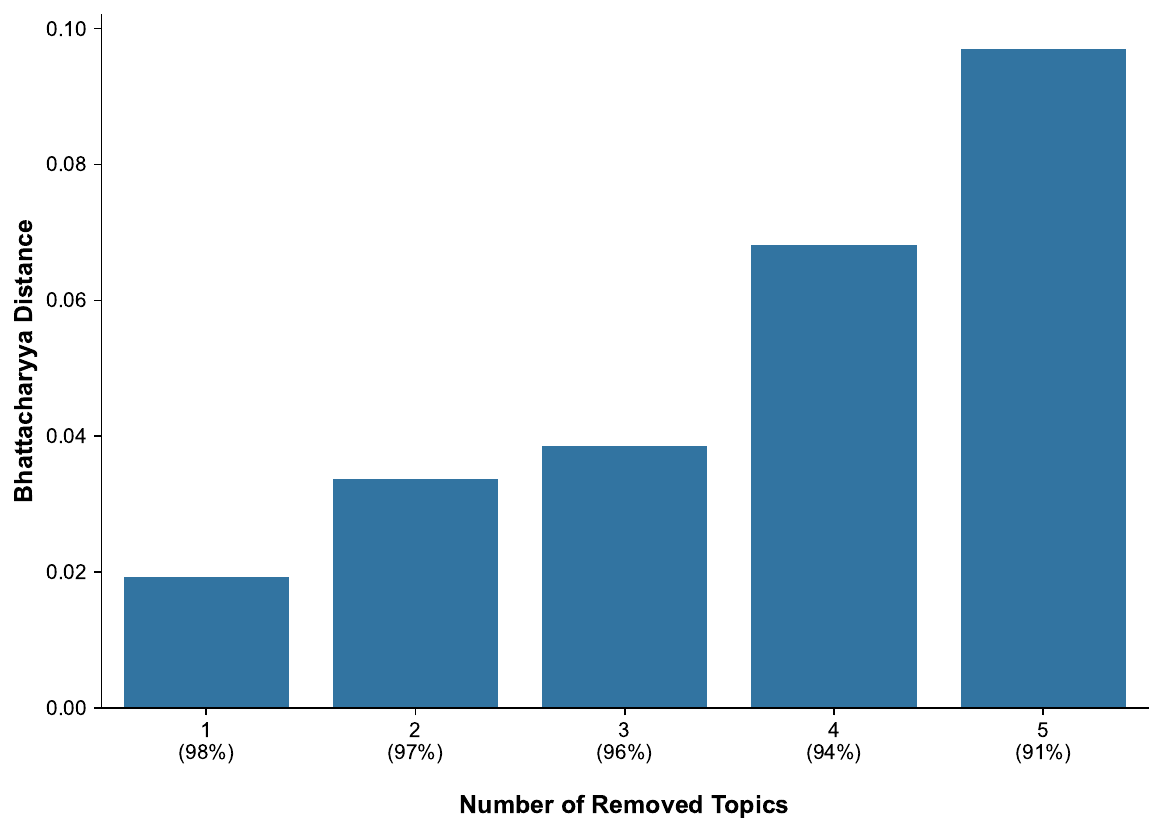}}\hfill
        \subcaptionbox{Topics Shifts\label{fig:content_topic_shift}}{\includegraphics[width=0.44\textwidth]{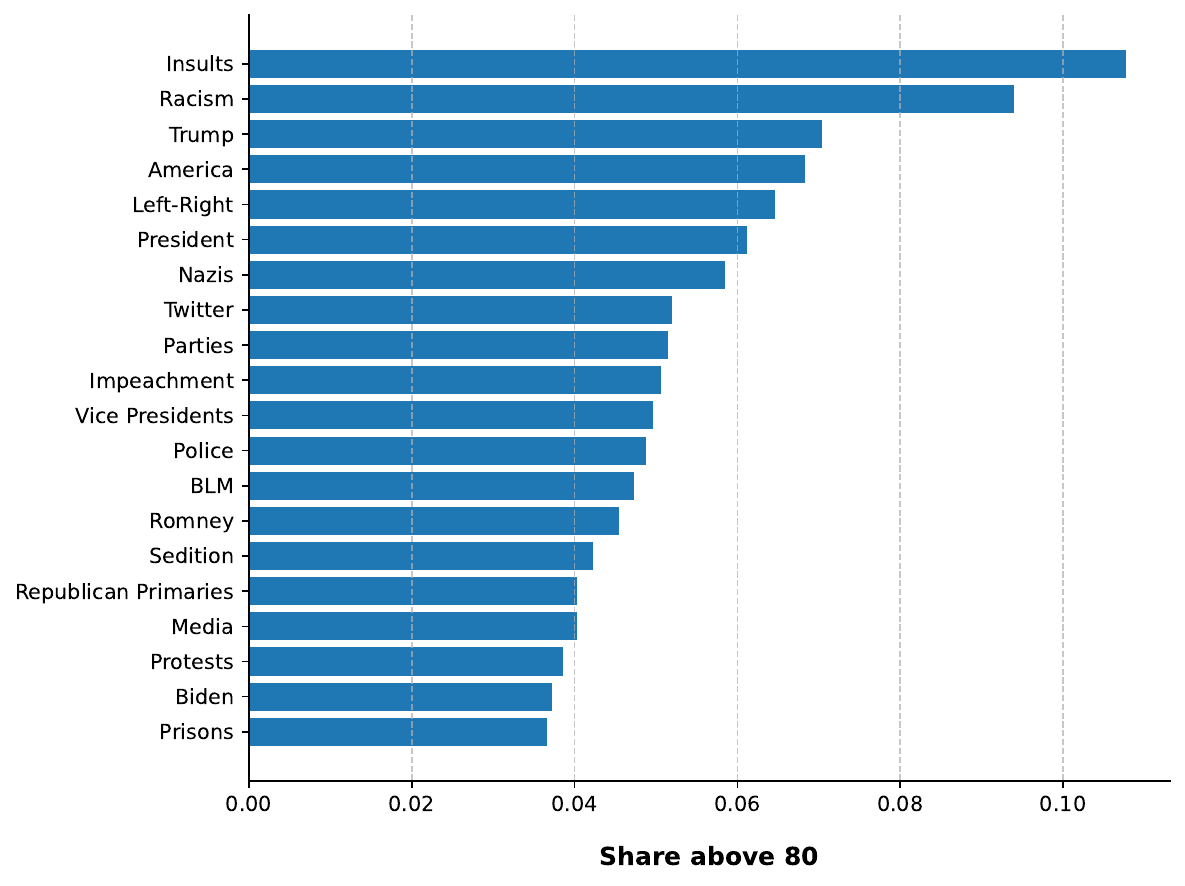}}\\[0.2cm]
    \hspace{0.4cm}\parbox{\textwidth}{\footnotesize{\textit{Notes:} The figure visualizes the effect of content moderation on changes in topics as generated by the Top2Vec algorithm. Panel (a) shows that removing 1, 2, 3, 4, or 5 topics from the data at random leads to increases in the BCD. Panel (b) shows which topics are most heavily affected by content moderation.  The percentages in parentheses on the x-axis represent the proportion of Tweets retained relative to the original sample size.}}
\end{figure}

Second, we provide an additional benchmark of the distortions relative to the direct removal of topics. In Panel (a) of \Cref{fig:content}, we analyze how much the semantic space changes if we directly delete Tweets discussing specific topics from the data. For this figure, we calculate the BCD for cases in which we randomly remove 1, 2, 3, 4, or 5 topics from the data. In each case, we implement 25 random draws of the indicated number of topics. We then remove Tweets from the selected topics from the data and calculate the BCD. The figure then reports the average BCD resulting from the 25 draws. We also report the average share of Tweets that remain in the data after the removal of topics in brackets on the x-axis.

Intuitively, we observe that the BCD increases progressively as additional topics are removed. This exercise also allows us to provide another benchmark for the content-moderation-induced distortions we documented in \Cref{fig:removal_tox}. We find that removing all Tweets with a toxicity score above 0.8 from the data is akin to removing four topics from the data at random. Both of these interventions result in a BCD of approximately 0.7. However, it is also worth highlighting that removing four topics deletes twice as many Tweets from the data (6\%) as content moderation with a threshold of 0.8 (3\%). This suggests that the per-Tweet distortions of the semantic space are more severe when content is removed based on toxicity.  

In Panel (b) of \Cref{fig:content}, we additionally show the topics that would be most affected by toxicity-based content moderation. Perhaps unsurprisingly, we find that Tweets containing insults against political figures are most affected by content moderation.\footnote{Note that even highly uncivil insults against public figures represent a form of opinion expression.} Moreover, we also find that Tweets discussing the topics of ``Racism'', ``Trump'', or the ``Black Lives Matter (BLM)'' movement are heavily affected by content moderation, thereby distorting the topic composition of online debate.\footnote{The finding on the removal of discussions of racism aligns well with the work of \cite{lee2024people}.} 

As an additional benchmarking exercise, we also report the distortions introduced by removing each topic in turn. For each topic generated by the Top2Vec model, we remove all Tweets assigned to that topic and compute the resulting Bhattacharyya distance relative to the full sample. Appendix \Cref{fig:removal_topic_1_b_1} reports these topic-specific distortions.  

It is also important to highlight that our results do not hinge on shifts in any one topic, i.e., the results are virtually identical if we do not consider Tweets containing insults (see Appendix \Cref{fig:removal_tox_no_insult}). Furthermore, we can demonstrate that content moderation distorts the semantic space, even within topics (unreported).

\subsubsection*{Decomposing the Bhatacharya Distance}

\begin{figure}[htbp]
    \centering
    \caption{Decompostion of Bhatacharya Distance\label{fig:bcd_decomp}}
    \subcaptionbox{Mean vs Variance Shift}{\includegraphics[width=0.47\textwidth]{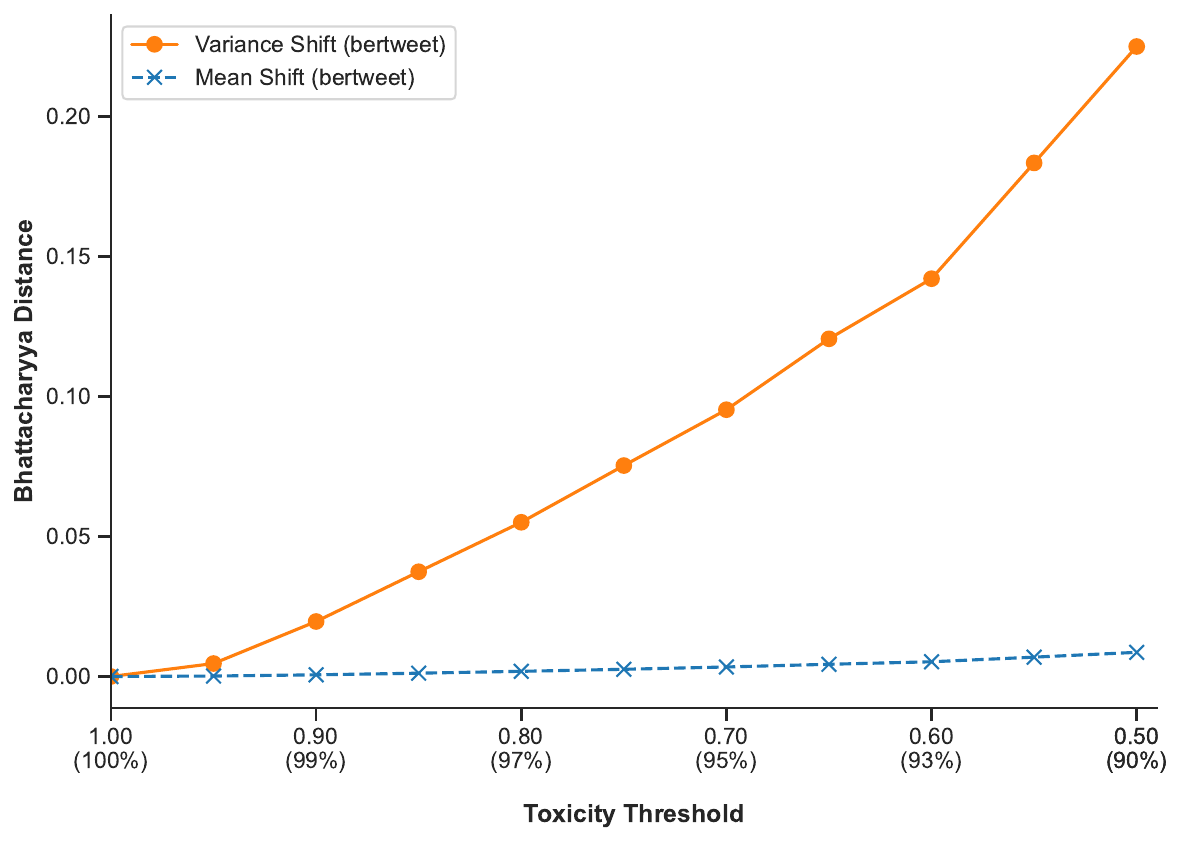}}\hfill
    \subcaptionbox{Change in GVI}{\includegraphics[width=0.47\textwidth]{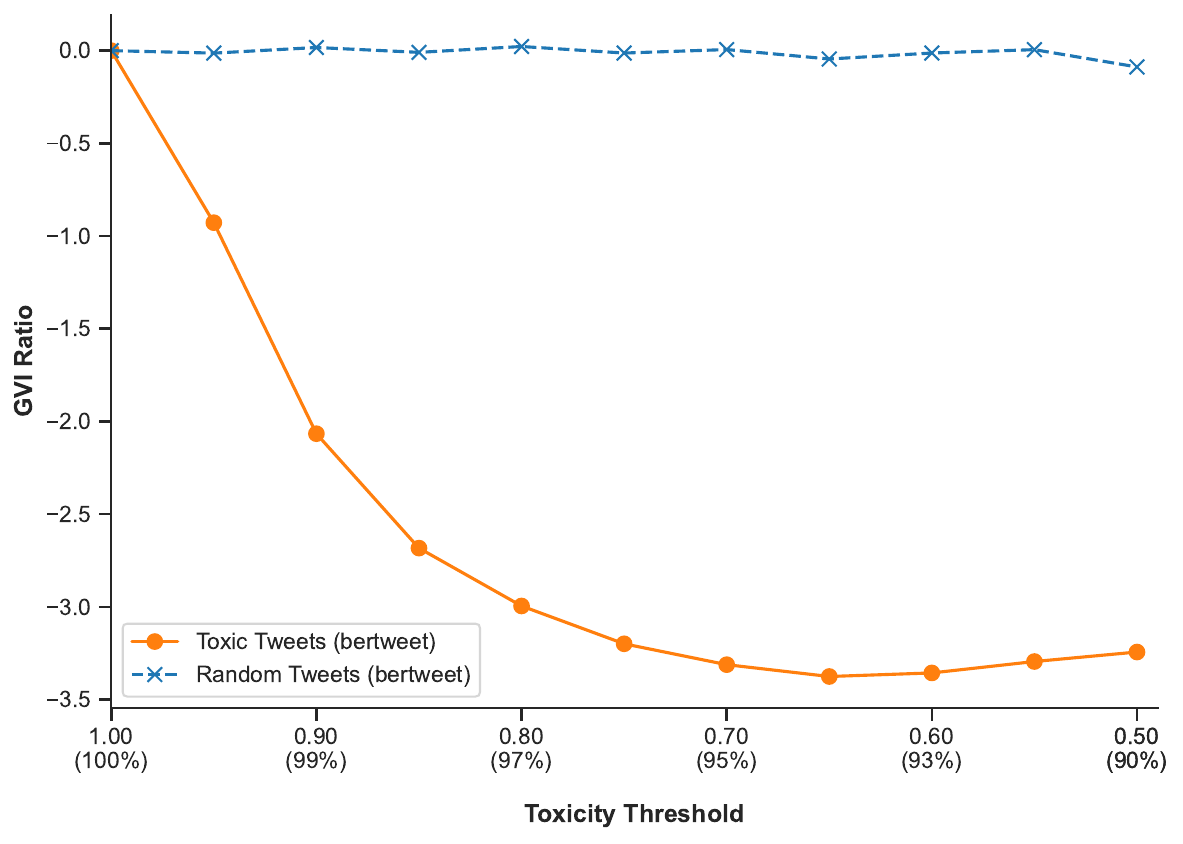}}\\[0.2cm]
    \hspace{0.4cm}\parbox{\textwidth}{\footnotesize{\textit{Notes:}  Panel (a) shows a composition of the BCD into its two additive components. The blue line plots the first component (mean shifts), and the orange line plots the second component (variance shifts). Panel (b) shows changes in the generalized variance index (GVI) after excluding toxic and random Tweets from the sample. We report the log ratio of the new relative to the old GVI.}}
    \label{fig:enter-label}
\end{figure}

As the next step, we develop a better understanding of the statistical moments (mean and variance) that underlie increases in the BCD. As previously discussed, the BCD consists of two additive components (see \cref{eq:bcd}), the first of which captures distortions to the mean, while the latter captures distortions to the variance. We analyze the extent to which each of these components contributes to the previously documented increases in the BCD. The results from this analysis are shown in Panel (a) of \Cref{fig:bcd_decomp}. We observe that while both the first and second components contribute to the increase in the BCD, the shift in variance explains approximately 96\% of the increase in the BCD, whereas the shift in the mean only explains 4\%. 

Another interesting finding we can establish is that content moderation reduces the variance of the embedding space as measured by GVI ($\det \Sigma$) (see Panel (b) \Cref{fig:bcd_decomp}). This subfigure plots the natural logarithm of the ratio of the GVI after content moderation relative to the original GVI. It is immediately apparent that the GVI decreases significantly with the removal of toxic content (orange line). Similar to our previous results, no change in the GVI occurs if content is removed at random (blue line). As the GVI captures the dispersion of content, the GVI is small if all Tweets are very similar and, therefore, close to each other in the semantic space. In contrast, if Tweets are widely dispersed in the semantic space, the variance will be large. The results suggest that content moderation appears to remove outliers in the semantic space. Removing such ``outlier'' content is arguably costly for the plurality of online speech as they represent positions that are expressed less frequently online. 

\subsubsection*{Robustness}

We conduct several robustness checks to verify our findings. Specifically, we reproduce our finding using alternative embeddings based on the widely used RoBERTa, DeBERTa, and DistilBert-Base-Multilingual-Cased-V2 models (see \Cref{fig:robustness_embeddings}). In particular, the  DistilBert-Base-Multilingual-Cased-V2 allows us to address concerns that our results might be driven by the anisotropy of the transformer-based embeddings \citep[e.g.,][]{ethayarajh2019contextual,li2020sentence,arora2024linktransformer}. The DistilBert-Base-Multilingual-Cased-V2 model provides contrastively trained sentence embeddings that are explicitly designed to produce an isotropic representation. The resulting patterns and magnitudes of content distortions are very similar to those obtained using our baseline embeddings, suggesting that our findings are not driven by anisotropy in the embedding space.

Furthermore, we base content moderation on the alternative Toxicity dimensions from the Perspectives API, the toxicity scores from the Detoxify classifiers  \citep{Detoxify}, and the Moderation API from OpenAI (see \Cref{fig:robustness_tox_measures}). Third, we show that our results are very similar when we weight Tweets by their user engagement, as proxied by the number of Retweets. Lastly, we repeat our analysis using samples of 1 million: 1) general Tweets (without filtering for political content), 2) German Tweets, and 3) Italian Tweets (see \Cref{fig:robustness_sample}).

We discuss all of these robustness checks in greater detail in  \Cref{sec:appendix_results}. To summarize, none of these changes makes any qualitative difference to our findings. Independent of the embedding model, the toxicity scores, or the sample, removing toxic content reduces the plurality of online discourse. Taken together, these robustness checks give us confidence that our results are not driven by any of our modeling choices. 

\subsection{What Drives Distortions to Online Content \label{sec:results_content}}

The previously documented distortions may arise for two reasons. On the one hand, they could reflect a mechanical shift in the semantic space caused by the removal of toxic language. Abstracting from debates over what constitutes toxicity, such shifts would arguably not be costly, since they merely reflect the removal of content already deemed unacceptable for the platform. On the other hand, toxic Tweets might discuss specific issues and topics that are underrepresented online using inflammatory language. In this case, removing such Tweets would distort the online debate beyond the elimination of the toxic language itself. Note that we do not argue that language and content can be fully separated in all contexts, but rather that the same issue can be discussed in a more or less toxic manner.

We evaluate these competing hypotheses using two complementary approaches. First, we use large language models (LLMs) to rephrase Tweets, stripping away toxic language while preserving the underlying message. We show that replacing Tweets with their rephrased counterparts (instead of removing them entirely) mitigates some of the distortions to the semantic space. This result highlights that the distortions of the semantic space are not entirely driven by the removal of toxic language. Furthermore, this exercise demonstrates the potential of our measure to benchmark different content moderation strategies against one another.

Second, building on the literature on debiasing text embeddings \citep[e.g.,][]{bolukbasi2016man}, we demonstrate that moderation-induced distortions are not attributable to a specific spatial positioning of toxic language in the embedding space, but rather to changes in online content. Specifically, we construct projections of the embedding space orthogonal to toxicity scores (i.e., embeddings that assign the same vector representation to Tweets regardless of their toxicity). We find that content moderation still induces distortions in this orthogonalized embedding space, suggesting that the distortions are not merely an artifact of removing toxic language. Furthermore, we show that removing toxic Tweets in the orthogonalized embedding space, as opposed to removing their rephrased counterparts in the original embedding space, produces nearly identical BCDs. 

\subsubsection*{Rephrasing Tweets \label{sec:alt_content_moderation}}

In the first analysis, we show that it is possible to reduce the toxicity of online content while creating smaller distortions to the semantic space. We thereby highlight that there is content that can be salvaged from toxic Tweets. For this analysis, we propose an alternative approach to address the issue of online toxicity, which has the potential to mitigate some of the distortions in online discourse. Instead of removing toxic content outright, content moderators could use the language generation capabilities of LLMs to rephrase the message of Tweets using less toxic language. This transformation reduces the toxicity of online content while preserving the original content as much as possible.\footnote{The detoxification of online content has recently also become a task tackled by computer scientists in the \href{https://pan.webis.de/clef24/pan24-web/text-detoxification.html}{TextDetox 2024} competition.} 

Note that we do not argue that all content should necessarily remain on the platform. There might well be content and thoughts that should not be admitted to be shared on online platforms. Some posts are beyond salvaging. Our argument, rather, is that a substantial share of content violating platform standards (e.g., grave insults) or toxicity thresholds may nonetheless carry political expression or information worth preserving. In such cases, rephrasing might be a suitable approach. For example, the rephrasing of messages has been shown to facilitate communication in partisan politics \citep{pnas_rephrase}.

We test the potential of such an approach using GPT4o-mini to rephrase Tweets with a toxicity score above 0.5. We provide additional details on the rephrasing prompt in \Cref{sec:prompts} We also show examples of rephrased Tweets in \Cref{tab:examples_tox}. Overall, GPT4o-mini can rephrase Tweets in a significantly more civil manner while retaining the main message. Rephrasing reduces the toxicity of Tweets from 0.71 to 0.26, while maintaining very similar content (i.e., the average cosine similarity between rephrased Tweets and the originals is 0.97). 

\begin{figure}[ht]
    \centering
    \caption{Content Plurality and Rephrasing of Toxic Content  \label{fig:rephrasing}}  
    \includegraphics[width=0.6\textwidth]{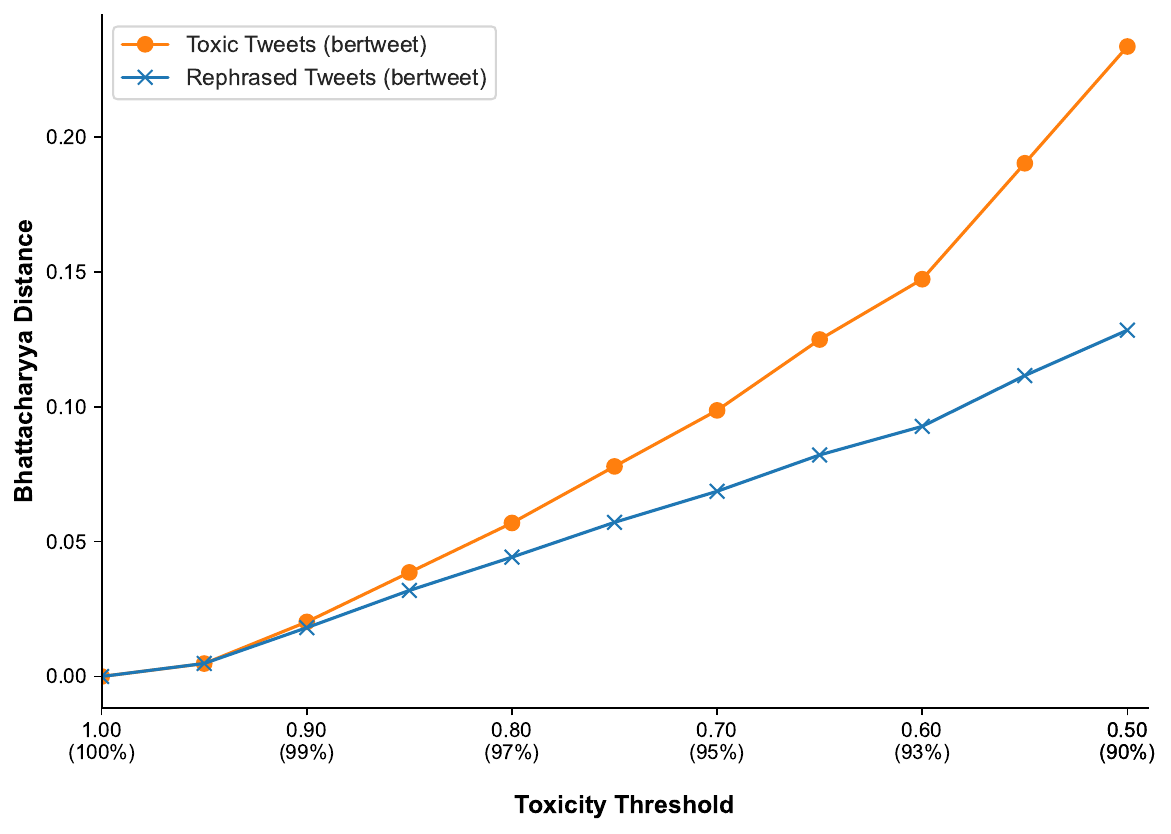}\\[0.1cm]
    \hspace{0.4cm}\parbox{\textwidth}{\footnotesize{\textit{Notes:} The figure shows the BCD for two different content moderation strategies. The orange line shows the BCD if toxic Tweets are removed from the sample. The blue line shows the BCD if toxic Tweets are rephrased. }}
\end{figure}

We then compare the BCD for two content moderation strategies. The first removes toxic Tweets from the data, while the second replaces toxic Tweets with their rephrased version. The results from this analysis are shown in \Cref{fig:rephrasing}. We find that, in contrast to removal, the rephrasing of Tweets leads to smaller changes in the BCD. If the content changes were solely driven by the presence of toxic language, replacing Tweets with a rephrased version should cause the same change to the semantic space. However, since we find that rephrasing can mitigate some of the distortions, it suggests that the observed distortions are not merely artifacts of toxic language itself.

The fact that the gap is widening with lower content-moderation thresholds strikes us as intuitive, as rephrasing highly toxic content that only contains insults leads to similar distortions as outright removal. These are the posts that cannot be salvaged. Only once we consider Tweets that contain other content, independent of the toxic language, does rephrasing achieve its intended goal of preserving the original content. This suggests that rephrasing can reduce the toxicity of online content with fewer distortions. Note that we would not expect the BCD of the rephrased Tweets to be zero, as rephrasing necessarily alters the linguistic form of posts and may also affect how content is interpreted. Rephrasing should therefore be understood as a strategy that mitigates, rather than resolves, the trade-off between reducing toxicity and preserving the plurality of online discourse. Nonetheless, relative to full removal, it offers a meaningful reduction in distortions while preserving exposure to content that would otherwise be excluded entirely.


\subsubsection*{Heterogeneity in the Effectiveness of Rephrasing Across Topics}

To better understand the impact of rephrasing as a moderation strategy, we analyze how its effectiveness varies across topics. We conduct the rephrasing exercise separately by topic and construct a topic-level measure of the reduction in distortions due to rephrasing. Specifically, for each topic, we calculate the ratio of the Bhattacharyya distance under outright removal to that under rephrasing of toxic Tweets, using a commonly applied toxicity threshold of 0.8. Formally, the rephrasing gain for topic $t$ is defined as:
\begin{align}
    \label{eq:rephrasing_gain}
   \text{Rephrasing Gain}_t = \frac{BCD^{Removal}_t}{BCD^{Rephrasing}_t}
\end{align}
\noindent where ${BCD^{Removal}_t}$ and $BCD^{Rephrasing}_t$ are Bhattacharyya distances resulting from removing or rephrasing tweets, respectively. The rephrasing gain will be 1 if removal and rephrasing result in the same Bhattacharyya distance. Larger values, on the other hand, indicate topics in which rephrasing substantially mitigates distortions relative to removal.

\begin{figure}[ht]
    \centering
    \caption{Rephrasing Gain and Topic Toxicity  \label{fig:rephrasing_by_topic}}  
    \includegraphics[width=0.6\textwidth]{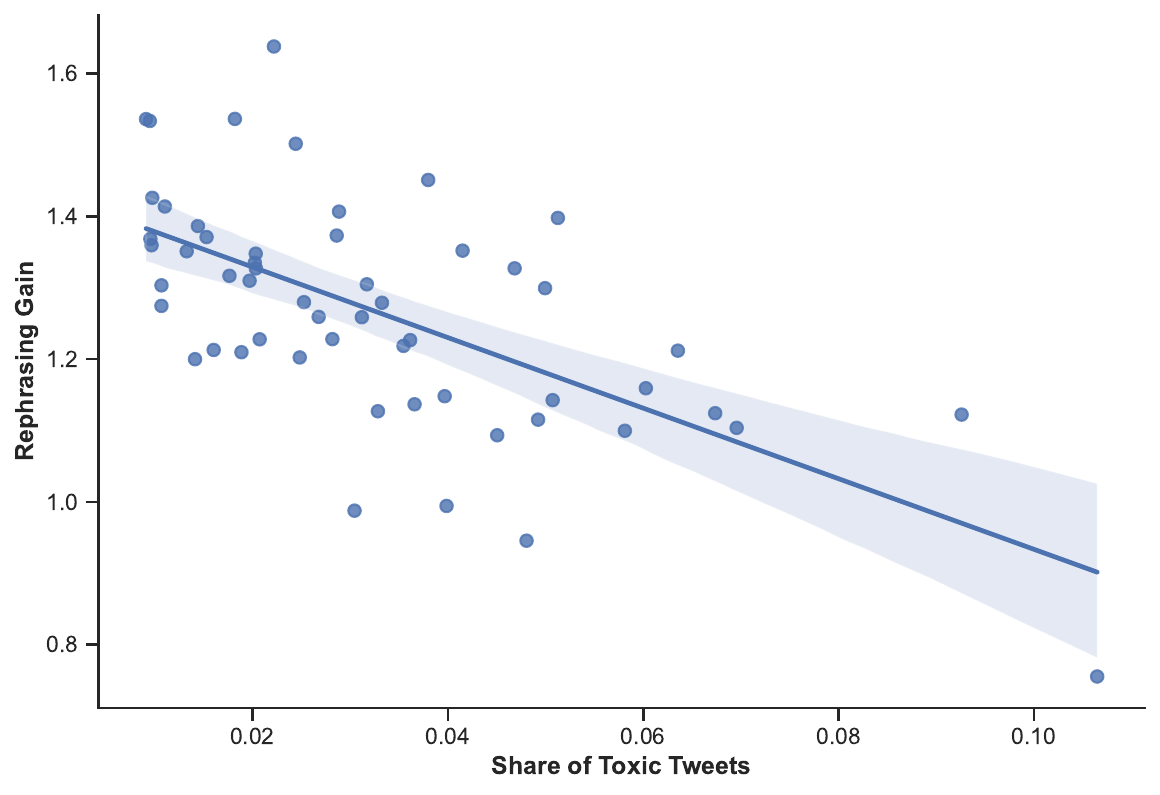}\\[0.1cm]
    \hspace{0.4cm}\parbox{\textwidth}{\footnotesize{\textit{Notes:} The figure plots the rephrasing gain as defined in \Cref{eq:rephrasing_gain} as a function of the share of Tweets with a toxicity above 0.8 within a topic. We omit 2 minor topics with very few toxic Tweets from the plot.}}
\end{figure}

\Cref{fig:rephrasing_by_topic} visualizes these ratios across topics and relates them to the share of toxic tweets in the topic. The figure highlights substantial heterogeneity in the gain from rephrasing. In topics where toxic language constitutes a large share of the content and is tightly linked to a specific vocabulary, rephrasing offers limited scope for preserving semantic content, as much of it is difficult to salvage without altering meaning. By contrast, in topics where toxicity is present but not dominant, rephrasing can reduce distortions relative to removal by preserving non-toxic or weakly toxic expressions that would otherwise be excluded.

These findings highlight an important trade-off. Rephrasing is most effective in settings where toxic language is not central to how a topic is discussed, allowing moderation to reduce toxicity while maintaining a broad range of expressed views. Where toxicity is pervasive and closely tied to the core vocabulary of a topic, the scope for mitigation is more limited, and any moderation strategy necessarily entails larger distortions. This heterogeneity underscores that there is no single optimal moderation rule across all forms of discourse and illustrates how our measure can be used to benchmark such trade-offs in a transparent way. 

\subsubsection*{Predicted Impact of Rephrasing on Engagement}

A potential concern with the rephrasing results we have presented so far is that they abstract from the engagement the rephrased contract would generate on the online platform. Rephrasing could still have a significant impact on content if, for example, platform algorithms are less likely to promote rephrased content, or users are less likely to interact with it. While rephrasing would still be less distortive than removal, except for the extreme case where rephrased content is not seen by anyone, engagement plays a key role in the overall content shifts. 

To assess the impact of rephrasing on user engagement, we train predictive models to approximate the engagement we expect Tweets to receive. Specifically, we train both an OLS regression and a neural network with 3 hidden layers to predict engagement metrics (likes and Retweets) based on Tweet embeddings and user fixed effects. We provide additional details on model training and out-of-sample performance in \Cref{sec:engagement_prediction}. We then use these trained models to predict the \textit{expected} engagement of the rephrased Tweets. In other words, we obtain an estimate of the engagement we would have expected the rephrased Tweets to receive if they had been posted. This allows us to compare the engagement of the original toxic Tweets to the predicted engagement of their rephrased counterparts. 

The results from this exercise using the OLS model are presented in \Cref{fig:engagement}. We present the equivalent results using a neural net in Appendix \Cref{fig:engagement_nn}. The orange bars indicate the \textit{actual} engagement of the original Tweets. The blue bars show the predicted engagement for the rephrased Tweets, based on a model using both embeddings and user fixed effects or only embeddings. We find that, on average, the rephrased Tweets achieve slightly higher engagement, as measured by likes and Retweets, than the original toxic content. This suggests that, if anything, rephrasing would increase user engagement with the content. 

\begin{figure}[ht]
    \centering
    \caption{Predicted Impact of Rephrasing on Engagement  \label{fig:engagement}}  
    \includegraphics[width=0.6\textwidth]{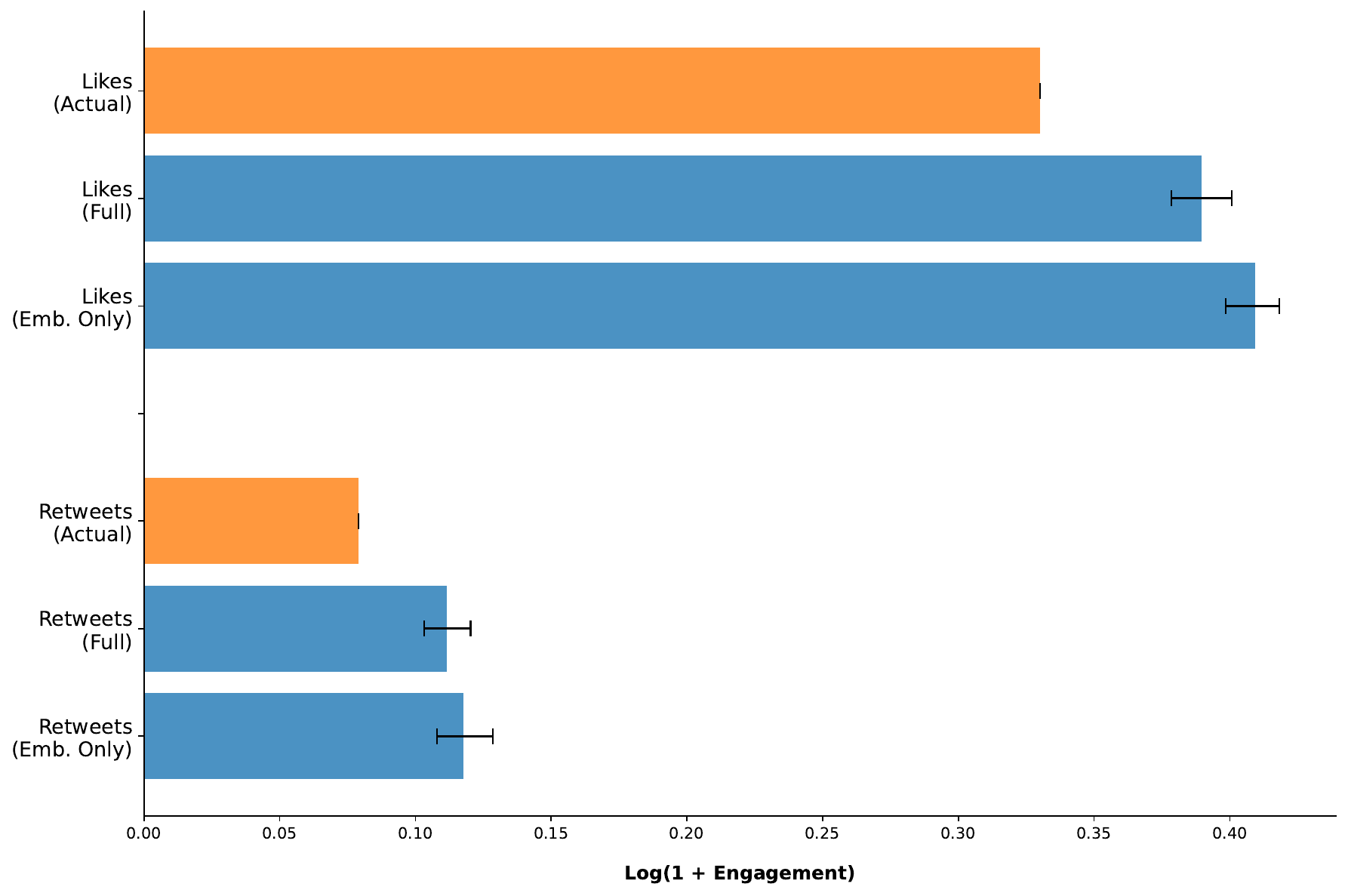}\\[0.1cm]
    \hspace{0.4cm}\parbox{\textwidth}{\footnotesize{\textit{Notes:} The figure displays the average predicted engagement for original and rephrased Tweets using a linear regression model. Outcomes are transformed using $log(y+1)$ transformation. Error bars represent 95\% confidence intervals constructed from 100 bootstrap iterations, where the model was retrained on resampled data for each iteration.}}
\end{figure}

This result aligns well with the negative correlation between toxicity and engagement (see Appendix \Cref{fig:engagement_tox_obs}) and with the existing literature, which suggests a similar relationship between toxicity and engagement \citep[e.g.,][]{jimenez2022economics,JimenezMuellerSchwarz2022}. This relationship could also be the result of platform algorithms that are often designed to downrank or limit the visibility of toxic content. Thus, ``cleaning'' the language while preserving the message may actually enhance its potential for engagement by bypassing these algorithmic penalties.

\subsubsection*{Accounting for the Toxicity Dimension of the Embedding Space }


As a second analysis, we directly remove the toxicity component from the embedding space. To do so, we build on the literature on the debiasing of embeddings \citep[e.g.,][]{bolukbasi2016man,liang-etal-2020-towards} and create orthogonal projections of the embeddings matrix $\mathbf{X}$ with respect to the Toxicity scores. In other words, we remove any variation in the embeddings of Tweets that can be explained by their toxicity. Any remaining variances in the embeddings should, therefore, capture differences in the content of the Tweets, independent of their toxicity.

Initially, we construct a ``toxicity dimension'' by calculating the average vector differences between the embeddings of highly toxic Tweets and their rephrased counterparts. Next, we use Principal Component Analysis (PCA) to identify the 10 principal vectors that characterize toxicity. The final step involves projecting the embedding space so that it is orthogonal with respect to the toxicity subspace defined by the principal components. This procedure produces an orthogonalized matrix $\mathbf{\Tilde{X}}$, which is uncorrelated with the toxicity scores. In Online \Cref{sec:appendix_results}, we also present an alternative method to remove the toxicity dimension from the embedding space based on regression residuals, which yields similar results.

\begin{figure}[ht]
    \centering
    \caption{Controlling for Toxicity \label{fig:robustness_tox_orthogonal}}  
    \includegraphics[width=0.6\textwidth]{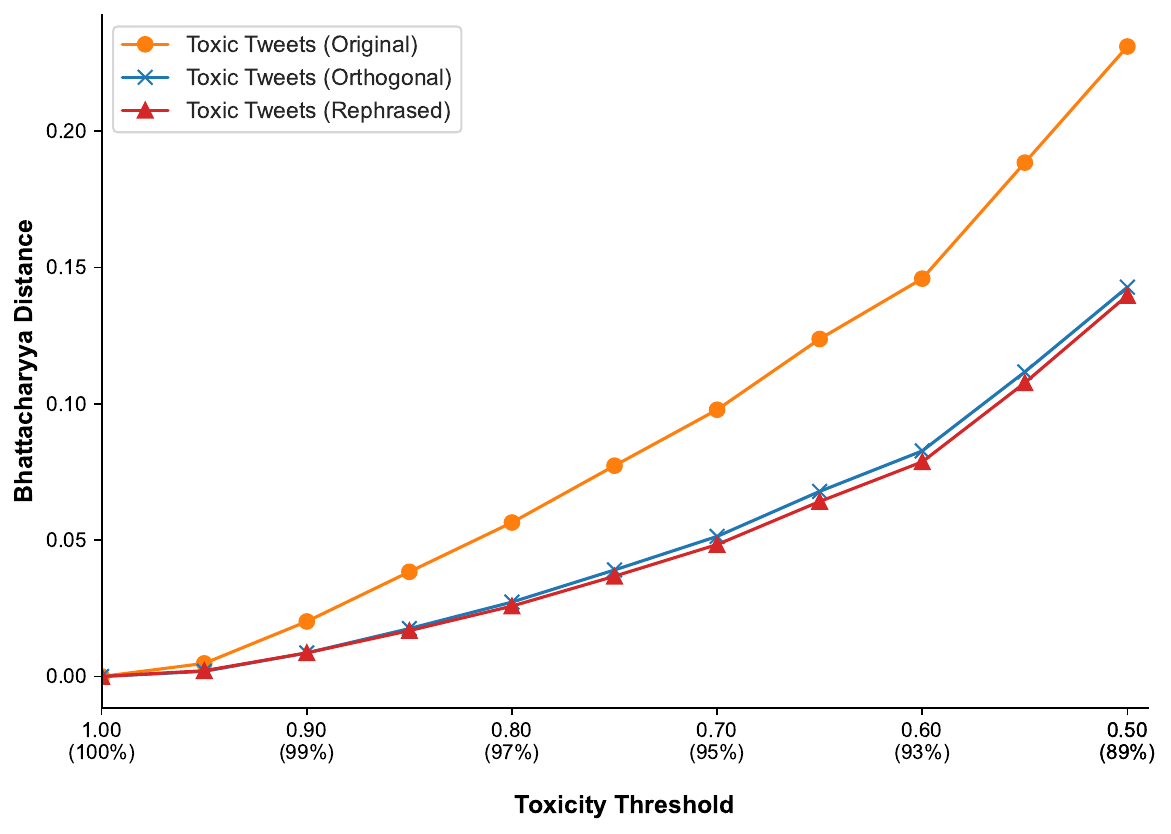}\\[0.1cm]
    \hspace{0.4cm}\parbox{\textwidth}{\footnotesize{\textit{Notes:}  The figure shows the BCD, derived from 
    the original embedding space (orange line) and the orthogonalized embedding space (blue line). We also report the BCD for the removal of rephrased Tweets (red line).}}
\end{figure}

We then repeat the previous analysis using the orthogonalized embedding matrix $\mathbf{\Tilde{X}}$. \Cref{fig:robustness_tox_orthogonal} compares the BCD obtained when removing Tweets from the original embedding space (orange line) and the orthogonalized embedding space (blue line). We also report the BCD for the removal of rephrased Tweets (red line).\footnote{In this exercise, all Tweets are first replaced with their rephrased versions to define the baseline sample. Afterwards, Tweets are removed based on their pre-rephrasing toxicity scores.}

Three key findings emerge from this analysis. 
First, even in the orthogonalized embedding space, we continue to observe substantial, albeit somewhat smaller, changes in the BCD. This again suggests that the observed shifts in BCD arise to a considerable extent from the underlying content of Tweets rather than their toxic language. 
Second, the BCD resulting from the removal of orthogonalized Tweets is virtually identical to that obtained by removing their rephrased counterparts. This demonstrates that our orthogonalization procedure effectively removed the toxicity dimension, bringing the embeddings close to those of the rephrased Tweets. 
Third, removing rephrased Tweets produces more pronounced changes to the embedding space than replacing toxic Tweets with their rephrased versions (see \Cref{fig:rephrasing}). Intuitively, this underscores that removing content introduces larger distortions to the embedding space than rephrasing.

\section{Conclusion}

This paper proposes and validates a new methodology for measuring content-moderation-induced distortions in online content. This new methodology enables us, for the first time, to quantify the impact of removing toxic content on online discourse. Given the crucial importance and heated nature of the debate surrounding this issue, it is important to be clear about what our measure achieves and what it does not capture. The BCD provides a heuristic measure for changes in the embedding space of content circulating on online platforms. As the embeddings measure many semantic characteristics of a text, changes in the mean and the variance of the embeddings provide insight into the extent of the semantic distortions. 

That being said, our measure does not provide a universal measure of online content and free speech (nor is it the goal of this paper). Given the complex and multifaceted nature of the concept of free speech, which encompasses different philosophical and legal standpoints \citep[see, for example,][for a review]{warburton2009free}, no single measure can ever capture all facets of online discourse. Similarly, despite extensive efforts by the research community, no measure of hate speech can ever fully capture the importance of cultural context and changing societal norms \citep[e.g.,][]{brown2017hate}.

Nonetheless, automated hate speech detection tools have proven helpful to rein in toxic speech and are hence widely deployed online. As online platforms and regulators inevitably face trade-offs when it comes to moderating online content, we believe it is essential to have measures that also capture the cost of content moderation. We believe that by shedding light on the trade-offs involved in content moderation, our measure represents a fundamental advancement in the application of NLP tools for content moderation and highlights a highly fruitful direction for future research. 

Our rephrasing analysis also highlights an important cost–benefit trade-off inherent in content moderation. On the benefit side, rephrasing reduces toxicity to a similar extent as outright removal while preserving substantially more semantic content. As our results show, rephrasing induces smaller distortions to the semantic space, maintains a broader range of expressed views, and preserves the possibility of engagement and exposure that would otherwise be eliminated under removal. From the perspective of information economics and market design, moderation rules can be viewed as platform design choices that shape the set of information products that remain available and their visibility \citep{bergemann2019information,kominers2024content}. Selective removal, therefore, distorts the information space on which learning takes place, with implications for beliefs, participation, and incentives across users of social media platforms.

At the same time, rephrasing entails nontrivial costs. Relative to automated removal, it requires additional computational and organizational resources, including the deployment of language models, monitoring of output quality, and safeguards against unintended changes in meaning. Rephrasing may also introduce residual distortions through changes in tone or specificity, and it may raise governance concerns if users perceive rewritten content as intrusive or manipulative. Moreover, rephrasing is not suitable for all types of content, as some material may be undesirable to retain in any form, such as defamation or direct threats. Rephrasing should therefore be understood not as a costless substitute for removal, but as a strategy that trades higher implementation costs for potentially lower informational and expressive costs. 

While our empirical application focuses on content moderation on social media platforms, the proposed measure could also be applied to quantify distortions to media content arising from other interventions, such as legal restrictions on speech, changes in defamation law, or regulatory shocks affecting media environments. The implications of our findings are likely to be even more pronounced in autocratic settings, where restrictions on freedom of expression are typically more extensive and where governments’ incentives often diverge from those of citizens \citep[e.g.,][]{besley2006handcuffs,egorov2009resource}. In such environments, censorship is frequently used to selectively reshape the information environment, thereby limiting learning about political and economic conditions \citep[e.g.,][]{enikolopov2011media,qin2017does,guriev2019informational}. From this perspective, our measure provides a tool to quantify how strongly such interventions distort the media landscape.” 

It is also worth highlighting that our approach is agnostic to the specific mechanism through which content moderation is implemented. In practice, moderation decisions may be made by automated classifiers, crowdsourced reporting, professional content moderators, or hybrid systems that combine these approaches. These approaches may differ in systematic ways, for example, because automated systems tend to rely on keyword- or model-based thresholds that may disproportionately flag certain forms of language, crowdsourced reporting may reflect the preferences or coordination of active user groups, and professional moderators may apply platform guidelines more contextually but at higher cost and lower scale. The proposed measure can be applied uniformly across these settings to quantify how different moderation strategies distort the semantic composition of online discourse.

\clearpage
\bibliography{bib_social_media_pluralism}
\bibliographystyle{chicago}

\clearpage
\appendix
\setcounter{page}{1}
\renewcommand{\thesubsection}{\thesection.\arabic{subsection}.}
\renewcommand{\thetable}{A.\arabic{table}}\setcounter{table}{0}
\renewcommand{\thefigure}{A.\arabic{figure}}\setcounter{figure}{0}
\renewcommand{\theequation}{A.\arabic{equation}}\setcounter{equation}{0}

\section*{\centering \LARGE Online Appendix}
The online appendix presents further details on data construction, methodology, and robustness exercises:
\begin{itemize}
    \item  \Cref{sec:appendix_data} provides additional details on the data.
    \item  \Cref{sec:appendix_methodology} provides additional details on the methodology.
    \item  \Cref{sec:appendix_results} provides additional results.
\end{itemize}


\section{Additional Details on Data \label{sec:appendix_data} }
\renewcommand{\thetable}{A.\arabic{table}}\setcounter{table}{0}
\renewcommand{\thefigure}{A.\arabic{figure}}\setcounter{figure}{0}
\renewcommand{\theequation}{A.\arabic{equation}}\setcounter{equation}{0}

\subsection{Representative Twitter Data \label{sec:appendix_data_sample}}
In our study, we initiated our data collection process with a cohort of 432,882%
 randomly selected American Twitter users, a dataset that was meticulously collected in 2015 by \citep{siegel2021trumping}. This particular sample selection bears two pivotal advantages. Firstly, it allows us to construct a comprehensive and representative overview of Twitter activity across the United States. Secondly, these users have been actively engaged on Twitter for several years, and we do not face problems with composition changes. Consequently, we were able to conduct our analysis based on a good approximation of the content that circulated on Twitter. The initial step in our data collection involved retrieving the Tweets posted by each user within the \cite{siegel2021trumping} sample. In totality, our dataset encompassed 399%
 Million Tweets spanning from 2014 to the commencement of 2022, along with corresponding user profile information. We provide some summary statistics in \Cref{tab:sum_stat}

\begin{table}[ht]
    \centering
    \caption{Summary Statistics \label{tab:sum_stat}}
    \begin{threeparttable}
    \begin{tabular}{lc}
    \toprule
    Number of Tweets & 5M\\
    Average Toxicity Perspectives API Score & 0.19\\
    Average Toxicity Detoxify API Score & 0.11\\
    Average OpenAI Moderation API Score & 0.12\\
    \bottomrule 
    \end{tabular}
    \begin{tablenotes}[para,flushleft] \textit{Notes:} This table provides summary statistics on the number of users, the number of Tweets, and the average Tweet toxicity, as generated by the Perspective API, Detoxify, and OpenAI's Moderation API. \end{tablenotes}
    \end{threeparttable}
\end{table}

\subsection{Filtering Political Tweets \label{sec:appendix_data_filtering_pol}}

We developed a political Tweet classifier by fine-tuning the BERTweet model on a human-labeled dataset of political and non-political Tweets. To create the training dataset, we first compiled a list of political keywords and accounts associated with political figures. This produced a subset of data with a higher density of political Tweets, though many non-political Tweets remained. We then randomly sampled 10,000 Tweets and distributed them among five undergraduate research assistants for manual labeling. Coders were provided with labeling criteria and instructed to categorize each Tweet in a binary true/false format. Each subset included 200 shared Tweets to evaluate inter-coder reliability. We evaluated agreement using Cohen's kappa coefficient, which ranges from 0 to 1, where 1 indicates perfect agreement and values near zero suggest agreement no better than chance. The average Cohen's kappa score across all coder pairs was 0.76. We then use this dataset to train and evaluate our political Tweet classifier. 

\begin{table}[htb]
    \centering
    \footnotesize
    \caption{Confusion Matrix \label{tab:confusion_matrix}}
    \begin{threeparttable}
    \begin{tabular}{lcccc}
        \toprule
        & & \multicolumn{2}{c}{\textbf{Predicted Label}}  \\
        & & Non-Political & Political & \textbf{Total}\\
        \cmidrule(r{2pt}){3-4} \cmidrule(l{2pt}){5-5}
 \multirow{2}{*}{\textbf{True Label}} & \multicolumn{1}{c|}{ Non-Political}  &  647 (0.94) & \multicolumn{1}{c|}{38 (0.06)}  & 685\\
                   &                \multicolumn{1}{c|}{Political}  & 59 (0.19)   & \multicolumn{1}{c|}{256 (0.81)}  & 315 \\
         \cmidrule(r{2pt}){3-4} \cmidrule(l{2pt}){5-5}
         & \textbf{Total} & 706 & 294 & 1,000 \\
         \midrule
         
        \multicolumn{5}{c}{\textbf{Accuracy:} 0.903, \textbf{Precision:} 0.871, \textbf{Recall:} 0.813, \textbf{F1-score:} 0.841} \\
    \bottomrule
    \end{tabular}
    \begin{tablenotes}[para,flushleft] \textit{Notes:} This table reports the confusion matrix for the political Tweet classifier. \end{tablenotes}
    \end{threeparttable}
\end{table}

\begin{figure}[htb]
    \centering
    \caption{ROC Curve \label{fig:roc_curve}}
    \includegraphics[width=0.5\linewidth]{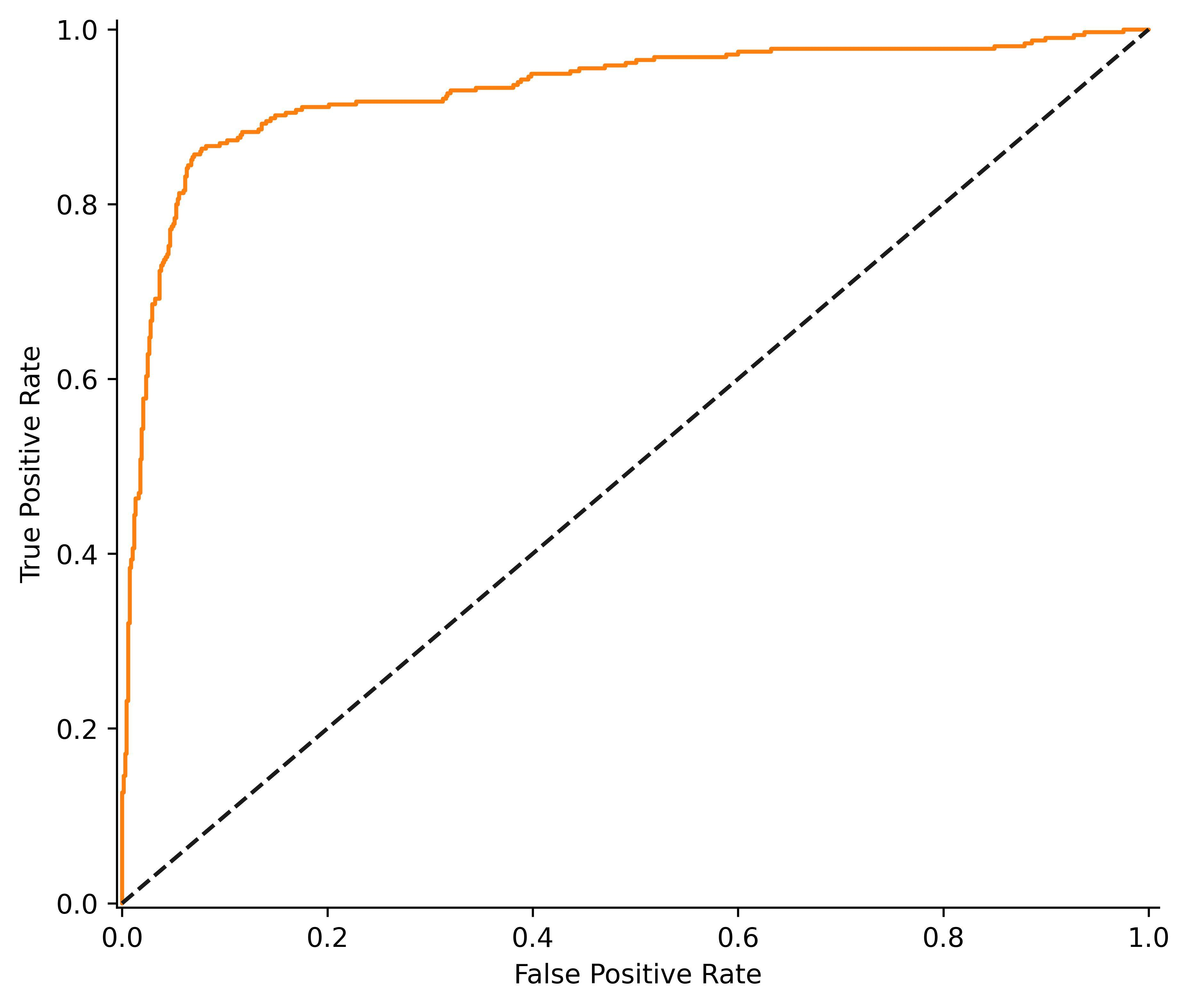}\\[0.2cm]
    \hspace{0.4cm}\parbox{\textwidth}{\footnotesize{\textit{Notes}: The figure reports the ROC curve for the political Tweet classifier.}}
\end{figure}

\noindent From the 10,000 unique labeled examples (we removed duplicates used for inter-coder reliability assessment), we allocated 8,000 for training, 1,000 for in-training evaluation, and 1,000 for the final test set. We fine-tuned the BERTweet model for binary classification over 10 epochs, implementing an early stopping criterion based on the F1 score of the evaluation set. On the test set, the model achieved an accuracy score of 0.90 and an F1 score of 0.84. \Cref{tab:confusion_matrix} presents the evaluation results on the test examples. \Cref{fig:roc_curve} illustrates the Receiver Operating Characteristic curve for the test set, with an area under the curve of 0.88. 

\subsection{Toxicity Measures \label{sec:appendix_data_tox_measures}}
To gauge the toxicity of these Tweets, we use Google's Perspective API, a tool widely acknowledged for its efficacy in identifying hate speech \citep{wulczyn2017ex, dixon2018measuring}. This state-of-the-art API assigns a toxicity score ranging from 0 to 1 across six distinct dimensions: general toxicity, severe toxicity, identity attacks, insults, profanity, and threats. The scores can be approximately interpreted as the probability that a randomly chosen user will classify content as toxic. For example, a score of 0.8 means that around 80\% of users would judge the content to be toxic. 

The Perspective API performs well in classifying toxic text and can assess Tweets in several languages, including English, Spanish, French, German, Portuguese, Italian, and Russian. In our dataset, English is, unsurprisingly, the predominant language, and we restrict our analysis to English-speaking Tweets. Instances where the API did not assign toxicity scores were primarily due to the absence of textual content, such as Tweets containing only hyperlinks. We provide examples of highly toxic Tweets in \Cref{tab:examples_tox}. The examples indicate that the Perspective API accurately identifies toxic content. As described in the main paper, we additionally use the Detoxify \citep{Detoxify} package and OpenAI's Moderation API \citep{openai} as an alternative toxicity classification algorithm. Overall, we also find that the different models broadly agree in their toxicity evaluation of Tweets. The correlation between toxicity scores from the Perspective API and Detoxify is 0.85, while the correlation between the Perspective API and the Moderation API is 0.72.\footnote{As the Moderation API does not report a direct Toxicity score, we always consider the maximum of all provided scores.}  

\begin{figure}[ht]
    \centering
    \caption{Comparison Toxicity Scores: Perspective API vs. Detoxify vs. OpenAI \label{fig:tox_scatter}}  
    \subcaptionbox{Detoxify}{\includegraphics[width=0.45\textwidth]{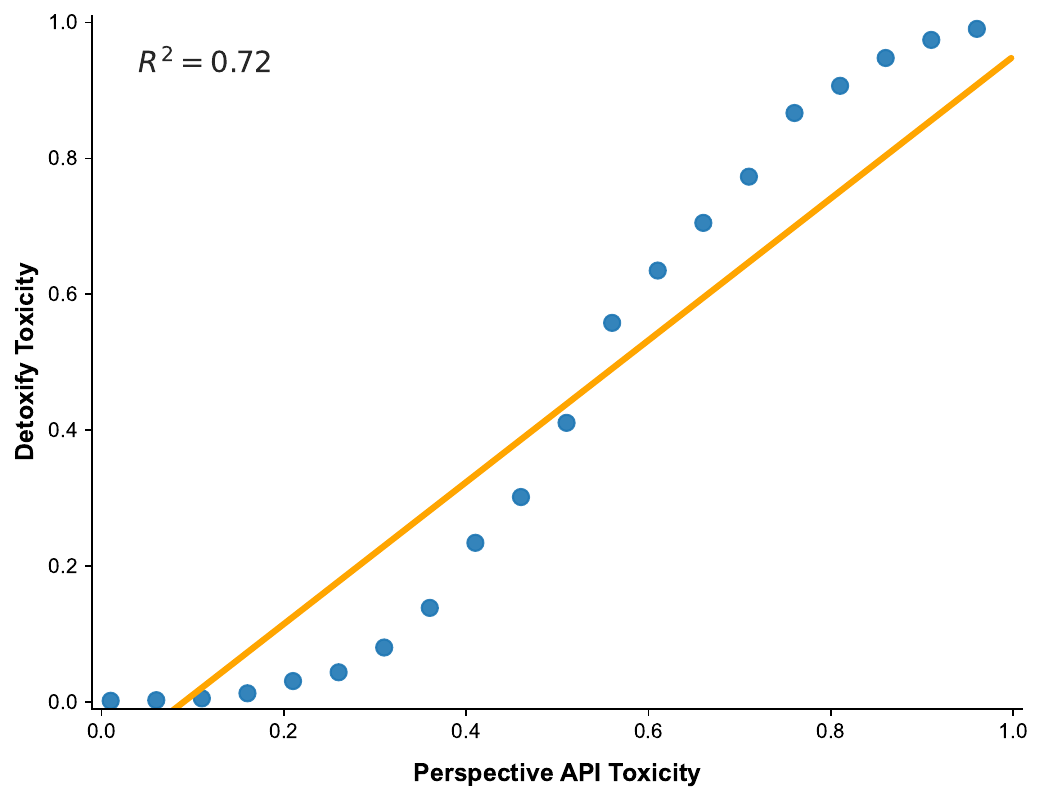}}\hfill
    \subcaptionbox{Moderation API}{\includegraphics[width=0.45\textwidth]{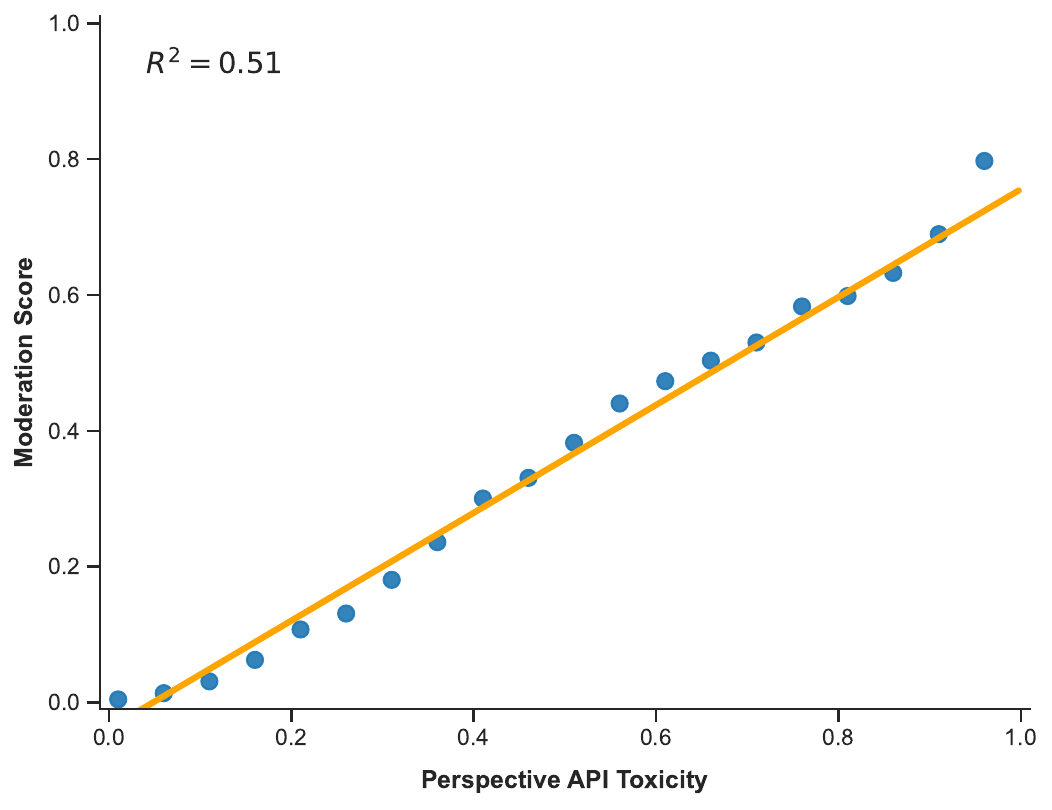}}
    \\[0.2cm]
    \hspace{0.4cm}\parbox{\textwidth}{\footnotesize{\textit{Notes}: The figure shows a binscatter plot of the toxicity scores from the Perspectives API relative to the scores from the Detoxify package and OpenAI's moderation API. The line was fitted based on a linear regression. Data points are grouped into 20 bins of equal size.}}
\end{figure}

\subsubsection*{Distribution of Toxicity Across Users}

An important question for interpreting the implications of toxicity-based moderation is how toxic content is distributed across users. In particular, if toxic language were produced primarily by a very small set of highly active accounts, moderation might disproportionately affect this minority rather than broadly constraining the expressive capacity of typical users. Conversely, if toxicity is more widely distributed across the user base, moderation would have broader implications for the composition of online discourse.

\begin{figure}[ht]
    \centering
    \caption{Distribution of Toxicity Across Users \label{fig:tox_dist_users}}  
    \subcaptionbox{Average Toxicity}{\includegraphics[width=0.45\textwidth]{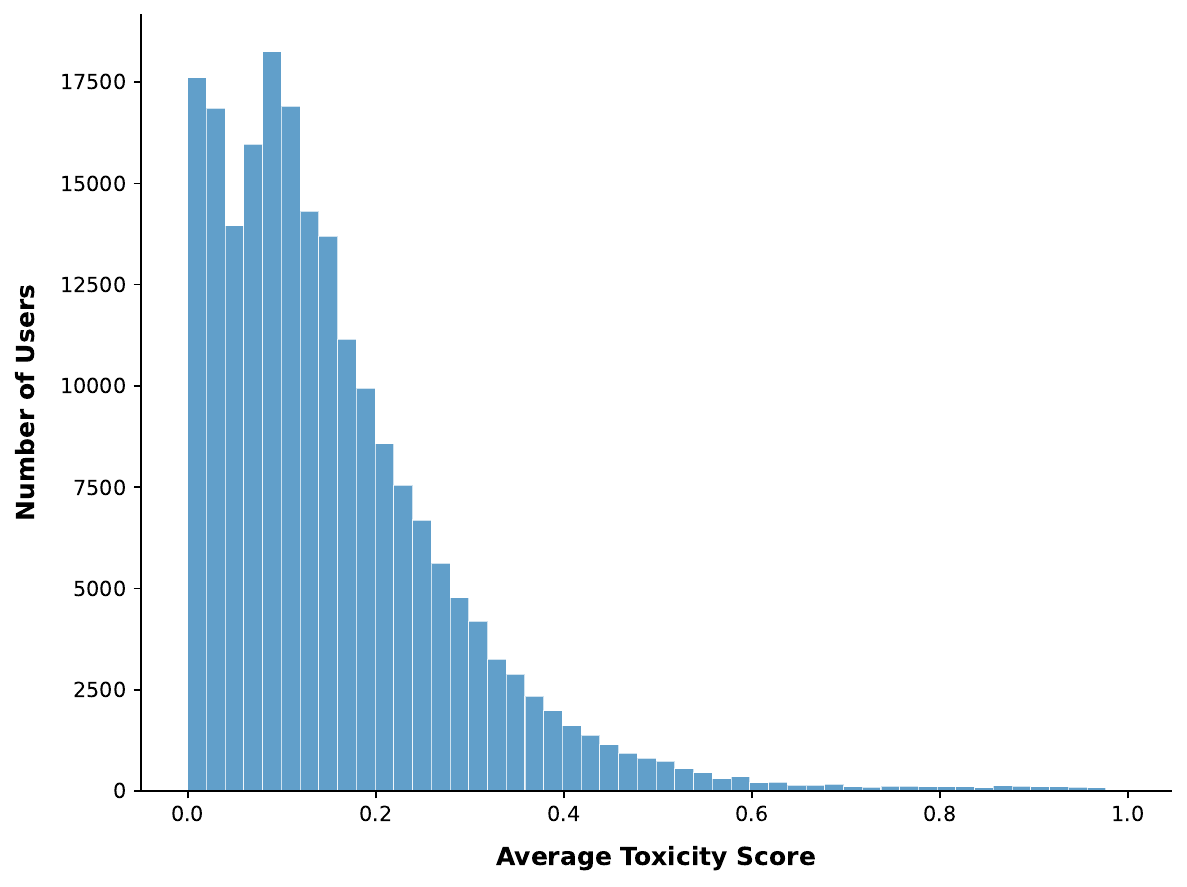}}\hfill
    \subcaptionbox{Number of Toxic Tweets}{\includegraphics[width=0.45\textwidth]{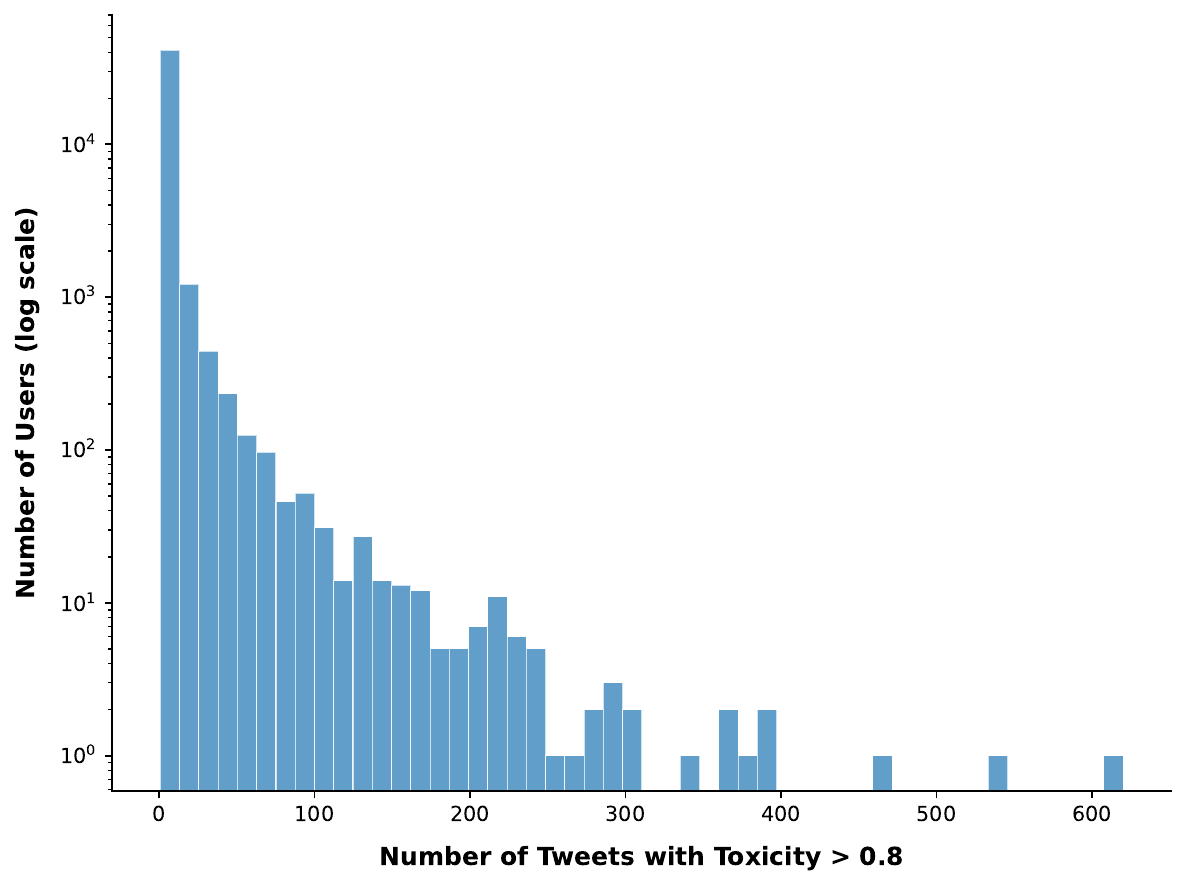}}
    \\[0.2cm]
    \hspace{0.4cm}\parbox{\textwidth}{\footnotesize{\textit{Notes}: The figure shows the toxicity distribution at the user level. Panel (a) shows the distribution of average toxicity scores by user. Panel (b) shows the distribution of the number of highly toxic tweets (toxicity $>$ 0.8) produced by each user, with the y-axis on a logarithmic scale. Toxicity scores are based on the Perspective API.}}
\end{figure}

\Cref{fig:tox_dist_users} illustrates the distribution of toxic Tweets across users in our sample. This exercise allows us to assess whether toxic content is primarily generated by a small group of highly active users or whether toxicity is more broadly distributed across the user population.  We find that the average toxicity scores have a skewed distribution, with most users producing content with an average toxicity below 0.125 (see Panel a). In Panel (b), we show the average number of highly toxic Tweets (toxicity $>$ 0.8) per user. While still right-tailed, the distribution shows that there is a significant fraction of users who produce at least some toxic content. 

Taken together, these figures suggest that toxicity-based moderation likely does not affect only a small group of fringe users but also shapes the overall composition of online discourse. This finding supports an interpretation of our results in which content moderation can alter the representativeness of expressed views, rather than merely suppressing the speech of a marginal subset of users.

\subsection{Twitter’s Content Moderation Policies on Hate and Toxicity, 2006–2022 \label{sec:appendix_data_twitter_policies}}

For most of the period covered by our data, Twitter maintained a set of content moderation policies that explicitly prohibited hate speech, abusive behavior, and other forms of toxic expression, even as the platform’s enforcement practices and conceptualizations of objectionable speech evolved. In the following, we provide a short outline of some of the important content moderation milestones. 

\begin{itemize}
  \item \textbf{Early Period 2006 to 2012.}
  From its launch in 2006 through the early 2010s, Twitter relied on a comparatively thin and reactive rule framework: policy and enforcement focused heavily on operational integrity (e.g., spam) and legal compliance, while protections against harassment and hate were less explicit and less systematized than in later years \citep{WiredHateSpeech2015}.
  
  \item \textbf{2012: Country-specific content withholding (jurisdictional compliance).}
  Twitter began using (and publicly discussing) mechanisms to withhold specific content in particular countries rather than remove it globally, reflecting growing cross-national legal constraints and the emergence of ``geo-blocked'' moderation as a distinct tool \citep{TNWWithholding2012}.

  \item \textbf{2012: Transparency reporting (public accountability for removals and requests).}
  Twitter launched a transparency reporting tool covering government requests for user data and content removal, and also began reporting copyright requests, providing a public record of state and private legal pressures that drive moderation outcomes \citep{ArsTransparency2012}.

  \item \textbf{2015: Anti-abuse pivot (expanded threat standards + new enforcement instruments).}
  Twitter broadened its approach to violent threats and abuse and introduced stronger enforcement mechanisms, including time-based account locks and early forms of reach-limiting for suspected abusive content. This marks a key shift from primarily complaint-driven removals to a broader enforcement toolkit \citep{TwitterAbuseUpdate2015}.

  \item \textbf{2015: Explicit ``hateful conduct'' and self-harm policies.}
  Twitter updated its rules to explicitly prohibit hateful conduct and added policy language and interventions concerning self-harm, signaling a wider definition of harm beyond direct threats and toward group-based and wellbeing-related harms \citep{TwitterHatefulConduct2015}.

  \item \textbf{2016: Institutionalization via Trust \& Safety governance.}
  Twitter created a Trust \& Safety Council, formalizing external stakeholder engagement and reinforcing the platform's framing of moderation as a balancing problem between safety and freedom of expression \citep{TechCrunchTSC2016}.

  \item \textbf{2020: COVID-19 misinformation policy (public-health information integrity).}
  During the pandemic, Twitter adopted explicit rules against COVID-19 misinformation, extending moderation beyond harassment/hate into health-related claims where harms arise via behavioral influence (e.g., discouraging expert guidance or promoting false cures) \citep{TechCrunchCOVID2020}.

  \item \textbf{2020: Expanded election/civic integrity enforcement.}
  Twitter strengthened its approach to election-related misinformation via more explicit civic integrity rules and enforcement (often combining labels and other distribution interventions), reflecting the platform-wide shift toward structured governance of high-stakes political information flows \citep{EngadgetElection2020}.

  \item \textbf{2022: ``Crisis misinformation'' policy.}
  Twitter introduced a framework aimed at reducing the spread and amplification of misleading claims during crises (e.g., wars and disasters), further entrenching the use of labeling and amplification controls as core moderation tools \citep{TechCrunchCrisis2022}.

\end{itemize}

\clearpage
\subsection{Additional Details on Topic Model \label{sec:topic_model} }

\begin{table}[htb]
    \centering
    \caption{Most Relevant Words by Topic\label{tab:topic_words}}
    \adjustbox{width=0.87\textwidth}{
    \begin{threeparttable}
    \begin{tabular}{ccl}
        \toprule
        \textbf{Number} & \textbf{Topic Label} & \textbf{Topic Words} \\
        \midrule
        0 & Insults &   idiot	hates	hahahahaha	stfu	moron	fuk	scum	yawn	dumbass	yikes	pathetic\\
        1 & Trump & trump	trumps	djt	trumpers	drumpf	trumpster	trumpy	trumpsters	potus	donald	nevertrump\\
        2 & Voting & voter	ballot	voted	voting	ballots	voters	vote	votes	disenfranchise	elections	election\\
        3 & Govenors & rauner	mccrory	mayor	gubernatorial	blasio	mayors	governor	lepage	rahm	redistricting	mayoral \\
        4 & Vaccines &vaccinate	vaccinating	unvaccinated	immunization	vaccinated	vaccine	vaccines	vaccinations	vaccination	cdc	vaxxers \\
        5 & Parties & gops	gop	repub	repubs	dems	republican	republicans	democrats	rinos	democrat	bipartisanship\\
        6 & Education & devos	defund	rauner	educators	privatize	underfunded	taxpayers	nea	education	educate	defunded\\
        7 & Legislators & wyden	durbin	legislator	cispa	grandstanding	legislate	lawmakers	toomey	rauner	lawmaker\\
        8 & President & potus	presidential	trump	prez	obummer	president	trumps	presidency	presidents	obamas	whitehouse\\
        9 & Special Council & manafort	indictments	mueller	comey	gowdy	kushner	grassley	dershowitz	colluded	trumpers	nunes\\
        10 & Media & msnbc	foxnews	cnn	maddow	hannity	olbermann	drudge	megyn	cspan	tyt	reuters\\
        11 & Obama & obamas	obama	obummer	nobama	barack	potus	prez	newsmax	presidential	teleprompter	romney\\
        12 & Media Personalities & fareedzakaria	govmikedewine	katrinapierson	krystalball	plf	johnbrennan	flapol	narendramodi	laurenboebert	bhive	mikepence\\
        13 & Prisons &jailing	jailed	exonerated	prosecutors	sentencing	sentenced	jails	indict	incarceration	imprison	correctional \\
        14 & Biden & biden	bidens	kaine	vp	palin	obummer	nobama	cheney	reince	joe	veep\\
        15 & Racism & racists	racist	racism	racial	sharpton	naacp	racially	divisiveness	klan	supremacist	blacks\\
        16 & Taxes & tax	taxation	taxing	fiscal	taxes	taxed	irs	taxable	deductions	redistribution	cbo\\
        17 & America & america	merica	murica	unamerican	amerikkka	americas	americans	american	patriotic	usa	unpatriotic\\
        18 & Police & policemen	police	cops	policeman	policing	acab	nypd	sheriffs	lapd	dorner	cop\\
        19 & Goverment & govt	government	gov	goverment	governmental	governments	feds	bureaucrats	privatize	privatized	govs\\
        20 & Healthcaree & obamacare	cbo	singlepayer	medicare	aca	healthcare	repeal	medicaid	ahca	trumpcare	uninsured\\
        21 & Politics & politics	political	politic	politicians	apolitical	politically	politician	politicized	pols	partisanship	divisiveness\\
        22 & Prayers & prayers	praying	prayer	pray	prayed	condolences	amen	blessings	thankful	salute	praise\\
        23 & Twitter & tweeting	tweets	tweet	tweeted	retweets	twitter	retweeted	tweeter	retweeting	retweet	covfefe\\
        24 & Senate & mcconnell	grassley	gops	rinos	senate	manchin	bipartisanship	toomey	senators	filibuster	obstructionist\\
        25 & Veterans & veterans	servicemen	veteran	salute	honoring	commemorate	soldiers	commemorating	vets	thanking	heros\\
        26 & Guns & nra	guns	giffords	firearms	shootings	gun	firearm	gunman	armed	massacres	feinstein\\
        27 & States &florida	fla	floridians	fl	ohioans	michiganders	broward	tallahassee	kentucky	hoosier	ohio \\
        28 & Deportation & deportations	illegals	deporting	deportation	immigration	deport	deported	amnesty	immigrants	migrant	immigrant\\
        29 & Environment & polluters	algore	renewables	epa	fracking	greenpeace	polluting	globalwarming	politicized	deniers	exxon\\
        30 & Russia & putin	kremlin	russia	crimea	kgb	russians	ukrainian	russian	ukraine	ukrainians	oligarchs\\
        31 & Clinton & hillary	clintons	hilary	clinton	hrc	killary	huma	trump	lewinsky	gillibrand	trumps\\
        32 & Congress & congressmen	congressional	congress	senate	lawmakers	constitutional	congressman	grassley	incumbents	wyden	senatorial\\
        33 & Left-Right & liberal	liberals	conservatives	conservative	conservatism	liberalism	leftists	leftist	libs	rightwing	libtard\\
        34 & Federal Reserve & bernanke	yellen	fomc	krugman	bullish	zerohedge	schiff	jpmorgan	cnbc	economist	recession\\
        35 & Protests & protestors	protesters	protesting	protester	protests	protestor	demonstrators	protest	rioting	rioters	protested\\
        36 & Democracy & democracy	democracies	undemocratic	democratically	oligarchy	democratic	dictatorship	dictatorships	tyranny	unelected	demagogue\\
        37 & Supreme Court & scotus	scalia	sotomayor	alito	ginsburg	kagan	gorsuch	justices	recuse	grassley	rbg\\
        38 & Impeachment & impeachment	impeach	impeached	impeachable	impeaching	indictments	grassley	dershowitz	pardons	sedition	pardoning\\
        39 & Civil Rights & mlk	confederate	confederacy	gettysburg	juneteenth	tubman	proclamation	secession	naacp	commemorating	divisiveness\\
        40 & Canada & trudeau scheer	ndp	harper	kenney	wynne	rudd	cpc	tories	conservatives	libdems\\
        41 & Terrorism & terrorism	terrorist	terrorists	jihadist	jihadists	radicalized	islamist	bombings	islamists	islamophobia	qaida\\
        42 & Budget & debt	deficits	cbo	krugman	debts	austerity	bernanke	trillion	deficit	bailout	defund\\
        43 & Syria & airstrikes	assad	syria	daesh	isil	bashar	mosul	jihadists	isis	jihadist	syrian\\
        44 & UK & libdems	miliband	farage	clegg	brexit	tories	corbyn	ukip	gove	snp	labour\\
        45 & Israel & israelis	netanyahu	zionist	zionists	palestinians	aipac	antisemitism	antisemitic	israel	zionism	idf\\
        46 & LGBTY &doma	lgbtq	gays	homophobic	gaymarriage	homophobia	bigots	legislating	dadt	homosexuality	gsa\\
        47 & Sanders & bernie	sanders	feelthebern	bern	hillary	dnc	hrc	electable	tyt	canvassing	disavow\\
        48 & Debate & debates	debate	debating	debated	discussions	cspan	rebuttal	controversy	argument	discourse	fiorina\\
        49 & India & bjp	kejriwal	modi	swamy	ndtv	gujarat	nawaz	bihar	aap	narendra	mlas\\
        50 & Sedition & sedition	inciting	facist	censorship	freedoms	fascist	incitement	fascism	ammendment	totalitarianism	authoritarians\\
        51 & Socialism & socialism	socialist	socialists	capitalist	marxism	marxist	communist	communism	capitalism	marxists	capitalists\\
        52 & Romney & romney	romneys	mitt	romneyryan	obummer	santorum	nobama	mittens	obama	gingrich	gops\\
        53 & Petitions & petition	petitions	signed	signatures	impeach	signing	amnesty	protest	oppose	cispa	sopa\\
        54 & China & china	jinping	chinese	taiwan	beijing	mao	gao	ccp	hk	reuters	hong\\
        55 & Africa & nigeria	nigerians	nigerian	mugabe	davidcorndc	niger	zuma	anc	lagos	boko	somali\\
        56 & Marijuana & legalization	decriminalize	legalize	legalizing	legalized	marijuana	mmj	cannabis	thc	hemp	weed\\
        57 & Abortion & prolife	prochoice	abortions	abortion	lifers	aborted	unborn	fetus	jimmykimmel	fetal	personhood\\
        58 & Southern States & bama	alabama	auburn	lsu	mizzou	mississippi	arkansas	clemson	birmingham	tennessee	kentucky\\
        59 & Iran &iran	iranians	ahmadinejad	iranian	khamenei	mullahs	tehran	rouhani	netanyahu	shah	kissinger\\
        60 & Republican Primaries & cruz	rubio	kasich	ted	nevertrump	fiorina	rino	reince	grassley	rinos	huckabee\\
        61 & Nazis & nazi	nazis	goebbels	nazism	hitler	holocaust	gestapo	adolf	reich	auschwitz	fascist\\
        62 & Afghanistan & taliban	afghanistan	afghan	afghans	kabul	qaeda	bergdahl	insurgents	qaida	osama	jihadist\\
        63 & BLM & blm	alllivesmatter	sharpton	naacp	protestors	antifa	disavow	supremacists	protestor	protesters	seiu\\
        64 & North Korea & dprk	nk	kim	seoul	jong	rodman	korea	missiles	wwiii	korean	nukes\\
        65 & Twitter & tweeted	tweets	retweeted	tweet	confirms	tweeting	withheld	twitpic	cancelled	retweets	retweet\\
        66 & Vice Presidents & pence	kaine	vp	impeaching	impeachment	impeach	djt	trumpers	mattis	impeached	nevertrump\\
        \bottomrule
    \end{tabular}
    \begin{tablenotes}[para,flushleft] \textit{Notes:} This table reports the 10 most important topic words for each of the 67 topics generated by the Top2Vec topic model. The labels were assigned by the authors based on the topic words. \end{tablenotes}
    \end{threeparttable}
    }
\end{table}

\begin{figure}[htb]
    \centering
    \caption{Toxicity Composition of Topics (1/2) \label{fig:topic_tox_composition}}
    \includegraphics[width=0.7\linewidth]{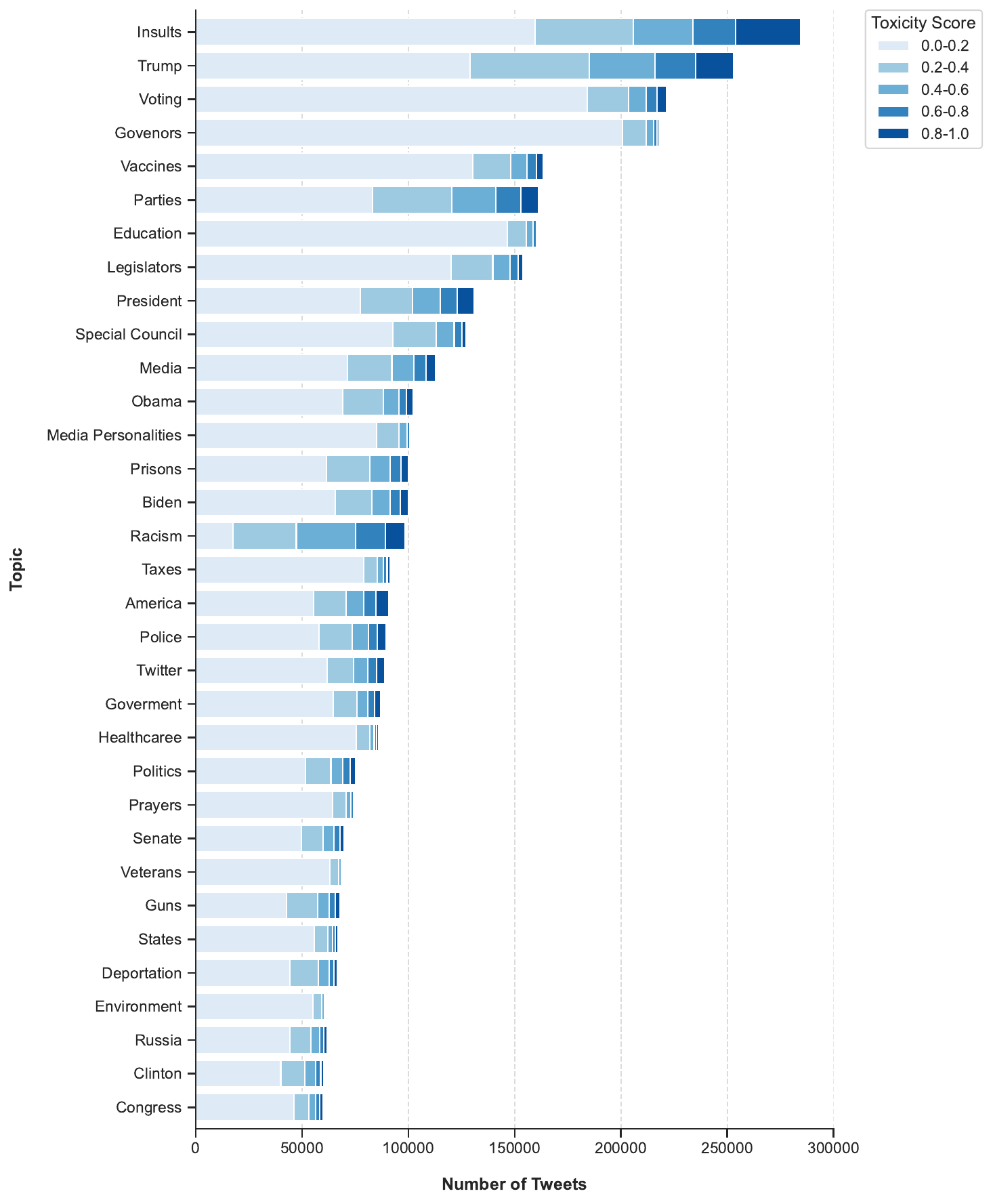}\\[0.2cm]
    \hspace{0.4cm}\parbox{\textwidth}{\footnotesize{\textit{Notes}: The figure shows the size and the toxicity composition of each of the topics created by the Top2Vec topic model.}}
\end{figure}

\begin{figure}[htb]
    \centering
    \ContinuedFloat
    \caption{Toxicity Composition of Topics (2/2) }
    \includegraphics[width=0.7\linewidth]{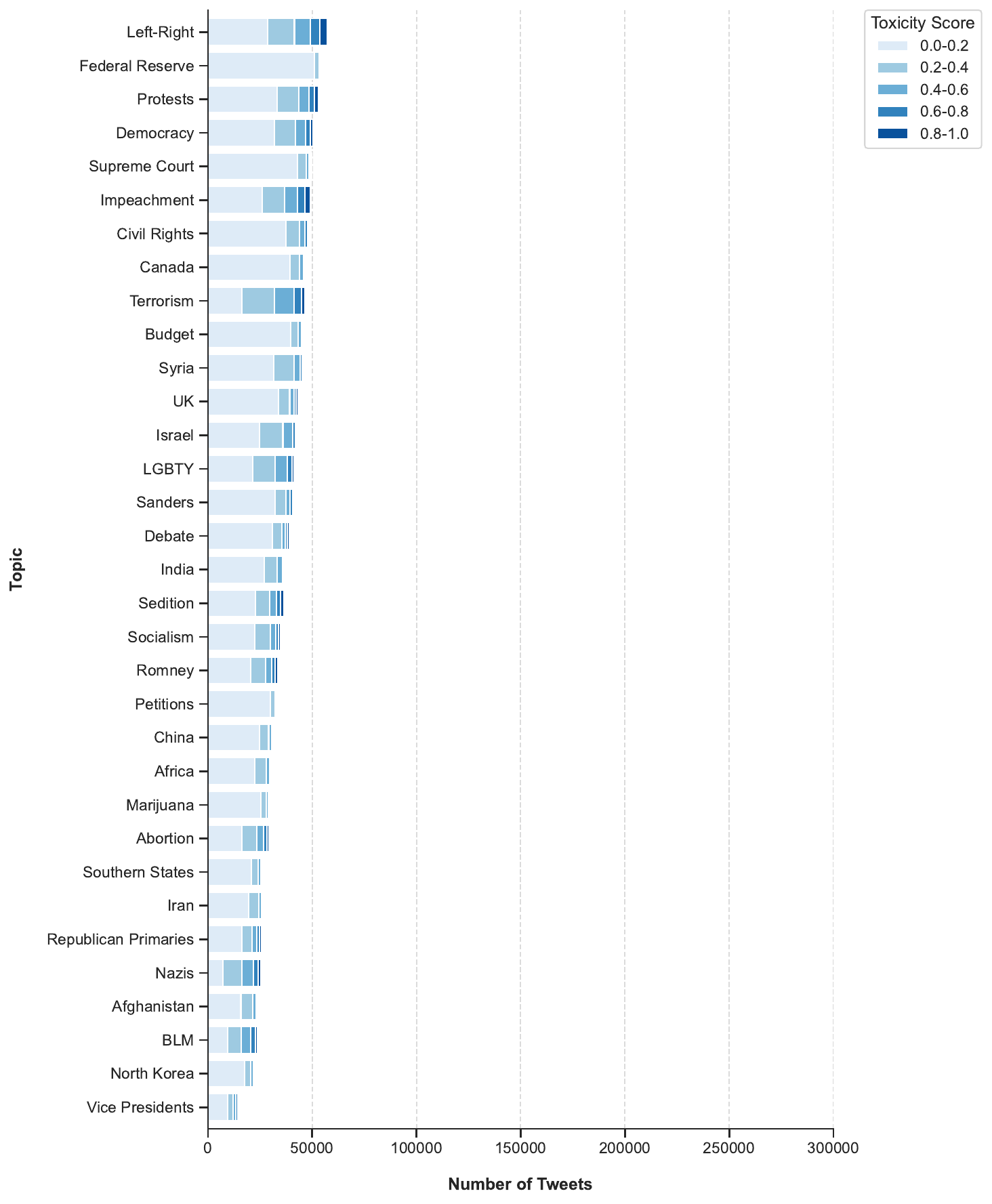}\\[0.2cm]
    \hspace{0.4cm}\parbox{\textwidth}{\footnotesize{\textit{Notes}: The figure shows the size and the toxicity composition of each of the topics created by the Top2Vec topic model.}}
\end{figure}

\clearpage
\section{Additional Details on Methodology \label{sec:appendix_methodology} }
\renewcommand{\thetable}{B.\arabic{table}}\setcounter{table}{0}
\renewcommand{\thefigure}{B.\arabic{figure}}\setcounter{figure}{0}
\renewcommand{\theequation}{B.\arabic{equation}}\setcounter{equation}{0}

\subsection{Additional Details on Embeddings \label{sec:appendix_embeddings} }
To distill information from the raw text of Tweets into quantifiable data, we use three pre-trained models: BERTweet \citep{nguyen_bertweet_2020}, RoBERTa \citep{liu_roberta_2019}, DeBERTa \citep{he_deberta_2021}, and DistilBert-Base-Multilingual-Cased-V2 \citep{reimers-2019-sentence-bert}. These models all use on the BERT-style Transformer architecture, which has become a staple in text classification and natural language understanding tasks. The strength of these models comes from their utilization of an ''attention'' mechanism, allowing them to evaluate words in the context of their surrounding text, thereby capturing the subtleties of language that are often lost in traditional analysis. We transform the raw Tweets into analyzable embeddings by tokenizing the text into its constituent word tokens. These tokens then serve as input to our models, which produce numerical representations—or embeddings—of each Tweet. These embeddings convey the semantic and syntactic nuances of the language used by Twitter users and form the backbone of our computational analysis. Further details on the specific attributes of these models within our study are provided in the following.

\subsubsection*{BERTweet}
In the rapidly expanding field of Natural Language Processing (NLP), the inception of BERT (Bidirectional Encoder Representations from Transformers) and its subsequent iterations have marked a significant milestone. Introduced by \citep{devlin_bert_2019}, BERT leverages an architecture known as Transformers \citep{vaswani_attention_2017} to process words in relation to all the other words in a sentence, which contrasts with prior models that viewed words in sequence. The BERT class of language models, with BERTweet as one example, are proficient in tasks such as part-of-speech tagging, named entity recognition, and text classification. The original BERT model was trained on an extensive corpus comprising sources like Wikipedia and books, known for their structured and formal English. However, the nature of Twitter's text, characterized by brevity and idiosyncratic language usage, presented a unique challenge for these models. To address this, BERTweet was specifically trained on an 80GB corpus containing 850 million English Tweets \citep{nguyen_bertweet_2020}. This vast training corpus allows BERTweet to learn about the distinctive language patterns on Twitter. 

To provide a brief overview of the data processing pipeline. Initially, the raw text of Tweets undergoes a tokenization process wherein the text is segmented into ``tokens,'' which are the basic units for the model to understand. Imagine tokenization as the breaking down of a sentence into individual words and symbols, which are then analyzed by the language model. Once tokenized, these Tweets are fed into the pre-trained BERTweet model. This model excels in interpreting each token in context, producing a vector of size 768 that captures not just the semantics of the individual token but also its relationship to others in the Tweet, all while considering the token's position. 

\begin{figure}[ht]
    \centering
    \caption{Histogram of Embeddings \label{fig:hist_embeddings}}  
    \includegraphics[width=0.5\textwidth]{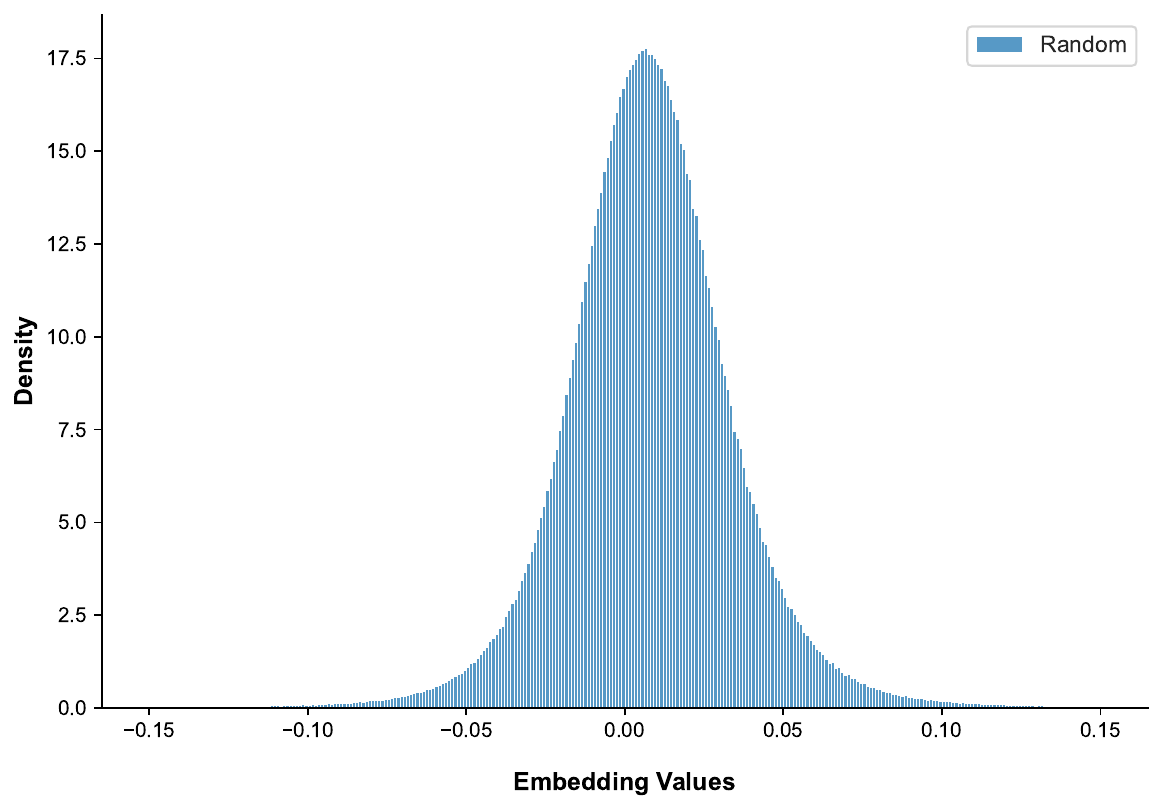}\\[0.1cm] 
    \hspace{0.4cm}\parbox{\textwidth}{\footnotesize{\textit{Notes}: The figure shows a histogram of the embedding dimensions. For this figure, we aggregate data across all dimensions of the embedding for a sample of 10,000 Tweets. }}
\end{figure}

Subsequently, we create an embedding (vector) that captures the content of the Tweet as a whole by taking a weighted average of all token embeddings based on their attention weights. Attention weights are a major component of Transformers that ensures that more influential tokens have a bigger impact on the final Tweet embedding. Put differently, the model distinguishes which words carry more weight in conveying the Tweet's overall message and adjusts the embedding accordingly. Ultimately, each Tweet is distilled into a unit-length vector within a 768-dimensional space, enabling nuanced interpretations and analyses. The 768-embedding dimensions are approximately normally distributed (see \Cref{fig:hist_embeddings}).

\subsubsection*{RoBERTa}
RoBERTa \citep{liu_roberta_2019} is another widely used model that we can use to generate Tweet embeddings. Building upon the BERT foundation, RoBERTa implements several modifications to improve performance. In particular, RoBERTa extended the training duration, increased batch sizes, and exposed the model to a broader spectrum of data, including the large CC-NEWS dataset. RoBERTa also modifies BERT's training process by discarding the next sentence prediction objective and by training on longer sequences. Moreover, RoBERTa introduces variability in the masking pattern of the input data during training, which prevents the model from merely memorizing fixed patterns and encourages a deeper comprehension of language nuances. Together, these modifications are beneficial for understanding the context more effectively and led RoBERTa to achieve state-of-the-art results on many NLP tasks and benchmarks.

\subsubsection*{DeBERTa}
The third model we are using in our analysis is  DeBERTa \citep{he_deberta_2021}. DeBERTa (Decoding-enhanced BERT with disentangled attention) introduced innovative mechanisms that refine the workings of BERT and RoBERTa models. Its distinctive feature lies in the disentangled attention mechanism, which considers the content and the position of words separately, offering a more nuanced understanding of the text. Each word is represented by dual vectors that capture what the word is and where it stands in a sentence. This allows for a better interpretation of language nuances. Furthermore, DeBERTa's enhanced mask decoder predicts masked tokens by using their absolute positions, an improvement that aids in the pre-training process. With these advancements, DeBERTa improved the model's pre-training efficiency and improved the performance across a variety of downstream tasks, like the generation of new text that mimics human speech.

\subsubsection*{DistilBert-Base-Multilingual-Cased-V2}

DistilBERT-Base-Multilingual-Cased-V2 \citep{reimers-2019-sentence-bert} is a lighter and faster variant of BERT, specifically trained to handle multiple languages in a single model. DistilBERT applies knowledge distillation during training. This process reduces the model’s size and increases inference speed. The multilingual-cased version was trained on multiple languages using cased text, ensuring that it preserves distinctions in capitalization, which can be semantically relevant in many languages. As a result, this model is particularly well-suited for cross-lingual applications.

\subsection{Additional Details the Bhattacharyya Distance \label{sec:appendix_bcd} }

This subsection provides additional details on the Bhattacharyya distance (BCD) as a measure of distortions in semantic space, and provides guidance on its interpretation and applicability. The BCD was developed to quantify the distance between two probability distributions. Let $P$ and $Q$ denote two probability distributions over a random vector $X \in \mathbb{R}^d$ with densities $p(x)$ and $q(x)$. The Bhattacharyya coefficient is defined as:
\begin{equation}
BC(P,Q) = \int \sqrt{p(x)\, q(x)} \, dx.
\end{equation}
The Bhattacharyya distance is then defined as:
\begin{equation}
BCD(P,Q) = -\ln \left( BC(P,Q) \right).
\end{equation}
The coefficient $BC(P,Q)$ measures the overlap between two probability distributions and takes values in $[0,1]$, where 1 indicates identical distributions. The BCD converts this overlap into a divergence measure, with larger values indicating greater separation.

The Bhattacharyya distance can be used to quantify differences between both unimodal and multimodal distributions, and it is not restricted to normal data-generating processes. Throughout this paper, we rely on the closed-form expression that applies under the assumption of multivariate normality:
\begin{align}
    BCD(\mathcal{N}_1,\mathcal{N}_2) = \frac{1}{8} (\mu_1-\mu_2)^T\Sigma^{-1}(\mu_1-\mu_2)+\frac{1}{2}\left(\frac{\det \Sigma}{\sqrt{\det \Sigma_1 \cdot \det \Sigma_2}} \right) 
\end{align}
In semantic embedding spaces, the distribution of values along each dimension of the high-dimensional space follows a normal distribution. Thus, for our application, the BCD over embedding spaces (which represent semantic content) intuitively captures the idea that two text corpora are more similar if their semantic spaces overlap. Importantly, the BCD is a distributional measure rather than a pairwise similarity measure. It does not compare individual texts to one another, but instead compares the overall geometry of the embedding space. Intuitively, this operation can be imagined as comparing the shape and density of two high-dimensional point clouds.

\noindent The key properties that make the BCD an attractive measure for our setting include:
\begin{itemize}
    \item Symmetry: $BCD(P,Q) = BCD(Q,P)$.
    \item Invariance under linear transformations.
    \item Computational tractability in high-dimensional settings.
\end{itemize}
\noindent For normal distributions:
\begin{itemize}
    \item If two distributions have identical means and covariance matrices, the BCD equals zero.
    \item The BCD increases when the means move apart.
    \item The BCD increases when the covariance matrices diverge.
\end{itemize}

\noindent The BCD has a long tradition in statistics and empirical classification problems as a measure of separability between probability distributions. Conceptually, it quantifies how difficult it is to distinguish two data-generating processes based on observed features. In that sense, it provides a natural measure of distributional change that is closely related to statistical hypothesis testing.

In classical pattern recognition, the Bhattacharyya distance is used to measure class separability and provides an upper bound on the Bayes classification error \citep[e.g.,][]{kailath1967divergence}. In computer vision, the Bhattacharyya coefficient is used to compare feature distributions, such as color histograms, or image segmentation \citep[e.g.,][]{michailovich2007image}. 

For researchers interested in applying this approach in other contexts, several practical considerations are worth highlighting. First, the BCD is most informative when the object of interest is a global change in the composition of content (such as moderation or a natural experiment setup), rather than local or pairwise similarities between individual texts. In other words, it does not measure why and where the meaning of the corpora changed, only that it did. Second, when strong departures from approximate normality are a concern, the BCD can be computed using nonparametric estimates of marginal distributions. Finally, the measure is not designed to replace topic models or similarity-based approaches, but rather to complement them by providing a scalable and interpretable summary of distributional shifts in high-dimensional semantic spaces.

\subsection{Additional Details on the Rephrasing of Tweets \label{sec:prompts} }

For the rephrasing of Tweets, we used OpenAI's ``gpt-4o-mini-2024-07-18'' model. Each time, the model was asked to rephrase a single Tweet using the following prompt:
\begin{quote}
``Your task is to rephrase a highly toxic Tweet and write a less toxic version of it while aiming to make minimal changes to the original Tweet. It's crucial to preserve the original wording, content, style, and tone in the Tweet. Keep the Twitter special elements such as RT and \@XXX unchanged.
Example: Original: `\@some\_user The system is so fucked up. What’s sad is they can do wtf they want.' Rephrased: `\@some\_user The system is so messed up. What’s sad is they can do whatever they want.' Please respond in JSON format with the key `RephrasedText'. Here is the Tweet to rephrase:''
\end{quote}

\subsection{Additional Details on Engagement Prediction \label{sec:engagement_prediction} }

We employ two models for the engagement predictions: 1) a linear regression and 2) a neural network with three hidden layers. All models are trained on a sample of 1 Million political Tweets, with engagement metrics winsorized at the 99\textsuperscript{th} percentile and log transformed with one added to account for the heavy-tailed distribution of social media engagement.\footnote{The results are very similar without winsorization.} 

The neural network is trained with three fully connected hidden layers of dimensions 256, 64, and 32, respectively. The model uses Rectified Linear Unit (ReLU) activation functions and includes dropout layers with a rate of 0.1 to prevent overfitting. We experimented with deeper architectures by adding additional hidden layers, but found that increasing model complexity did not noticeably improve performance. The model was trained for 10 epochs with a batch size of 256 and an Adam optimizer with a learning rate of 0.001.

\begin{figure}[ht]
\centering
\caption{Engagement Prediction Performance: OLS vs. Neural Network\label{fig:engagement_comparison}}

\subcaptionbox{Likes (OLS)\label{fig:ols_likes}}{\includegraphics[width=0.4\textwidth]{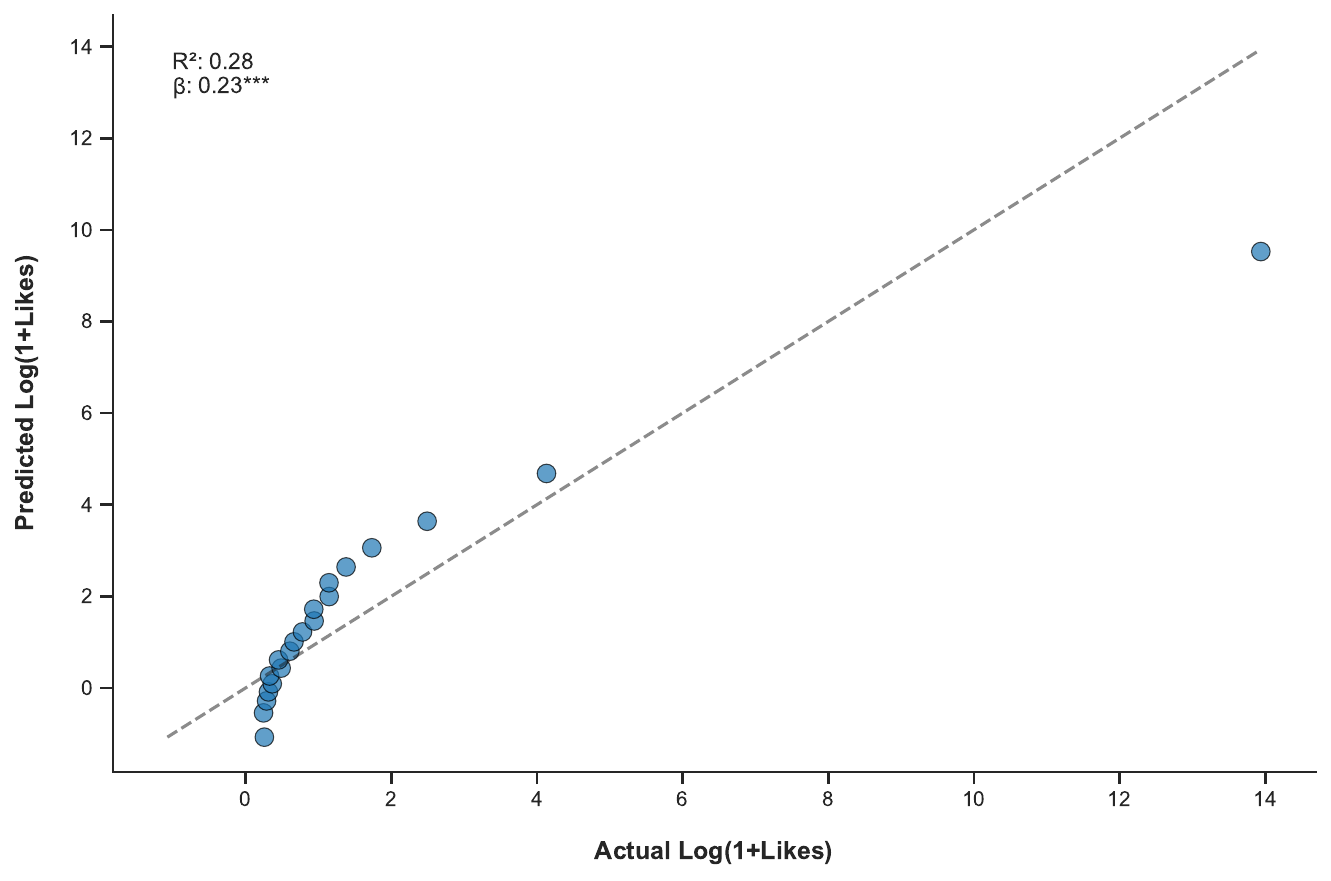}}\hfill
\subcaptionbox{Retweets (OLS)\label{fig:ols_retweets}}{\includegraphics[width=0.4\textwidth]{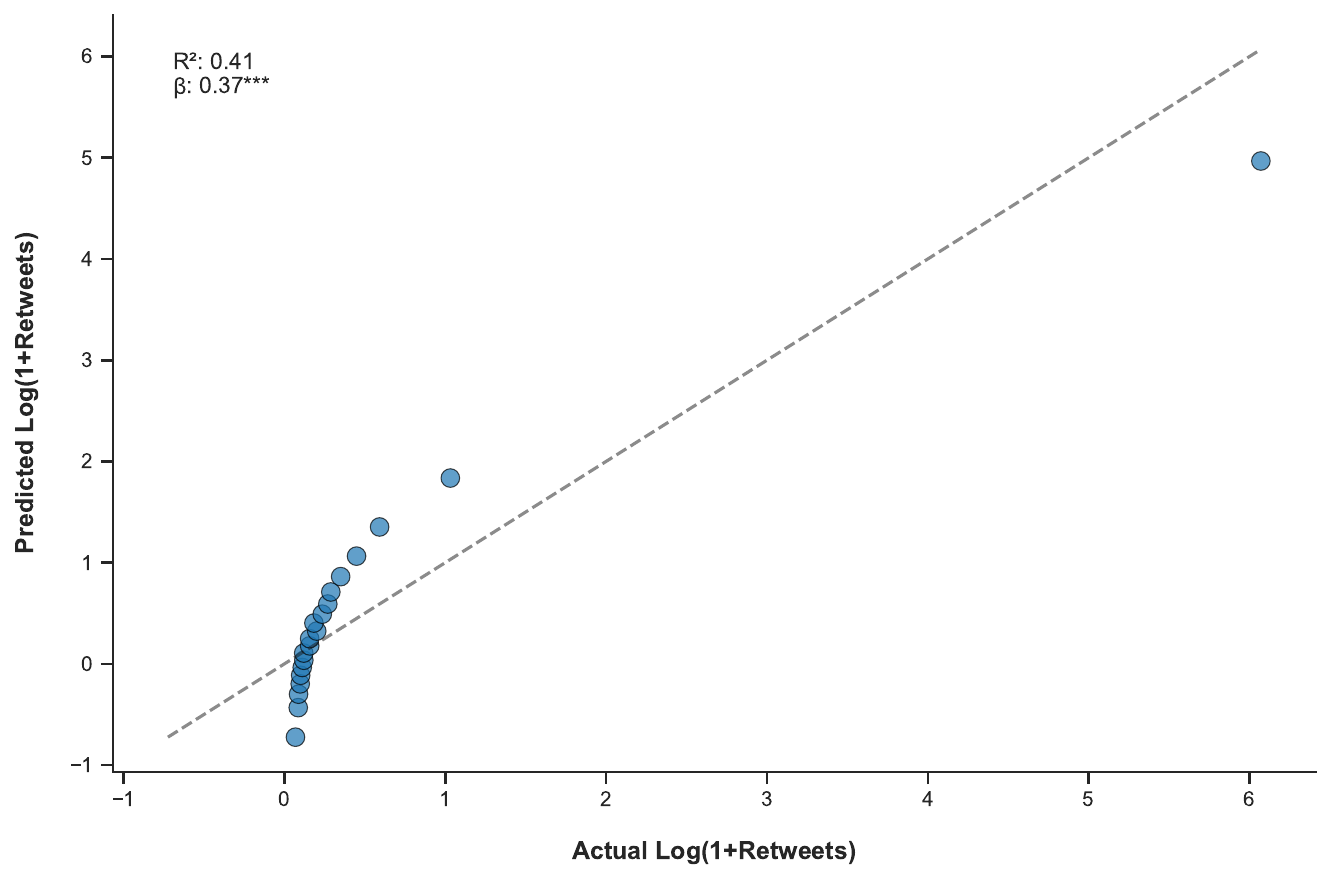}}\\

\subcaptionbox{Likes (NN)\label{fig:nn_likes}}{\includegraphics[width=0.4\textwidth]{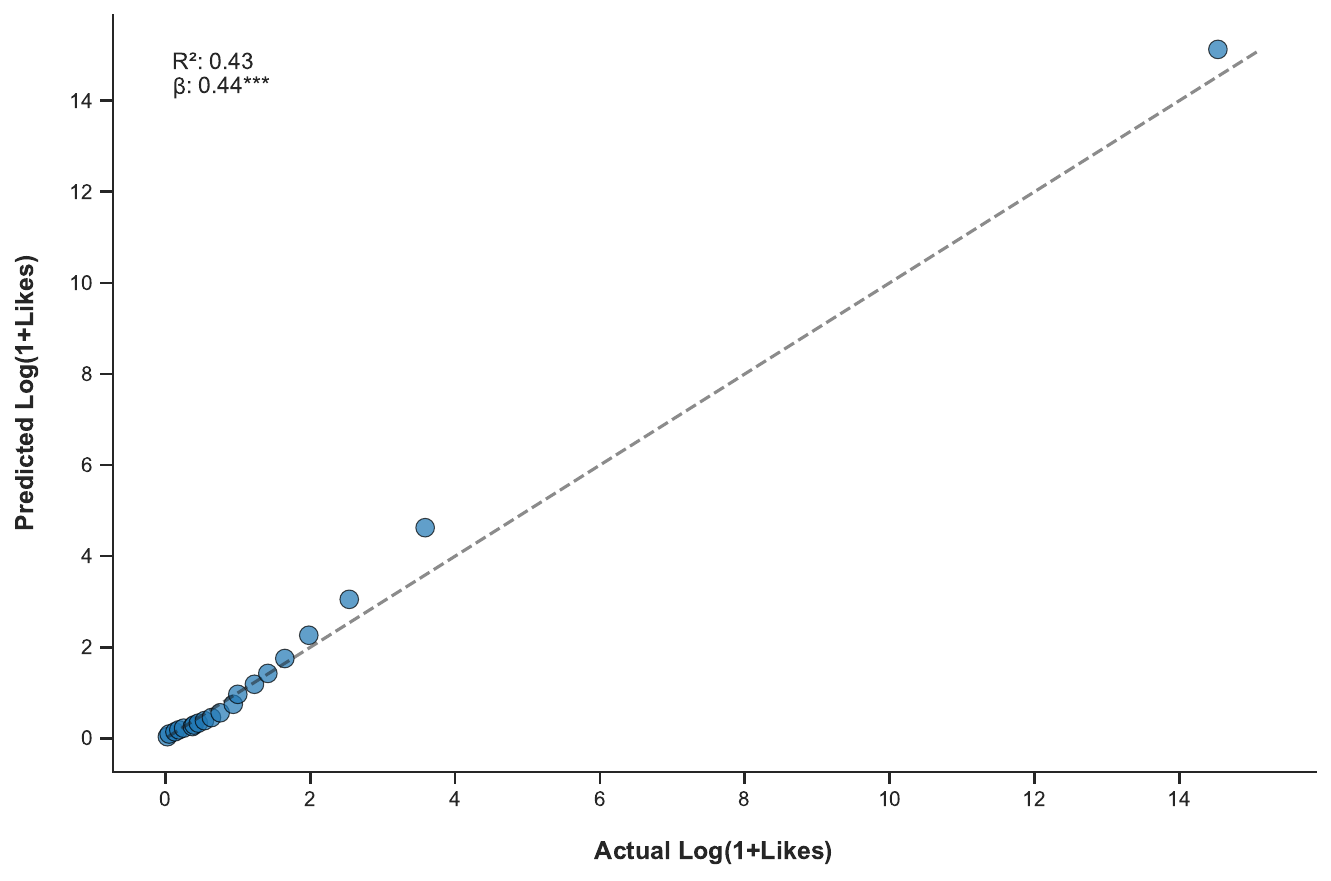}}\hfill
\subcaptionbox{Retweets (NN)\label{fig:nn_retweets}}{\includegraphics[width=0.4\textwidth]{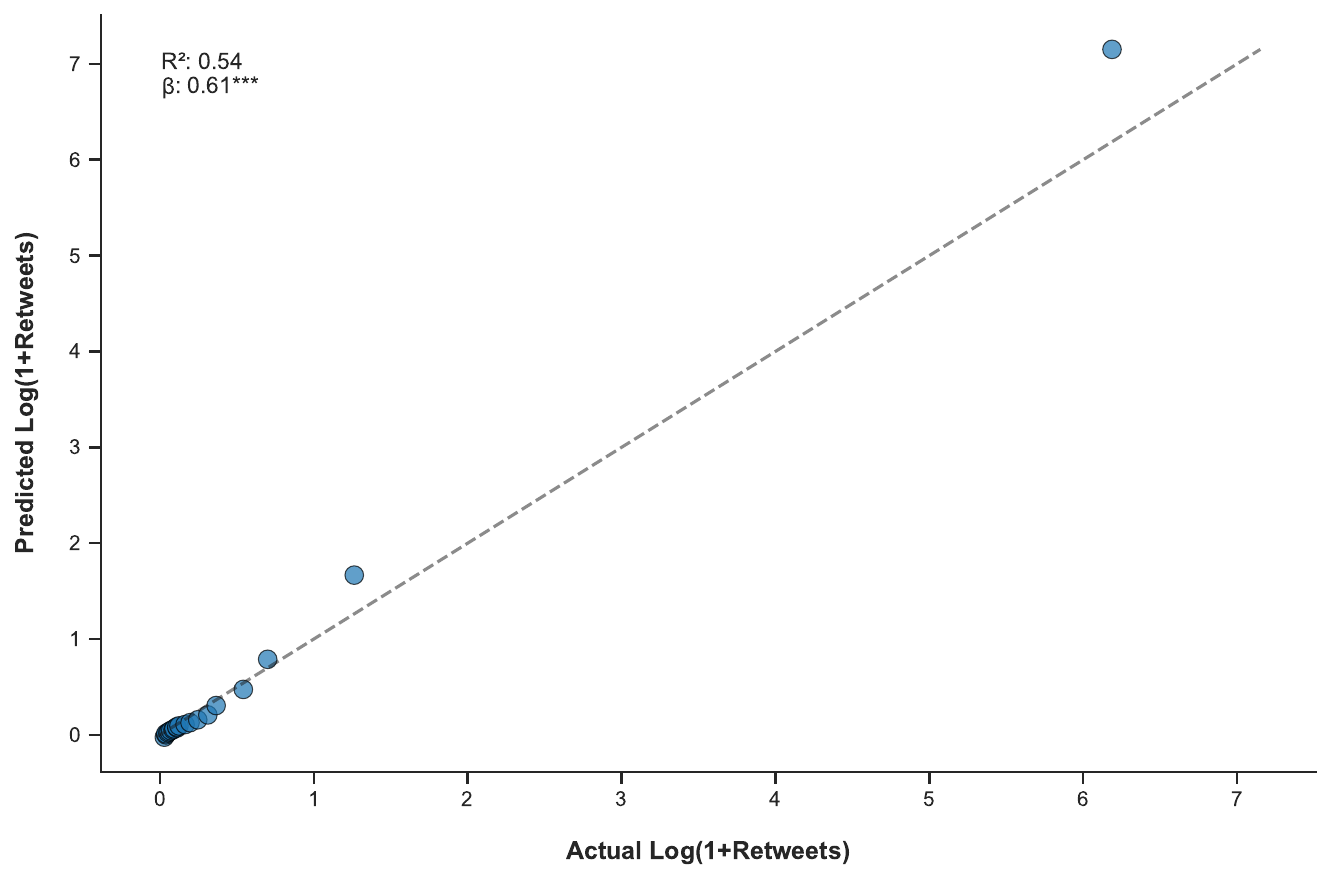}}\\

\parbox{\textwidth}{\footnotesize{\textit{Notes:} The figure displays binscatter plots comparing actual vs. predicted log-engagement for a holdout sample of 100,000 tweets. Panels (a) and (b) show results for the linear regression (OLS) model. Panels (c) and (d) show results for a neural network model featuring three hidden layers. Both models are trained on tweet embeddings and incorporate user fixed effects.}}
\end{figure}

We evaluate the predictive power of our models on a holdout sample of 100,000 Tweets. \Cref{fig:engagement_comparison} shows binscatter plots comparing the actual engagement (x-axis) and the predicted engagement (y-axis) for this holdout set. We find that both models achieve a high correlation between predicted and actual engagement. The models achieve a high $R^2$ of 0.41 for retweets and 0.28 for likes. Both plots indicate that our models effectively capture the relationship between semantic content and user engagement. 

\begin{figure}[ht]
    \centering
    \caption{Predicted Impact of Rephrasing on Engagement (Neural Network)  \label{fig:engagement_nn}}  
    \includegraphics[width=0.6\textwidth]{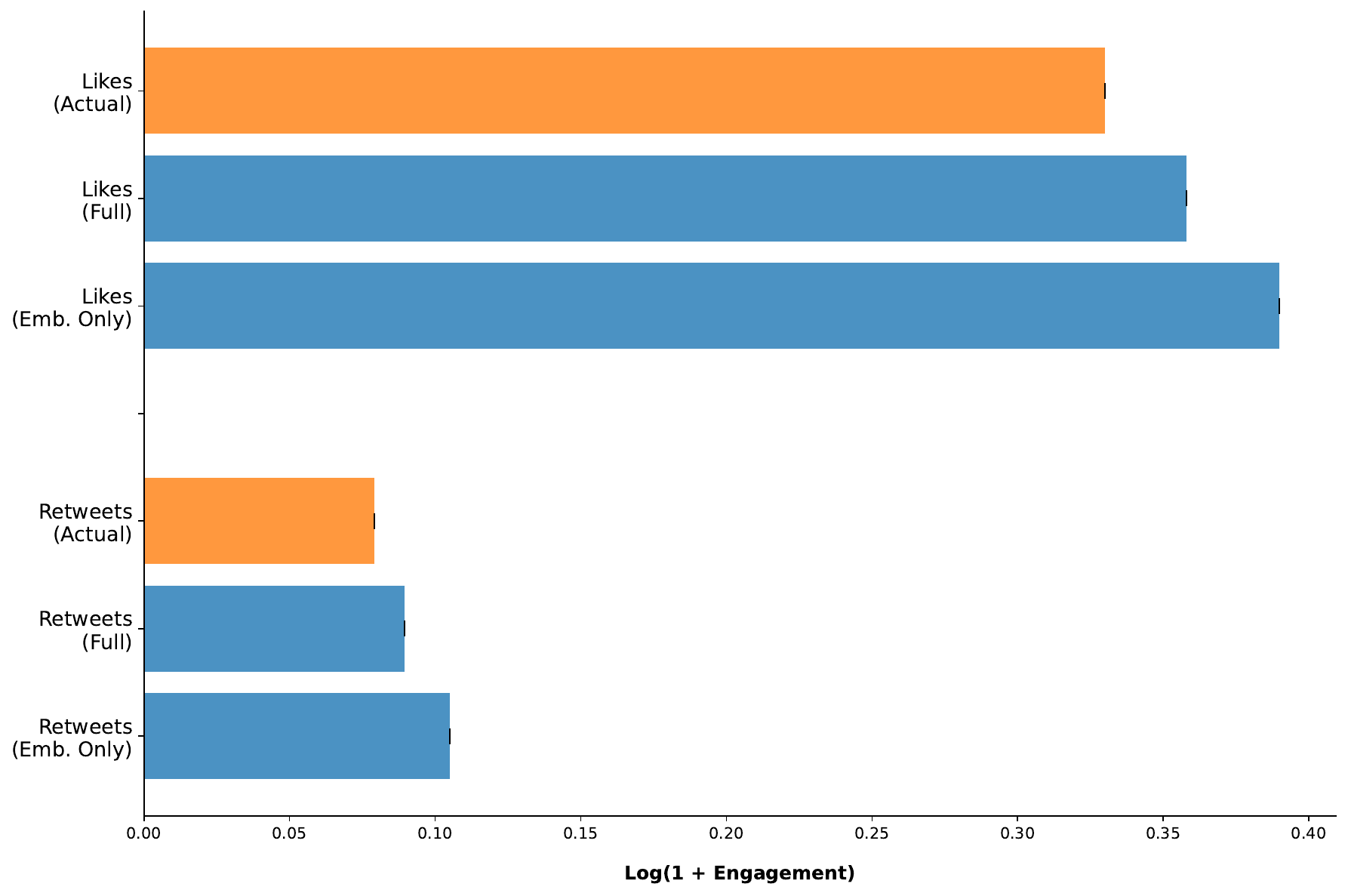}\\[0.1cm]
    \hspace{0.4cm}\parbox{\textwidth}{\footnotesize{\textit{Notes:} The figure displays the average predicted engagement for original and rephrased Tweets using a neural network model. Outcomes are transformed using $log(y+1)$ transformation. }}
\end{figure}

\begin{figure}[ht]
\centering
\caption{Raw Engagement vs. Toxicity\label{fig:engagement_tox_obs}}
\subcaptionbox{Likes\label{fig:engagement_likes}}{\includegraphics[width=0.45\textwidth]{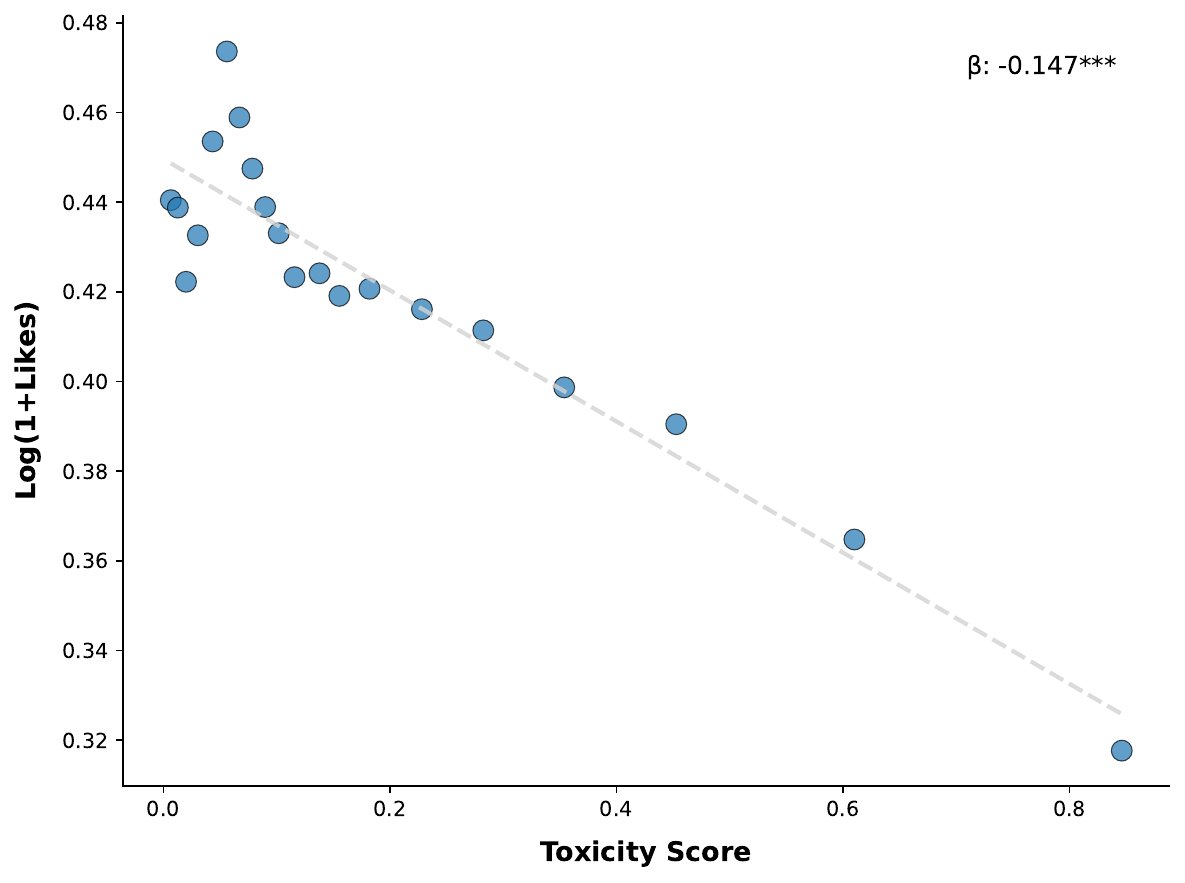}}\hfill
\subcaptionbox{Retweets \label{fig:engagement_retweets}}{\includegraphics[width=0.45\textwidth]{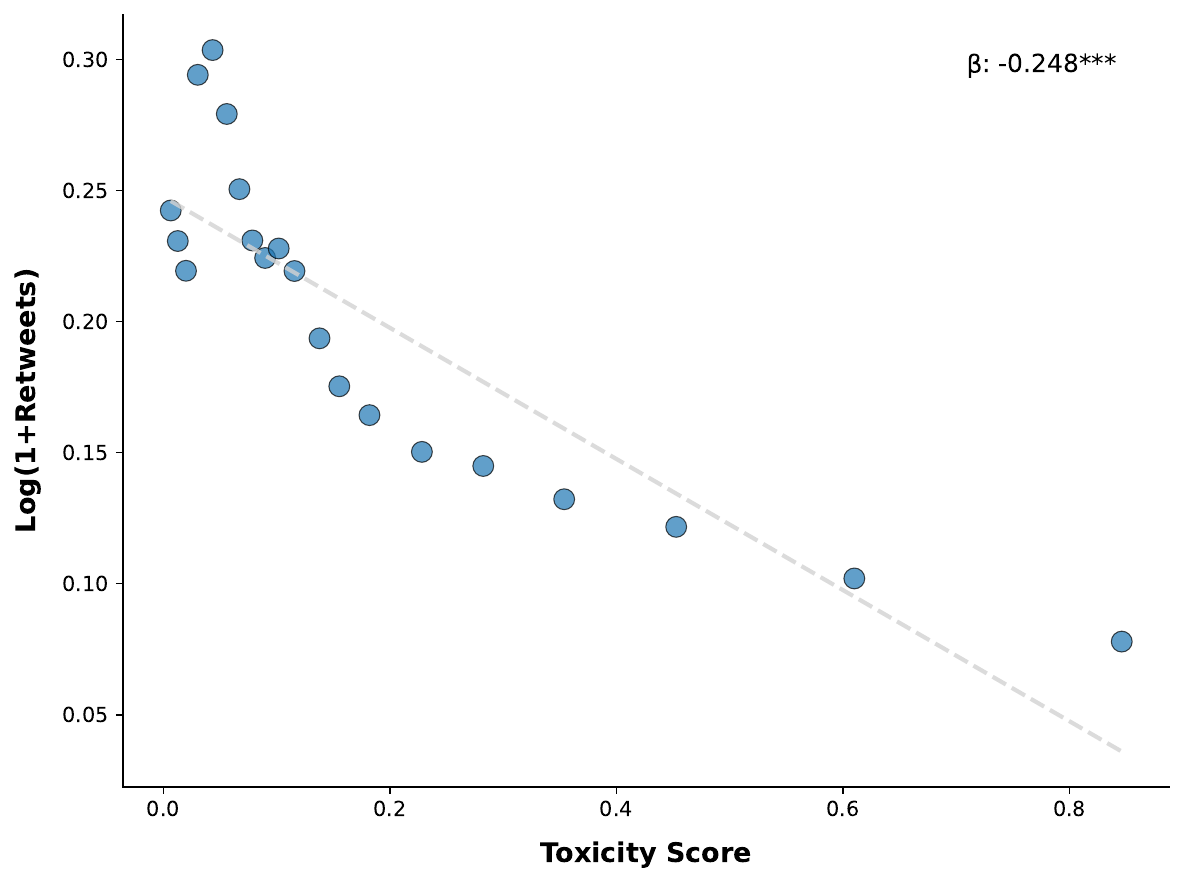}}
\hspace{0.4cm}\parbox{\textwidth}{\footnotesize{\textit{Notes:} The figure shows a binscatter plot of engagement as a function of the toxicity scores as created by the Perspective API. Engagement is transformed using $\log$ with one added inside. }}
\end{figure}

\clearpage
\section{Additional Results \label{sec:appendix_results} }
\renewcommand{\thetable}{C.\arabic{table}}\setcounter{table}{0}
\renewcommand{\thefigure}{C.\arabic{figure}}\setcounter{figure}{0}
\renewcommand{\theequation}{C.\arabic{equation}}\setcounter{equation}{0}

The following section describes additional results and robustness checks. We begin by showing that the average similarity of Tweets is not affected by content moderation and provide additional benchmarks based on the removal of topics. Further, we repeated our main analysis using alternative 1) embedding models, 2) toxicity scores, 3) engagement weighting, and 4) samples. We also demonstrate that our results are not driven by Tweets belonging to the ``Insult'' topic as created by the Top2Vec topic model. Lastly, we provide evidence from an alternative method to account for the toxicity dimension of the embedding space.

\subsubsection*{Removal of Toxic Content and Cosine Similarity}
In \Cref{fig:robustness_sim}, we report the average cosine similarity of Tweets after sequentially removing those above varying toxicity thresholds. Given the very high computational requirements of the similarity calculation, we conduct this analysis for a random subset of 500,000 Tweets. The results show that average similarity remains largely unaffected by the exclusion of toxic content. In unreported analyses, we confirm that this pattern holds when we focus on a subset of Tweets with the highest and lowest pairwise similarity.

\begin{figure}[ht]
    \centering
    \caption{Removal of Toxic Content and Average Cosine Similarity \label{fig:robustness_sim}}  
    \includegraphics[width=0.8\textwidth]{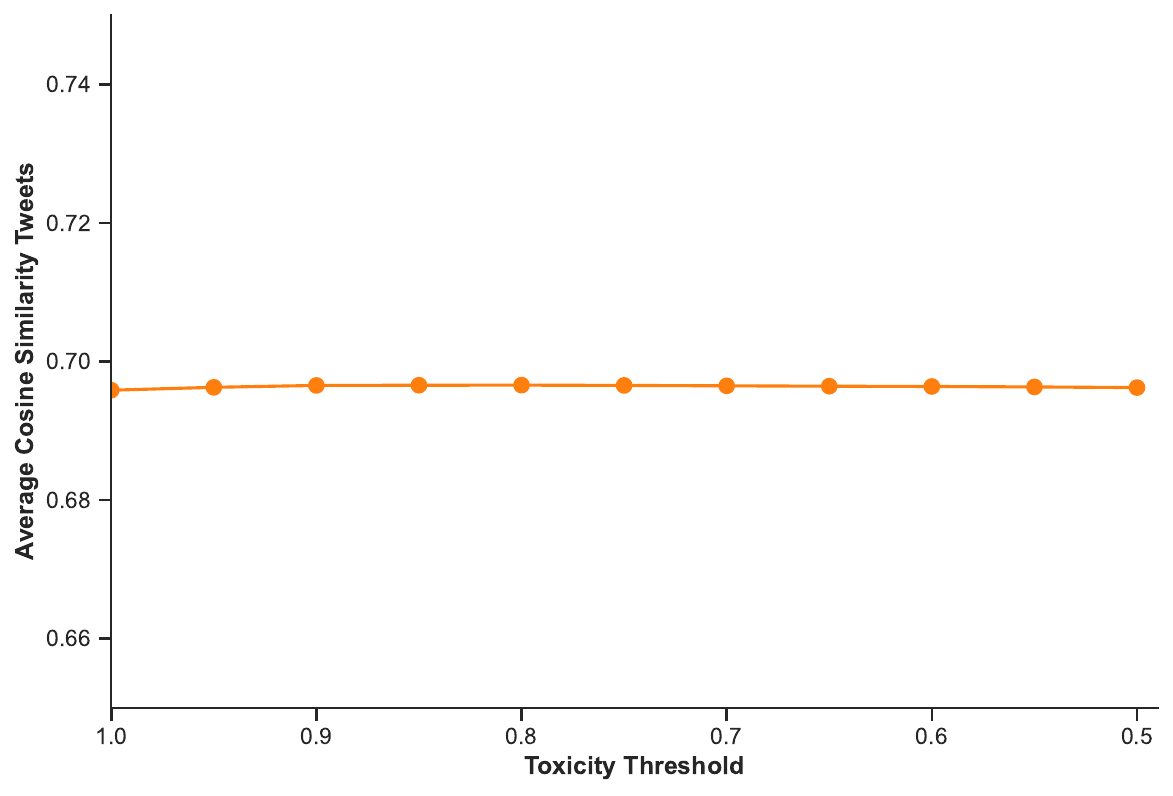}\\[0.2cm]
        \hspace{0.4cm}\parbox{\textwidth}{\footnotesize{\textit{Notes:}  The figure shows the average cosine similarity of Tweets after excluding Tweets with a toxicity score exceeding the threshold shown on the x-axis.}}
\end{figure}

\clearpage
\subsubsection*{Additional Benchmarks}

\begin{figure}[ht]
    \centering
    \caption{BCD: Removal of Individual Topics \label{fig:removal_topic_1_b_1}}
    \ContinuedFloat
    \includegraphics[width=0.66\textwidth]{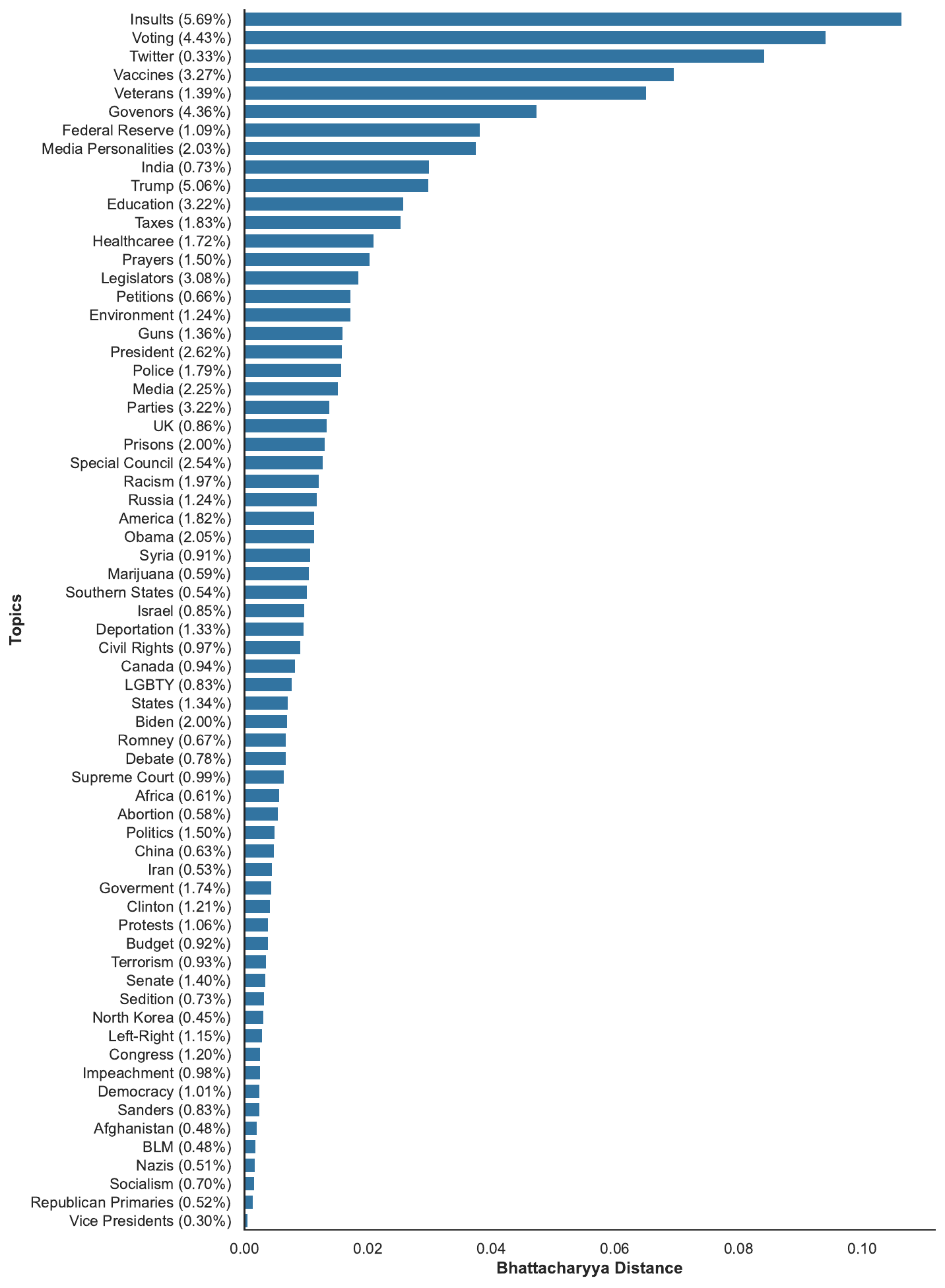}\\[0.2cm]
    \hspace{0.4cm}\parbox{\textwidth}{\footnotesize{\textit{Notes}: The figure plots the BCD obtained when each of the topics created by the Top2Vec topic model is removed from the data one by one. We also report the topic's share among all Tweets in brackets after the topic label.}}
\end{figure}

\clearpage
\subsubsection*{Alternative Embeddings}
\begin{figure}[ht]
    \centering
    \caption{Robustness: Alternative Embeddings \label{fig:robustness_embeddings}}  
    \includegraphics[width=0.6\textwidth]{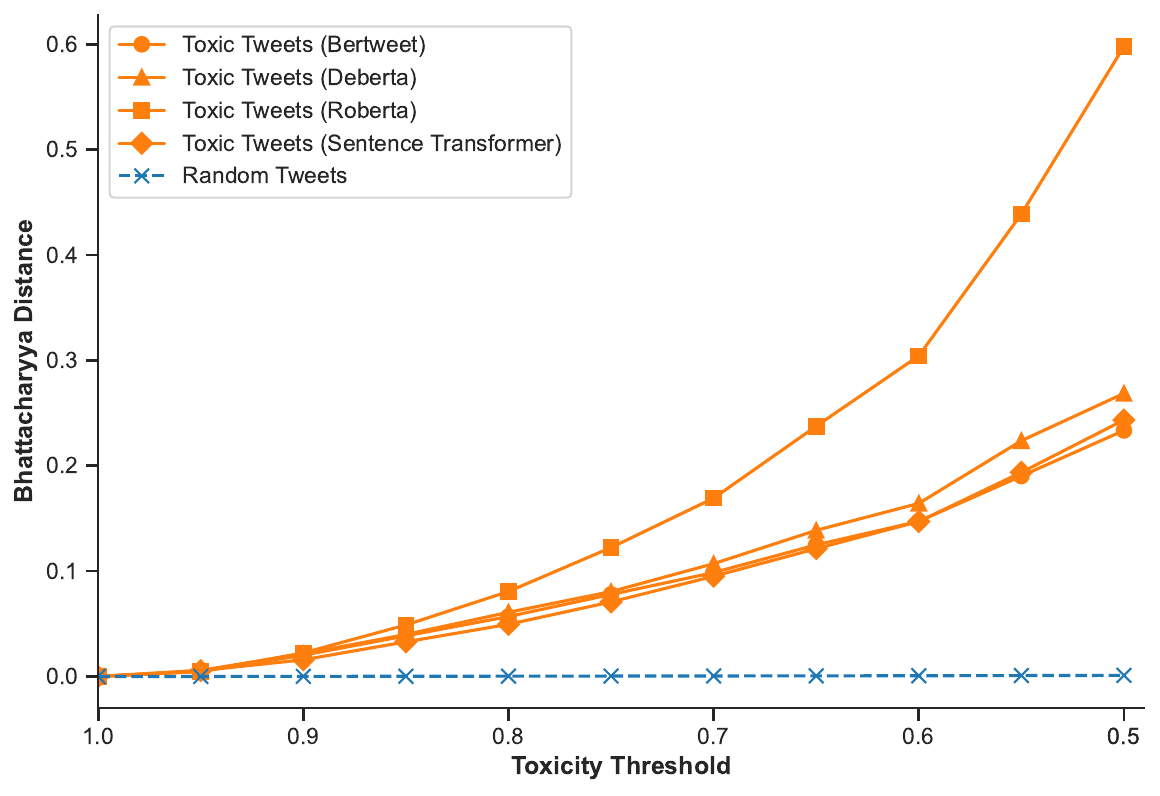}\\[0.2cm]
        \hspace{0.4cm}\parbox{\textwidth}{\footnotesize{\textit{Notes:}  The figure shows the BCD computed using embeddings generated by DeBerta and RoBerta, and DistilBert-Base-Multilingual-Cased-V2 after the exclusion of toxic and random Tweets from the data.}}
\end{figure}

As a robustness check, we reproduce our findings using the alternative embeddings described in \Cref{sec:appendix_embeddings} This test rules out that our findings are driven by the particularities of the specific transformer model we have chosen, even though BERTweet is one of the standard choices for the analysis of English-speaking Twitter data. For this robustness exercise, we created new embeddings based on the RoBERTa, DeBERTa, and DistilBert-Base-Multilingual-Cased-V2 models and reconstructed the BCD based on these embeddings. The findings in \Cref{fig:robustness_embeddings} highlight that the findings are remarkably similar not only with regards to the overall patterns but also the magnitudes of the content-moderation-induced increases in the BCD.

\subsubsection*{Alternative Toxicity Measures}
\begin{figure}[ht]
    \centering
    \caption{Robustness: Alternative Toxicity Measures\label{fig:robustness_tox_measures}}
    \subcaptionbox{Alternative Toxicity Dimension\label{fig:robustness_tox_dimensions}}{\includegraphics[width=0.45\textwidth]{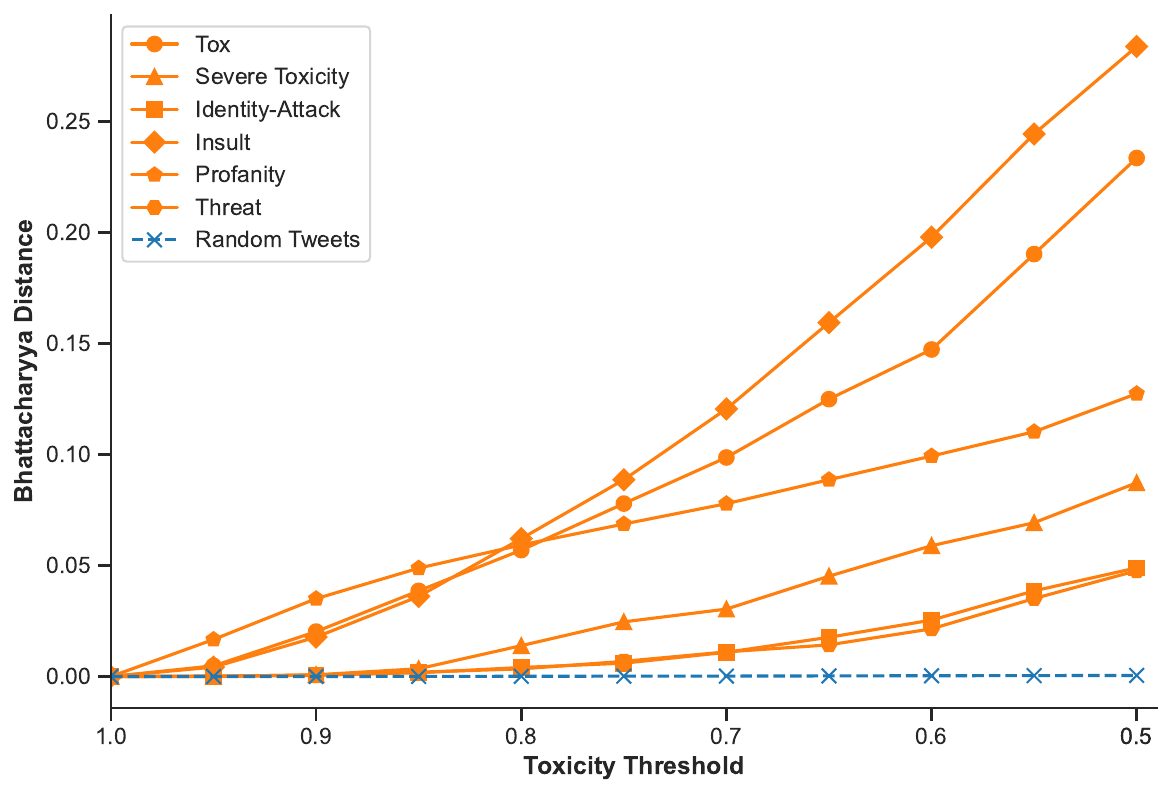}}\hfill
    \subcaptionbox{Detoxify and OpenAI Scores \label{fig:robustness_tox_detoxify}}{\includegraphics[width=0.45\textwidth]{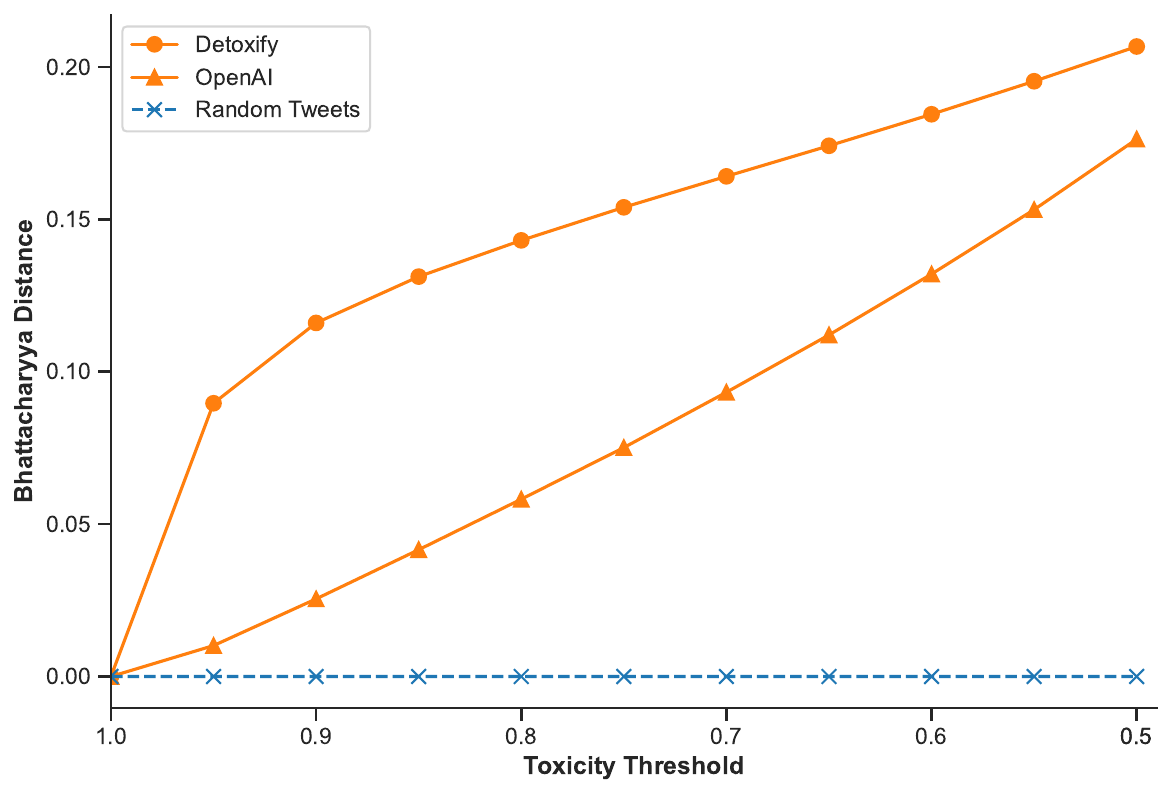}}\\[0.2cm]
    \hspace{0.4cm}\parbox{\textwidth}{\footnotesize{\textit{Notes:} Panel (a) shows the BCD following the exclusion of Tweets characterized by high levels of Severe Toxicity, Profanity, and Insult as identified by the Perspective API. Conversely, Panel (b) shows this measure after the removal of toxic Tweets, using toxicity scores generated by Detoxify and the OpenAI Moderation API.}}
\end{figure}

As another robustness check, we use the alternative Toxicity dimensions from the Perspectives API (see \Cref{fig:robustness_tox_dimensions}) as well as other toxicity scores based on the classifiers from Detoxify \citep{Detoxify} or OpenAI's Moderation API \citep{openai} (see \Cref{fig:robustness_tox_detoxify}). We find that the BCD is increasing independently of the toxicity measure that we are using. 

\subsubsection*{Engagement Weigting of Content}
\begin{figure}[ht]
    \centering
    \caption{Robustness: Engagement Weighting \label{fig:robustness_engagement}}  
    \includegraphics[width=0.54\textwidth]{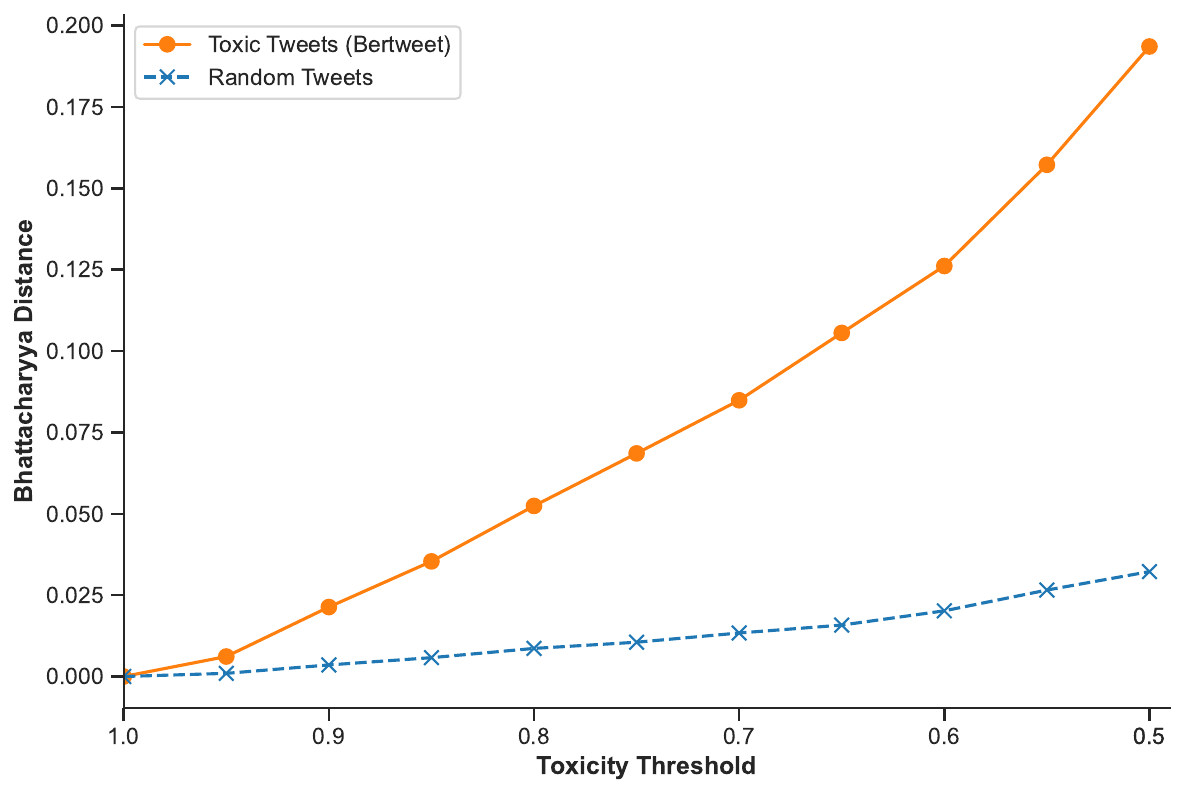}\\[0.1cm] 
    \hspace{0.4cm}\parbox{\textwidth}{\footnotesize{\textit{Notes:} The figure shows the BCD after excluding Tweets with a toxicity score exceeding the threshold shown on the x-axis. We weight Tweets by the number of retweets when calculating the BCD. The blue line illustrates the BCD when an equivalent number of Tweets is excluded from the dataset at random.}}
\end{figure}

As an additional robustness test, we weight the Tweets in our data by their engagement as measured by the number of Retweets when calculating the BCD. This allows us to arguably better account for the frequency with which users would encounter the toxic tweets. The results from this analysis are shown in \Cref{fig:robustness_engagement}. We find that weighting Tweets by their engagement has no bearing on our results and leads to very similar magnitudes for the BCD. 

\subsubsection*{Alternative Samples}

\begin{figure}[htb]
    \centering
    \caption{Bhattacharyya Distance Alternative Samples \label{fig:robustness_sample}}
    \subcaptionbox{Non-Political Tweets \label{fig:robustness_sample_all}}{\includegraphics[width=0.45\textwidth]{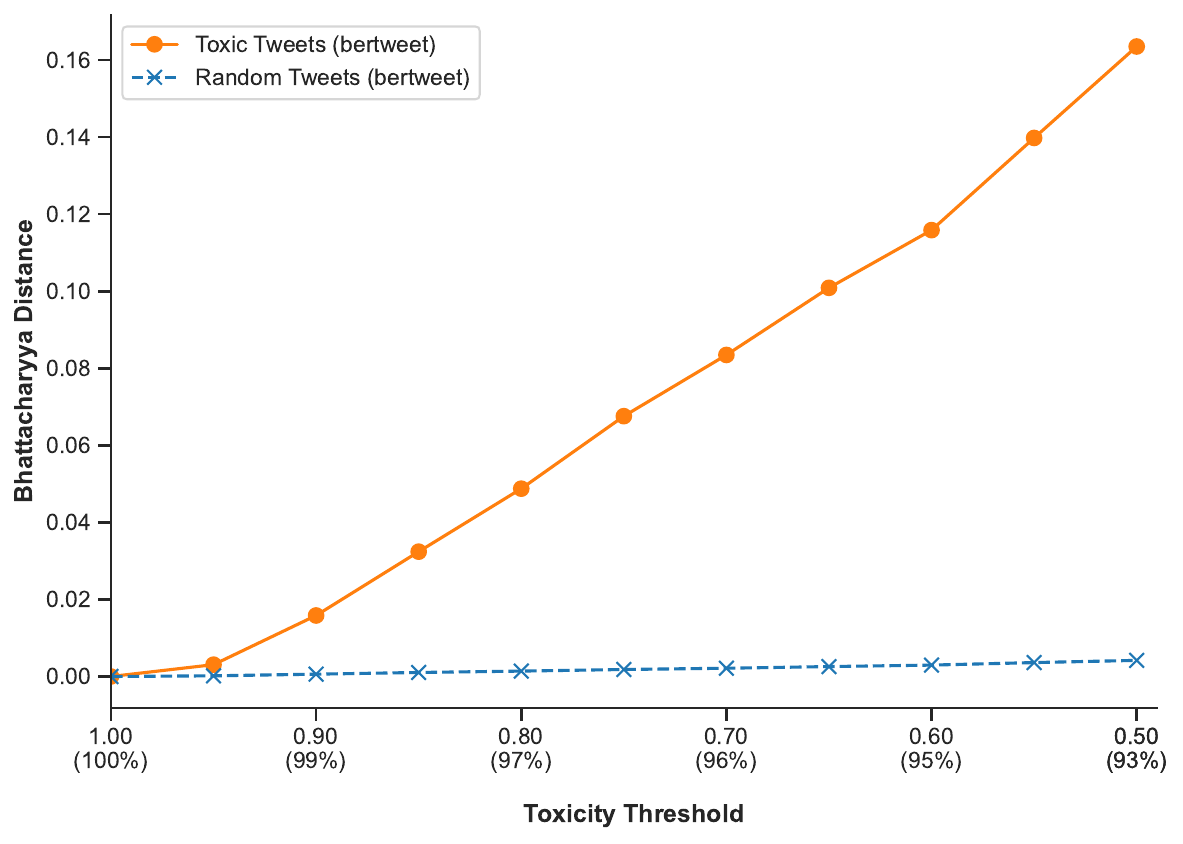}}\hfill
    \subcaptionbox{German Tweets \label{fig:robustness_sample_german}}{\includegraphics[width=0.45\textwidth]{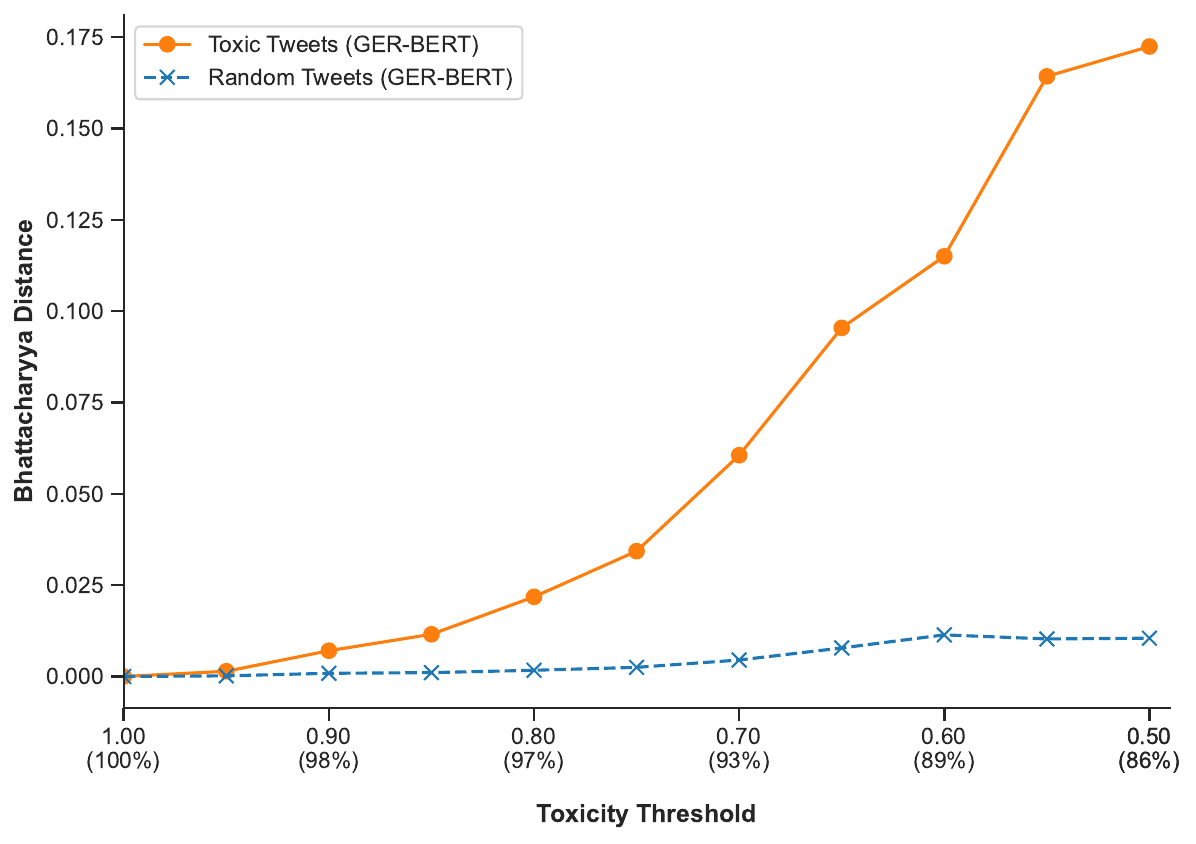}}\\[0.5cm]
    \subcaptionbox{Italian Tweets \label{fig:robustness_sample_italian}}{\includegraphics[width=0.45\textwidth]{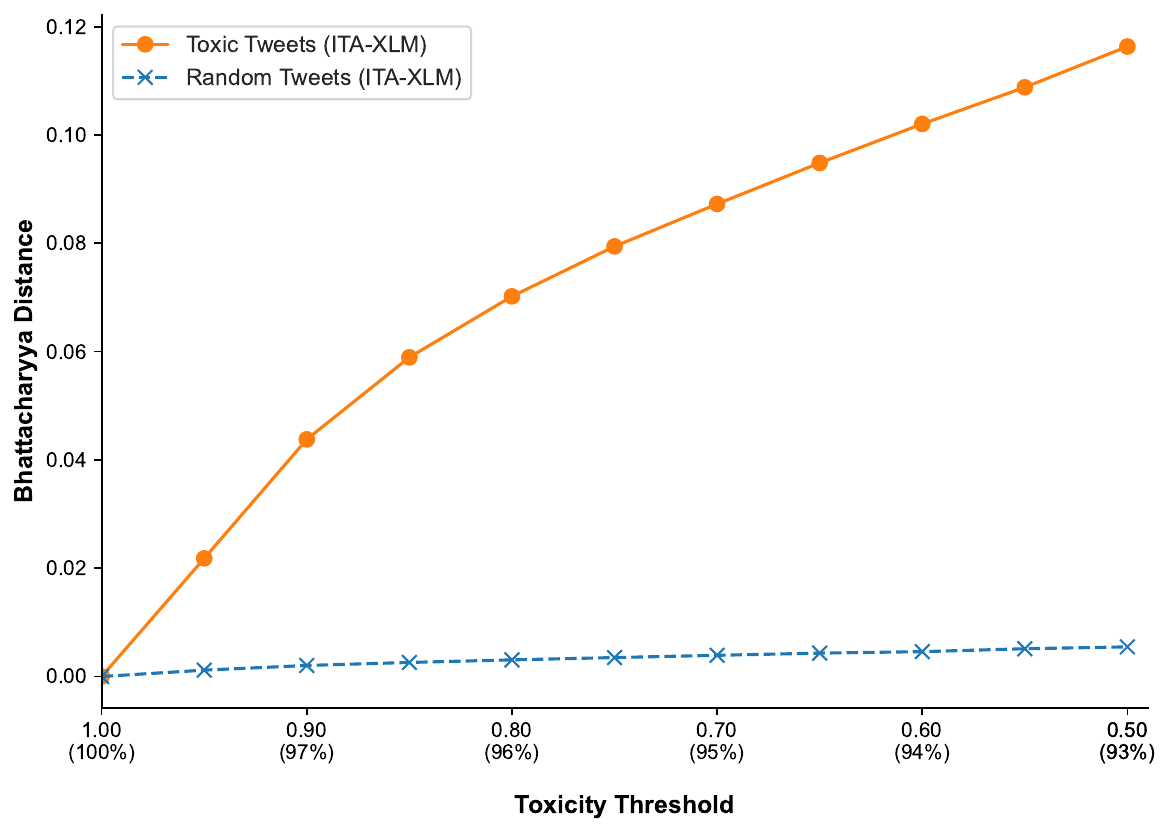}}\\[0.2cm]
    \hspace{0.4cm}\parbox{\textwidth}{\footnotesize{\textit{Notes:} The figure shows the BCD after the exclusion of toxic and random Tweets from the data. Panel (a) uses a sample of 1 million English Tweets without filtering for political content, Panel (b) uses a sample of 1 million German Tweets, and Panel (c) uses a sample of 1 million Italian Tweets.}}
\end{figure}

As a last robustness check, we repeat our main analysis based on three alternative samples of Tweets. First, we use 1 million randomly drawn Tweets from our English data without filtering for political content. Second, we use 1 million randomly drawn Tweets from a sample of politically interested German Twitter users. Third, we use 1 million randomly drawn Tweets from a sample of politically interested Italian Twitter users. To create embeddings for the German and Italian samples, we use the ``bert-base-german-uncased'' and ``twitter-xlm-roberta-base'' models, respectively. Toxicity in all cases was coded using the Perspectives API. For each of these samples, we repeat the analysis from \Cref{fig:removal_tox}. The results are presented in \Cref{fig:robustness_sample}. We find that independent sample content moderation appears to introduce distortions to the semantic space as measured by the BCD.

\subsubsection*{Excluding Tweets from the ``Insult'' Topic}

The figure below reproduces \Cref{fig:removal_tox} when we remove all Tweets belonging to the ``Insult'' topic in advance. The results are virtually identical, suggesting that content moderation also introduces distortions of the semantic space beyond simply removing uncivil content.  

\begin{figure}[ht]
    \centering
    \caption{Content Distortions and Removal of Toxic Content (No Insults Topic) \label{fig:removal_tox_no_insult}}  
    \includegraphics[width=0.6\textwidth]{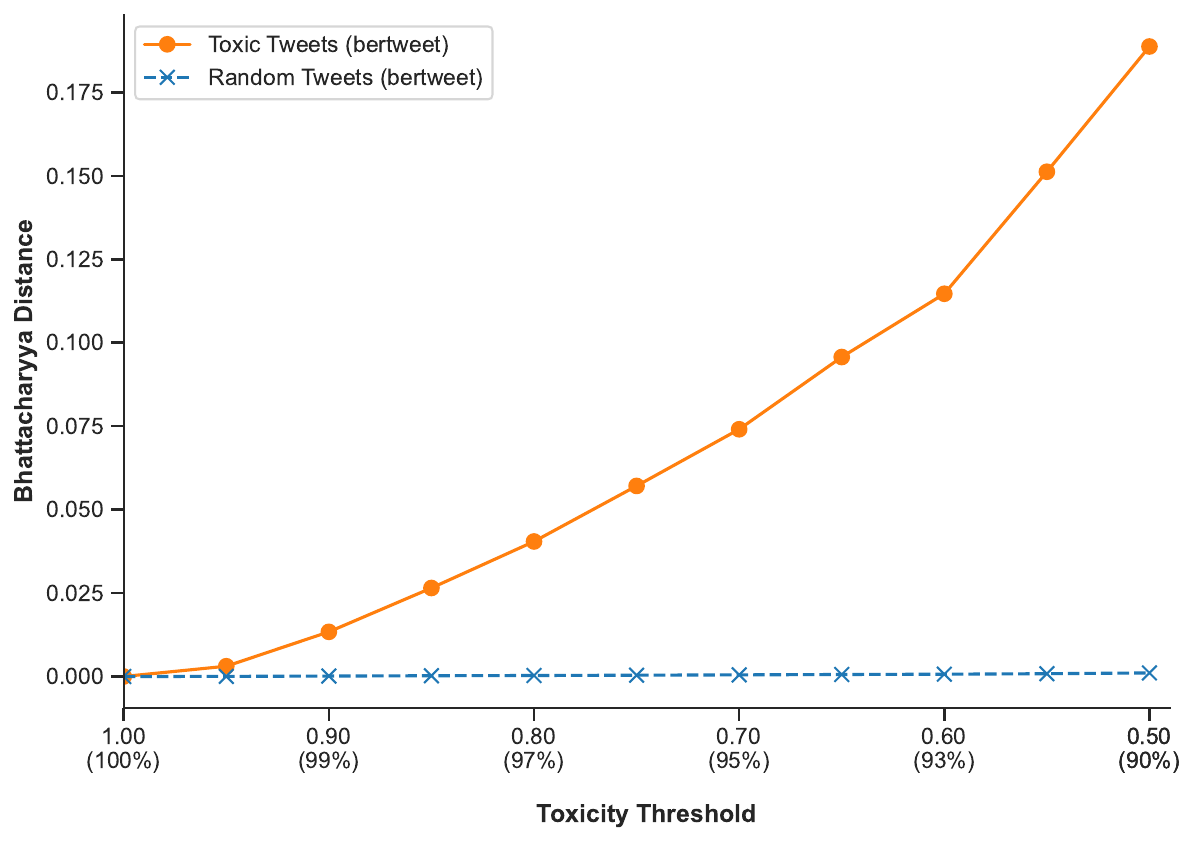}\\[0.2cm] 
    \hspace{0.4cm}\parbox{\textwidth}{\footnotesize{\textit{Notes:} The figure shows the BCD after excluding Tweets with a toxicity score exceeding the threshold shown on the x-axis. For this figure, we do not consider Tweets belonging to the ``Insult'' topic. The blue line illustrates the BCD when an equivalent number of Tweets is excluded from the dataset at random. The percentages in parentheses on the x-axis represent the proportion of Tweets retained relative to the original sample size.}}
\end{figure}

\clearpage
\subsubsection*{Additional Evidence Rephrasing}

\begin{figure}[htb]
    \centering
    \caption{Rephrasing Distortion across Top Topics \label{fig:topic_tox_bcd}}
    \includegraphics[width=0.9\linewidth]{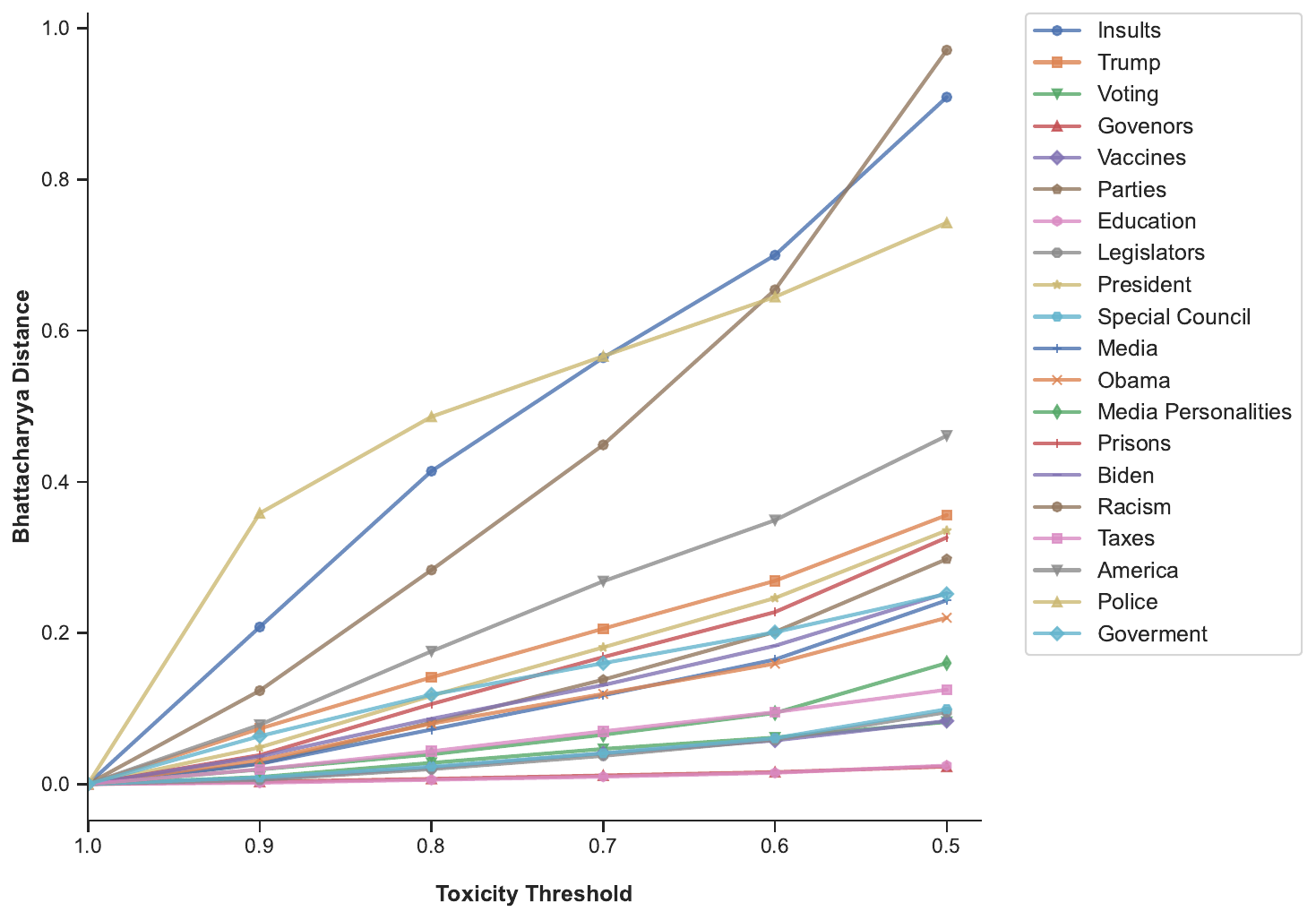}\\[0.2cm]
    \hspace{0.4cm}\parbox{\textwidth}{\footnotesize{\textit{Notes}: The figure shows the Bhattacharyya distance resulting from rephrasing tweets when conducted within the 20 most frequent topics as created by the Top2Vec topic model. The x-axis indicates the toxicity scores above which Tweets were rephrased.}}
\end{figure}

\Cref{fig:topic_tox_bcd} shows the degree of semantic distortion, measured by Bhattacharyya distance, resulting from replacing toxic tweets with their rephrased versions. The results highlight a trade-off between toxicity reduction and semantic preservation that varies by topic. Socially polarized topics with prevalent toxic content (see \Cref{fig:topic_tox_composition})(e.g., Racism, Police) suffer the greatest distortion, implying that toxicity is embedded in these specific debates. In contrast, broad political discourse (e.g., Trump, Government) shows moderate distortion, while policy-specific topics (e.g., Education, Voting) are largely unaffected. This suggests that while automated rephrasing is a viable strategy for policy debates, it poses a higher risk of semantic distortion for sensitive social discourses.

\clearpage
\subsubsection*{Alternative Method to Account for the Toxicity Dimension of the Embeddings}

Our alternative approach for the orthogonalization of the embeddings with respect to the toxicity scores uses linear regressions of the following form:
\begin{align*}
    x_d = \alpha + \mathbf{Tox'\beta} + \epsilon_d
\end{align*}
\noindent where $x_d\in \mathbf{X}$ is one of the $D$ embedding dimensions. $\mathbf{Tox}$ is a matrix containing 1000 indicators for the permilles of the toxicity distribution.\footnote{We chose a non-parametric transformation of the toxicity scores to flexibly account for any potential non-linearities. As we show in a robustness check, the findings are almost identical if we instead residualize linearly with regard to the toxicity score.} We estimate these regressions for all embedding dimensions $d\in D$ and replace the embedding dimensions with the regression residuals. The resulting residualized matrix $\mathbf{\Tilde{X}}$ is orthogonal to the toxicity scores. 

\begin{figure}[htb]
    \centering
    \caption{Controlling for Toxicity \label{fig:robustness_tox_residual}}  
    \includegraphics[width=0.6\textwidth]{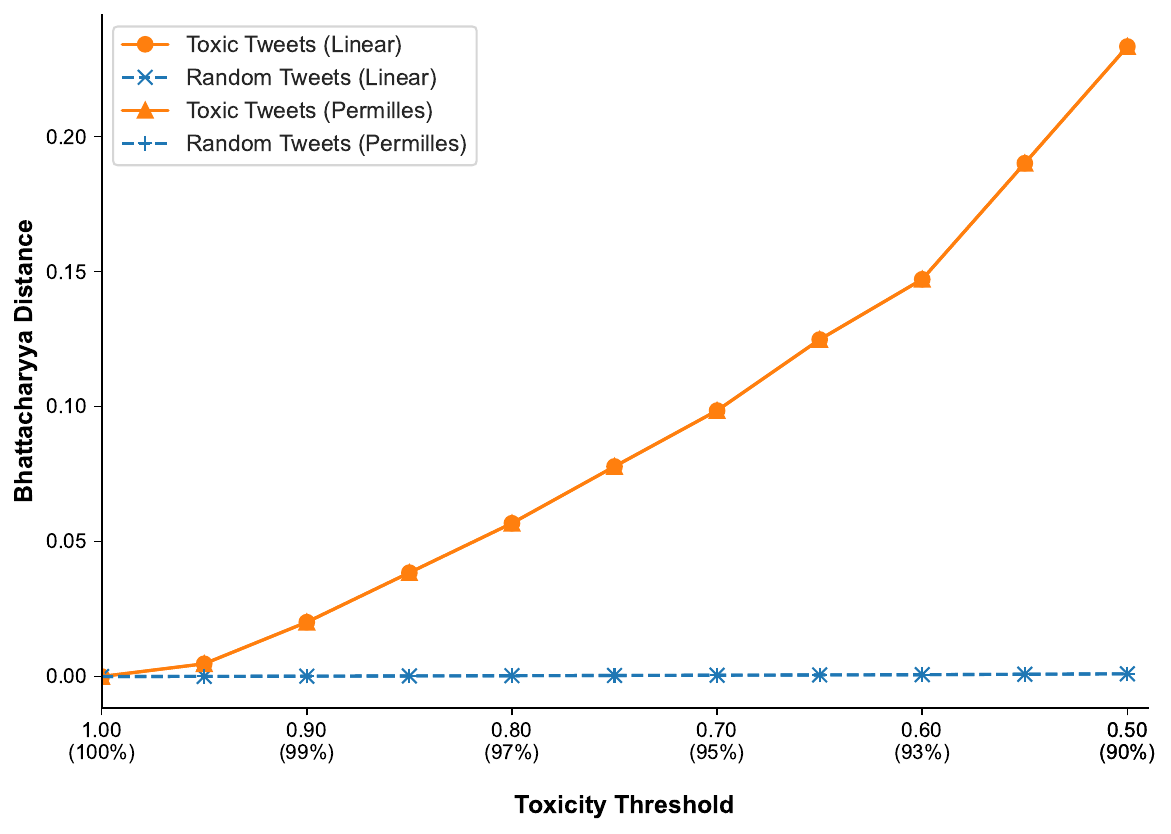}\\[0.1cm]
    \hspace{0.4cm}\parbox{\textwidth}{\footnotesize{\textit{Notes:}  The figure shows the BCD, derived from toxicity-debiased Tweet embeddings, after the exclusion of toxic and random Tweets from the sample. Linear and Percentile Residualization adjust each dimension of Tweets' embeddings by using the residuals of the regression of the embedding values against the toxicity score and toxicity permilles of Tweets. }}
\end{figure}

We then repeat our previous analysis based on the residualized embedding matrix $\mathbf{\Tilde{X}}$. \Cref{fig:robustness_tox_residual} visualizes the two different approaches to account for a Tweet's toxicity. More specifically, we use regressions and either residualize the embeddings using permilles or the linear toxicity score. We again find that removing the toxicity component from the embedding space has little impact on our findings. The overall patterns are close to our original results. The BCD sharply increases once toxic Tweets are removed from the data.

\end{document}